\documentclass[11pt, a4paper]{article}
\usepackage{jheppub}
\input{epsf.tex}
\usepackage{graphics} 
\usepackage{epsfig}
\usepackage{amsfonts}
\usepackage{amsmath}
\usepackage{transparent, colortbl}
\usepackage{booktabs}
\usepackage{multirow}
\usepackage{anysize}
\usepackage{array}
\usepackage{latexsym, amscd, amsthm}
\usepackage{pdfsync}
\usepackage{epsf}
\usepackage[all]{xy}
\usepackage[usenames,dvipsnames]{xcolor}
\usepackage{enumerate}
\usepackage{float}
\restylefloat{table}
\usepackage{upgreek}
\usepackage{subcaption}
\usepackage{array}
\usepackage{multirow}
\usepackage{graphicx,color}
\DeclareSymbolFont{extraup}{U}{zavm}{m}{n}
\DeclareMathSymbol{\varheart}{\mathalpha}{extraup}{86}
\DeclareMathSymbol{\vardiamond}{\mathalpha}{extraup}{87}
\newcommand{\bdiamond}{{\scriptstyle \vardiamond}}

\usepackage{fancyhdr}

\makeatletter
\def\CT@@do@color{%
  \global\let\CT@do@color\relax
        \@tempdima\wd\z@
        \advance\@tempdima\@tempdimb
        \advance\@tempdima\@tempdimc
        \kern-\@tempdimb
\transparent{0.6}%
        \leaders\vrule
                \hskip\@tempdima\@plus  1fill
        \kern-\@tempdimc
        \hskip-\wd\z@ \@plus -1fill }
\makeatother

\makeatletter
\newcommand{\thickhline}{%
    \noalign {\ifnum 0=`}\fi \hrule height 1pt
    \futurelet \reserved@a \@xhline
}
\newcolumntype{"}{@{\hskip\tabcolsep\vrule width 1pt\hskip\tabcolsep}}
\makeatother

\newtheorem{Theorem}{Theorem}[section]
\newtheorem{Lemma}[Theorem]{Lemma}
\newtheorem{Proposition}[Theorem]{Proposition}
\theoremstyle{definition}
\newtheorem{Definition}[Theorem]{Definition}
\theoremstyle{remark}

\newcommand{\Hom}{{\rm Hom}}

\def\ad{\mathrm{ad}}
\newcommand{\D}{\mathbb{D}}

\newcommand{\im}{\mathrm{im}}

\newcommand{\bcE}{\boldsymbol{\check{E}}}

\newcommand{\bp}{\begin{Proposition}}
\newcommand{\ep}{\end{Proposition}}
\newcommand{\bl}{\begin{Lemma}}
\newcommand{\el}{\end{Lemma}}
\newcommand{\bt}{\begin{Theorem}}
\newcommand{\et}{\end{Theorem}}
\newcommand{\bd}{\begin{Definition}}
\newcommand{\ed}{\end{Definition}}
\newcommand{\End}{\mathrm{End}}

\newcommand{\Mat}{\mathrm{Mat}}
\newcommand{\ev}{\mathrm{ev}}
\newcommand{\eqdef}{\stackrel{{\rm def.}}{=}}

\newcommand{\cinf}{{{\cal \cC}^\infty(M,\R)}}

\DeclareFontFamily{U}{rsf}{}
\DeclareFontShape{U}{rsf}{m}{n}{<5> <6> rsfs5 <7> <8> <9> rsfs7 <10-> rsfs10}{}
\DeclareMathAlphabet\Scr{U}{rsf}{m}{n}

\newcommand{\KA}{K\"{a}hler-Atiyah~}

\def\Z{\mathbb{Z}}
\def\C{\mathbb{C}}
\def\R{\mathbb{R}}

\def\H{\mathbb{H}}
\def\K{\mathbb{K}}
\def\S{\mathbb{S}}
\def\rk{{\rm rk}}

\def\dd{\mathrm{d}}

\def\vol{\mathrm{vol}}
\def\AdS{\mathrm{AdS}}

\def\w{\Omega}
\def\wf{\hat{\w}}

\def\ad{\mathrm{ad}}

\def\D{\mathbf{D}}

\newcommand{\be}{\begin{equation*}}
\newcommand{\ee}{\end{equation*}}
\newcommand{\ben}{\begin{equation}}
\newcommand{\een}{\end{equation}}
\newcommand{\beqa}{\begin{eqnarray*}}
\newcommand{\eeqa}{\end{eqnarray*}}
\newcommand{\beqan}{\begin{eqnarray}}
\newcommand{\eeqan}{\end{eqnarray}}

\newcommand{\nn}{\nonumber}

\newcommand{\id}{\mathrm{id}}

\newcommand{\tr}{\mathrm{tr}}

\newcommand{\transp}{\mathrm{transp}}

\def\cC{{\mathcal C}}
\def\cB{\Scr B}

\def\Cl{\mathrm{Cl}}

\def\cK{\mathrm{\cal K}}

\def\odd{\mathrm{odd}}
\def\Spin{\mathrm{Spin}}

\def\tcD{\tilde{\mathcal{D}}}

\def\cI{\mathcal{I}}
\def\cP{\mathcal{P}}

\def\cC{\mathcal{C}}

\def\G_2{\mathrm{G_2}}
\def\cO{\mathcal{O}}

\def\cS{\mathcal{S}}

\def\btu{\bigtriangleup}

\newcommand{\twopartdef}[4]
{
	\left\{
		\begin{array}{ll}
			#1 & \mbox{if } #2 \\
			#3 & \mbox{if } #4
		\end{array}
	\right.
}

\title{Geometric algebra techniques in flux compactifications}

\author{Calin-Iuliu Lazaroiu,$^{1}$ Elena-Mirela Babalic,$^2$
 and Ioana-Alexandra Coman$^3$ }

\affiliation{
 $^1$~Institute for Basic Science, Center for Geometry and Physics, 
Pohang 790-784, Republic of Korea\\
$^2$~Horia Hulubei National Institute for Physics and Nuclear Engineering, Department of Theoretical Physics,
Str. Reactorului no.30, P.O.BOX MG-6, Magurele 077125, Romania\\
$^3$~DESY, Theory Group, Notkestrasse 85, Bldg. 2a, D-22607 Hamburg, Germany
}

\emailAdd{calin@ibs.re.kr, mbabalic@theory.nipne.ro, ioana.coman@desy.de} 

\abstract{We study `constrained generalized Killing
(s)pinors', which characterize supersymmetric flux compactifications of
supergravity theories. Using geometric algebra techniques, we give
conceptually clear and computationally effective methods for
translating supersymmetry conditions into differential and algebraic
constraints on collections of differential forms. In particular, we
give a synthetic description of Fierz identities, which are an
important ingredient of such problems. As an application, we show how our
approach can be used to efficiently treat ${\cal N}=1$ compactifications
of M-theory on eight-manifolds and prove that we recover results previously 
obtained in the literature.}

\begin{document}

\maketitle 

\pagebreak

\vskip .6in

\section{Introduction}
\label{sec:intro}

A fundamental problem in the study of flux compactifications of M- and
string theory is to give efficient geometric descriptions of
supersymmetric backgrounds in the presence of fluxes. This leads, in
particular cases, to beautiful connections \cite{GauntlettWaldram,
Agricola} with the theory of G-structures, while in more general
situations it translates to
difficult mathematical problems involving novel geometric realizations
of supersymmetry algebras (see \cite{MartelliSparks, Tsimpis, WittGenG2,
GabellaSparks} for some examples).

When approaching this subject, one may be struck by the
somewhat ad-hoc nature of the methods usually employed, which signals
a lack of unity in the current understanding of the subject. This is
largely due to the intrinsic difficulty in finding unifying principles
while keeping computational complexity under control. In particular,
one confronts the lack of general and structurally clear descriptions
of Fierz identities, the fact that phenomena and methods which are
sometimes assumed to be `generic' turn out, upon closer inspection, to
be relevant only under simplifying assumptions and the insufficient
mathematical development of the subject of `spin geometry
\cite{SpinGeometry} in the presence of fluxes'.

The purpose of this paper is to draw attention to the fact that many
of the issues mentioned above can be resolved using ideas inspired by
a certain incarnation of the theory of Clifford bundles known as
`geometric algebra', which goes back to \cite{Chevalley} and \cite{Riesz} 
(see also \cite{Graf} and \cite{GA0, GA1, GA2} for an introduction) --- an approach
which provides a powerful language
and efficient techniques, thus affording a more unified and systematic
description of flux compactifications and of supergravity and string
compactifications in general. In particular, we show that the
geometric analysis of supersymmetry conditions for flux backgrounds
(including the algebra of those Fierz identities relevant for the
analysis) can be formulated efficiently in this language, thereby
uncovering structure whose implications have remained largely unexplored.
We mention here that our methods have a (non-trivial) 
connection with the G-structure and exceptional generalized geometry approaches, 
which were previously shown to be useful when studying flux compactifications. 
This connection will be discussed at length in a different publication.

Though the scope and applications of our approach are much wider, we
shall focus here on the study of what we call `constrained generalized
Killing (CGK) (s)pinor equations', which distill the mathematical
description of supersymmetry conditions for flux backgrounds.  A
constrained generalized Killing (s)pinor is simply a (s)pinor satisfying
conditions of the type $D_m\xi=Q_1\xi=\ldots =Q_\chi\xi=0$,
where $D_m=\nabla_m^S+A_m$ is some connection on a bundle $S$ of (s)pinors
(which generally differs from the connection $\nabla_m^S$ induced on $S$ by the
Levi-Civita connection $\nabla_m$ of the underlying pseudo-Riemannian manifold)
while $Q_j$ are some globally-defined endomorphisms of $S$.
Such equations are abundant in flux compactifications of
supergravity (see, for example, \cite{PT1,PT2}),
where $\xi$ is the internal part of a supersymmetry
generator while the equations themselves are the conditions that the
compactification preserves the supersymmetry generated by $\xi$. The
quantities $A_m$ and $Q_j$ are then certain algebraic combinations of
gamma matrices with coefficients dependent on the metric and fluxes.
An example with a single algebraic constraint $Q\xi=0$ (arising in a
compactification of eleven-dimensional supergravity) is discussed in
Section \ref{sec:application}, which the reader can consult
first as an illustration motivating the formal developments taken up
in the rest of the paper.

Using geometric algebra techniques, we show how such supersymmetry
conditions can be translated efficiently and briefly into a system of
differential and algebraic constraints for a collection of
inhomogeneous differential forms expressed as (s)pinor bilinears, thus
displaying the underlying structure in a form which is conceptually
clear as well as highly amenable to computation. The conditions which
we obtain on differential forms provide a generalization of the
well-known theory of Killing forms, which could be studied in more
depth through methods of K\"{a}hler-Cartan theory \cite{KahlerCartan}
--- even though we will not pursue that avenue in the present work.
We also touch on our implementation of this approach using various
symbolic computation systems.

As an example, Section \ref{sec:application} applies such
techniques to the study of flux
compactifications of M-theory on eight-manifolds preserving ${\cal N}=1$
supersymmetry in 3 dimensions --- a class of solutions which was analyzed
through direct methods in \cite{MartelliSparks} and \cite{Tsimpis}.  In that setting, we have a single algebraic condition
$Q\xi=0$, with $Q=\frac{1}{2}\gamma^m\partial_m\Delta
-\frac{1}{288}F_{m p q r}\gamma^{m p q r}- \frac{1}{6}f_p \gamma^p
\gamma^{(9)} -\kappa\gamma^{(9)}$ and $A_m=\frac{1}{4}f_p\gamma_{m}{}^
{p}\gamma^{(9)}+ \frac{1}{24}F_{m p q r}\gamma^{ p q r}+\kappa
\gamma_m\gamma^{(9)}$.  We show how our methods can be used to recover the
results of \cite{MartelliSparks} in a synthetic and computationally efficient
manner, while giving a more complete and general analysis. We express all equations in terms of
certain combinations of iterated contractions and wedge products which
are known as `generalized products' and whose conceptual role and
origin is explained in Section \ref{sec:KA}. The reader can, at this
point, pause to take a look at Subsection \ref{sec:applCGK}, which
should provide an illustration of the techniques developed in this
paper. 

The paper is organized as follows. In Section \ref{sec:CGK}, we define and discuss
constrained generalized Killing (s)pinors. In Section \ref{sec:KA}, we
recall the geometric algebra description of Clifford bundles as \KA
bundles while in Section \ref{sec:pin} we explain how pinor bundles are described
in this approach. Using our realization of spin geometry, Section \ref{sec:fierz}
presents a synthetic formulation of Fierz rearrangement identities for
pinor bilinears, which encodes identities involving four
pinors through certain quadratic relations holding in (a certain
subalgebra of) the \KA algebra of the underlying manifold.  We also
reformulate the constrained generalized Killing pinor equations in
this language and discuss some aspects of the differential and
algebraic structure resulted from this analysis, thereby
extending the well-known theory of Killing forms. In Section
\ref{sec:application}, we apply this formalism to the study of
${\cal N}=1$ compactifications of M-theory on eight-manifolds. We
conclude in Section \ref{sec:conclusions} with a few remarks on further directions. The
Appendices summarize various technical details and make contact with
previous work. The physics-oriented reader can start with Section
\ref{sec:application}, before delving into the technical and
theoretical details of the other sections.

\paragraph{Notations.} We let $\K$ denote one of the fields $\R$ or $\C$ of
real or complex numbers. We work in the smooth
differential category, so all manifolds, vector bundles, maps,
morphisms of bundles, differential forms etc. are taken to be
smooth. We further assume that our connected and smooth manifolds $M$
are paracompact and of finite Lebesgue dimension, 
so that we have partitions of unity of finite covering dimension subordinate to
any open cover. If $V$ is a $\K$-vector bundle over $M$, we let
$\Gamma(M,V)$ denote the space of smooth ($\cC^\infty$) sections of
$V$. We also let $\End(V)=\Hom(V,V)=V\otimes V^\ast $ denote the
$\K$-vector bundle of endomorphisms of $V$, where
$V^\ast=\Hom(V,\cO_\K)$ is the dual vector bundle to $V$ while
$\cO_\K$ denotes the trivial $\K$-line bundle on $M$. The unital ring
of smooth $\K$-valued functions defined on $M$ is denoted by
$\cinf=\Gamma(M,\cO_\K)$. The tensor product of $\K$-vector spaces
and $\K$-vector bundles is denoted by $\otimes$, while the tensor
product of modules over $\cC^\infty(M,\K)$ is denoted by
$\otimes_{\cinf}$; hence $\Gamma(M,V_1\otimes
V_2)=\Gamma(M,V_1)\otimes_{\cinf} \Gamma(M,V_2)$.  Setting $T_\K
M\eqdef T M\otimes \cO_\K$ and $T^\ast_\K M\eqdef T^\ast M\otimes
\cO_\K$, the space of $\K$-valued smooth inhomogeneous
globally-defined differential forms on $M$ is denoted by
$\Omega_\K(M)\eqdef \Gamma(M,\wedge T^\ast_\K M)$ and is a $\Z$-graded
module over the commutative ring $\cinf$.  The fixed rank
components of this graded module are denoted by
$\Omega^k_\K(M)=\Gamma(M,\wedge^k T^\ast_\K M)$ ($k=0\ldots d$, where
$d$ is the dimension of $M$).

The kernel and image of any $\K$-linear map
$T:\Gamma(M,V_1)\rightarrow \Gamma(M,V_2)$ will be
denoted by $\cK(T)$ and $\cI(T)$; these are $\K$-linear subspaces of
$\Gamma(M,V_1)$ and $\Gamma(M,V_2)$, respectively.  In the particular
case when $T$ is a $\cinf$-linear map (i.e. when it is a morphism of
$\cinf$-modules), the subspaces $\cK(T)$ and $\cI(T)$ are 
$\cinf$-submodules of $\Gamma(M,V_1)$ and $\Gamma(M,V_2)$,
respectively --- even in those cases when $T$ is not induced by any
bundle morphism from $V_1$ to $V_2$. We always denote a morphism
$f:V_1\rightarrow V_2$ of $\K$-vector bundles and the 
$\cinf$-linear map $\Gamma(M,V_1)\rightarrow \Gamma(M,V_2)$ induced by
it between the modules of sections by the same
symbol. Because of this convention, we clarify that the notations
$\cK(f)\subset \Gamma(M,V_1)$ and $\cI(f)\subset \Gamma(M,V_2)$ denote
the kernel and the image of the corresponding map on sections
$\Gamma(M,V_1)\stackrel{f}{\rightarrow} \Gamma(M,V_2)$, which in this
case are $\cinf$-submodules of $\Gamma(M,V_1)$ and $\Gamma(M,V_2)$,
respectively.  In general, there does {\em not} exist any sub-bundle
$\ker f$ of $V_1$ such that $\cK(f)=\Gamma(M,\ker f)$ nor any
sub-bundle $\im f$ of $V_2$ such that $\cI(f)=\Gamma(M,\im f)$ ---
though there exist sheaves $\ker f$ and $\im f$ with the corresponding
properties.

Given a pseudo-Riemannian metric $g$ on $M$ of signature $(p,q)$,
we let $(e_a)_{a=1\ldots d}$ (where $d=\dim M$) denote a local frame
of $T M$, defined on some open subset $U$ of $M$. We let $e^a$ be the
dual local coframe ($=$ local frame of $T^\ast M$), which satisfies
$e^a(e_b)=\delta^a_b$ and ${\hat g}(e^a,e^b)=g^{ab}$, where $(g^{ab})$
is the inverse of the matrix $(g_{ab})$. The contragradient frame
$(e^a)^\sharp$ and contragradient coframe $(e_a)_\sharp$ are given by:
\be
(e^a)^\sharp=g^{a b}e_b~~,~~(e_a)_\sharp=g_{ab}e^b~~,
\ee
where the $\sharp$ subscript and superscript denote the (mutually
inverse) musical isomorphisms between $T_\K M$ and $T^\ast_\K M$ given
respectively by lowering and raising indices with the metric $g$.  We
set $e^{a_1\ldots a_k}\eqdef e^{a_1}\wedge \ldots \wedge e^{a_k}$ and
$e_{a_1\ldots a_k}\eqdef e_{a_1}\wedge \ldots \wedge e_{a_k}$ for any
$k=0\ldots d$. A general $\K$-valued inhomogeneous form $\omega\in
\Omega_\K(M)$ expands as:
\ben
\label{FormExpansion}
\omega=\sum_{k=0}^d\omega^{(k)}=_{U}\sum_{k=0}^{d}\frac{1}{k!}\omega^{(k)}_{a_1\ldots
a_k} e^{a_1\ldots a_k}~~,
\een
where the symbol $=_{U}$ means that the equality holds only after
restriction of $\omega$ to $U$ and where we used the expansion:
\ben
\label{HomFormExpansion}
\omega^{(k)}=_U\frac{1}{k!}\omega^{(k)}_{a_1\ldots a_k} e^{a_1\ldots a_k}~~.
\een
The locally-defined smooth functions $\omega^{(k)}_{a_1\ldots a_k}\in
\cC^\infty(U,\K)$ (the `strict coefficient functions' of $\omega$) are
completely antisymmetric in $a_1\ldots a_k$. Given a pinor bundle on
$M$ with underlying fiberwise representation $\gamma$ of the Clifford
bundle of $T^\ast_\K M$, the corresponding gamma `matrices' in the
coframe $e^a$ are denoted by $\gamma^a\eqdef \gamma(e^a)$, while the
gamma matrices in the contragradient coframe $(e_a)_\sharp$ are denoted
by $\gamma_a\eqdef \gamma((e_a)_\sharp)=g_{ab}\gamma^b$. We will
occasionally assume that the frame $(e_a)$ is {\em pseudo-orthonormal}
in the sense that $e_a$ satisfy:
\be
g(e_a,e_b)~(=g_{ab})~=\eta_{ab}~~,
\ee
where $(\eta_{ab})$ is a diagonal matrix with $p$ diagonal
entries equal to $+1$ and $q$ diagonal entries equal to $-1$.

\section{Constrained generalized Killing (s)pinors}
\label{sec:CGK}

\paragraph{The basic set-up.}
Let $(M,g)$ be a connected pseudo-Riemannian manifold (assumed to be smooth and
paracompact) of dimension $d=p+q$, where $p$ and $q$ are,
respectively, the numbers of positive and negative eigenvalues of
$g$. We endow the cotangent bundle $T^\ast M$ with the metric ${\hat
g}$ induced by $g$. Setting $\K=\R$ or $\C$, we similarly endow the bundle
$T^\ast _\K M \eqdef T^\ast M\otimes  \cO_\K$ with the metric ${\hat g}_\K$
induced by extension of scalars. Of course, we have
$T^\ast _\R M=T^\ast M$ and ${\hat g}_\R={\hat g}$.  Let $\Cl(T_\K^\ast M)=\Cl(T^\ast
M)\otimes \cO_\K$ be the Clifford bundle defined by $T^\ast_\K M$ --- when the latter
is endowed with the metric given above. The fiber of $\Cl(T^\ast_\K M)$ at a point $x\in
M$ is the Clifford algebra $\Cl(T^\ast_{\K,x} M)=\Cl(T^\ast_{x} M)\otimes_\R \K$ of the quadratic
vector space $(T^\ast_{\K,x}, {\hat g}_{\K,x})$, where $T^\ast_{\K,x}\eqdef T^\ast_x M\otimes_\R \K$
and ${\hat g}_{\K,x}$ denotes the $\K$-valued bilinear pairing induced by ${\hat g}_x$.
The {\em even Clifford bundle} $\Cl^\ev(T_\K^\ast M)$ over $\K$ is the
sub-bundle of algebras of $\Cl(T^\ast_\K M)$ whose fibers are the even
subalgebras $\Cl^\ev (T^\ast_{\K,x} M)\subset \Cl(T^\ast_{\K,x} M)$.
Our point of view on (s)pinor bundles is that taken in
\cite{Trautman}.  Namely, we define a bundle of {\em $\K$-pinors} over
$M$ to be a $\K$-vector bundle $S$ over $M$ which is a bundle of
modules over the Clifford bundle $\Cl(T^\ast_\K M)$. Similarly, a
bundle of {\em $\K$-spinors} is a bundle of modules over the even
Clifford bundle $\Cl^\ev(T^\ast_\K M)$. Of course, a bundle of
$\K$-pinors is automatically a bundle of $\K$-spinors. Hence any pinor
is naturally a spinor but the converse need not hold. In this paper,
we focus on the case of {\em pinors}. A pinor bundle $S$ will be
called a {\em pin bundle} if the underlying fiberwise representation
of $\Cl(T^\ast_\K M)$ is irreducible, i.e. if each of the fibers of
$S$ is a simple module over the corresponding fiber of the Clifford
bundle. Similarly, a {\em spin bundle} is a spinor bundle for which
the underlying fiberwise representation of $\Cl^\ev(T^\ast_\K M)$ is
irreducible. Later on, we shall sometimes denote $g_\K$ by $g$ etc. 
in order to simplify notation.

\paragraph{Remark.} Physics terminology is often
imprecise with the distinction between spinors and pinors which we are
making here and throughout this paper.  Physically, one typically
assumes that $(M,g)$ is both oriented and time-oriented and one is
concerned with objects transforming in representations of the
orthochronous part $\Spin^\uparrow(p,q)$ of the spin group
$\Spin(p,q)$ and thus in vector bundles associated with a principal
bundle with fiber $\Spin^\uparrow(p,q)$ which is a double cover of the
principal ${\rm SO}^\uparrow(p,q)$-bundle consisting of those
pseudo-orthonormal frames of $(M,g)$ which are both oriented and
time-oriented.  Due to the issue of time-orientability, what matters
in many physics applications is not a spin structure in the standard
mathematical sense (see \cite{Sati} for a recent discussion with
applications to string theory) but rather a `time-oriented' spin
structure.

\paragraph{Constrained generalized Killing (s)pinors.}
Let us fix a $\K$-pinor or $\K$-spinor bundle $S$ over $M$, a
linear connection $D$ on $S$ and a finite collection of bundle endomorphisms
$Q_1,\ldots, Q_\chi\in \Gamma(M,\End(S))$.

\paragraph{Definition.} A {\em constrained generalized Killing (CGK) (s)pinor}
over $M$ is a section $\xi\in \Gamma(M,S)$ which satisfies the {\em
constrained generalized Killing (s)pinor equations}
$D\xi=Q_1\xi=\ldots =Q_\chi\xi=0$. We say that $D\xi=0$ is the {\em
$D$-flatness} or {\em generalized Killing (GK) (s)pinor} equation
satisfied by $\xi$ while $Q_1\xi=\ldots =Q_\chi \xi=0$ are the
{\em algebraic constraints} (or {\em $Q$-constraints}) satisfied by $\xi$.

\

\noindent When the algebraic constraints are trivial ($\chi=0$ or, equivalently,
when all $Q_j$ vanish), one deals with the generalized Killing (GK)
spinor equation $D\xi=0$. Since $D$ can be written as the sum
$\nabla^S+A$ of the spinorial connection $\nabla^S$ induced on $S$ by
the Levi-Civita connection of $(M,g)$ and an $\End(S)$-valued one-form
$A$ on $M$, the GK (s)pinor equations can be viewed as a deformation of
the {\em parallel (s)pinor} equation $\nabla^S\xi=0$, the deformation
being parameterized by $A$. Our terminology is inspired by the fact
that the choice $A_m=-\lambda \gamma_m$ (with $\lambda$ a real
parameter and $\gamma^m\in \Gamma(M,\End(S))$ the gamma `matrices' in
some local coframe of $(M,g)$) leads to the ordinary Killing (s)pinor
equations $\nabla_m\xi=\lambda \gamma_m\xi$. 

\paragraph{Remark.} In flux compactifications of supergravity, 
backgrounds admitting constrained generalized Killing spinors can be used 
to construct supersymmetric compactifications, provided that the equations 
of motion for all fields present in the backgound are also satisfied. 

\paragraph{Connection to supergravity and string theories.}
Constrained generalized Killing (s)pinors arise naturally in
supergravity and string theory.  In particular, they arise in
supersymmetric flux compactifications of string theory, M-theory and
various supergravity theories. In such setups, $\xi$ is a (s)pinor of
spin $1/2$ defined on the background pseudo-Riemannian manifold and
corresponds to the generator of supersymmetry transformations of the
underlying supergravity action (or string theory effective action)
while the constrained generalized Killing (s)pinor equations are the
conditions that the supersymmetry generated by $\xi$ is preserved by
the background. The connection $D$ on $S$ and the endomorphisms $Q_j$
are fixed by the precise data of the background, i.e. by the metric
and fluxes defining that background. For example, the supersymmetry
equations of eleven-dimensional supergravity involve the
supercovariant connection $D$, which acts on sections of the bundle
$S$ of Majorana spinors (a.k.a. real pinors) defined in eleven dimensions --- this
corresponds to the differential constraint $D\xi=0$, without any
algebraic constraint.  When considering a compactification of
eleven-dimensional supergravity down to a lower-dimensional space
admitting Killing (s)pinors, the internal part (which now plays the role
of $\xi$) of the generator of the supersymmetry variation is a section
of some bundle of (s)pinors (which now plays the role of $S$) defined
over the internal space, while the condition of preserving the
supersymmetry generated by the tensor product of this internal
generator and some Killing (s)pinor of the non-compact part of the
background induces a differential (generalized Killing) constraint as
well as an algebraic constraint for the internal part of the
supersymmetry generator. A specific example arising from
eleven-dimensional supergravity is discussed in Section
\ref{sec:application} below. Similarly, the supersymmetry equations
for IIA supergravity in ten dimensions (with Minkowski signature) can
be written in terms of a supercovariant connection $D$ defined on the
real vector bundle $S$ of Majorana spinors (a.k.a. real pinors) in ten
dimensions and an endomorphism $Q$ of $S$; we have $S=S^+\oplus S^-$
where $S^\pm$ are the bundles of Majorana-Weyl spinors of positive and
negative chirality.  The condition $D\xi=0$ for the supersymmetry
generator (a section $\xi$ of $S$, with positive and negative
chirality components $\xi_\pm$ --- which are sections of $S^\pm$ ---
such that $\xi=\xi_++\xi_-$) is the requirement that the supersymmetry
variation of the gravitino vanishes, while the condition $Q\xi=0$
encodes vanishing of the supersymmetry variation of the dilatino. When
considering compactifications on some internal space down to some
space admitting Killing (s)pinors, $\xi$ is replaced by its internal
part (a section of some (s)pinor bundle --- which now plays the role of
$S$ --- defined on the compactification space) while $D$ induces a
connection defined on this internal (s)pinor bundle as well as a further
algebraic constraint --- thereby leading once again to a system of
equations of constrained generalized Killing type, which is now
defined on the internal space.  Finally, the supersymmetry equations
for type IIB supergravity in ten dimensions (with Minkowski signature)
can be formulated\footnote{They can, of course, also be formulated
using {\em complex} Weyl spinors.} (see, for example, \cite{PT2}) in
terms of sections of the real vector bundle $S=S^+\oplus S^+$ of
Majorana-Weyl spinor doublets, with a supercovariant connection $D$
defined on this bundle as well as two endomorphisms $Q_1,Q_2$ of $S$.
The condition $D\xi=0$ for sections $\xi$ of $S$ is the requirement
that the supersymmetry variation of the gravitino vanishes in the
background, while the conditions $Q_1\xi=Q_2\xi=0$ are, respectively,
the requirements that the supersymmetry variations of the axionino and
dilatino vanish. When considering a compactification down from ten
dimensions, the constraints $Q_1\xi=Q_2\xi=0$ descend to similar
constraints for the internal part of $\xi$, while the constraint
$D\xi=0$ induces both a differential and an algebraic constraint for
the internal part --- hence the compactification procedure produces a
differential constraint while increasing the number of algebraic
constraints, the resulting equations being again of constrained
generalized Killing type, but formulated for sections of some bundle
of (s)pinors defined over the internal space of the compactification.

\paragraph{Some mathematical observations.}
Let us for simplicity consider the case of a single algebraic
constraint ($Q\xi=0$).  Let $\cK(Q)$ denote the $\cinf$-submodule of
smooth solutions to the equation $Q\xi=0$ and $\cK(D)$ denote the $\K$-vector
subspace of smooth solutions to the equation $D\xi=0$. Then the $\K$-vector
subspace $\cK(D,Q)$ of smooth solutions to the CGK spinor equations equals
the intersection $\cK(D)\cap \cK(Q)$.  In general, the dimension of
the subspace $\ker (Q_x)\subset S_x$ of the fiber of $S$ at a point
$x$ may jump as $x$ varies inside $M$, so $Q$ does not admit a sub-bundle
of $S$ as its kernel (in fact, this is one reason why smooth vector
bundles do not form an Abelian category) --- even though it does admit
a kernel in the category of sheaves over the ringed space associated
with $M$. On the other hand, a simple argument\footnote{Indeed, the
values of $\xi_1,\ldots,\xi_s$ at two points $x,y$ of $M$ are related
through the parallel transport of $D$ along some curve connecting $x$
and $y$ in $M$. Since the parallel transport gives a linear
isomorphism between the fibers $S_x$ and $S_y$ of $S$, it follows that
any linear dependence relation between $\xi_1(x),\ldots,\xi_s(x)$ also
holds --- with the same coefficients --- between
$\xi_1(y),\ldots,\xi_s(y)$. Since $x$ and $y$ are arbitrary, this
would give a linear dependence relation over $\K$ (i.e. with constant
coefficients) between the globally-defined sections $\xi_1,\ldots,
\xi_s$, which contradicts our assumptions. } using parallel transport
shows that any linearly-independent (over $\K$) collection of smooth
solutions $\xi_1,\ldots,\xi_s$ of the generalized Killing (s)pinor
equation must be linearly independent everywhere, i.e. the vectors
$\xi_1(x),\ldots,\xi_s(x)$ must be linearly-independent in the fiber
$S_x$ for any point $x$ of $M$. In particular, there exists a
$\K$-vector sub-bundle $S_D$ of $S$ such that $\rk_\K S_D=\dim_\K
\cK(D)$ and such that $\Gamma(M,S_D)=\cK(D)\otimes_\K \cinf$; in
fact, any basis of the space of solutions $\cK(D)$ of the generalized
Killing (s)pinor equations provides a global frame for $S_D$ (which,
therefore, must be a trivial vector bundle). Since the restriction of
$D$ to $S_D$ is flat, the bundle $S_D$ is sometimes referred to as
`the $D$-flat vector sub-bundle of $S$'. The condition that the
generalized Killing (s)pinor equations admit exactly $s$ linearly
independent solutions over $\K$ (i.e., the condition $\dim_\K
\cK(D)=s$) amounts to the requirement that $S_D$ has rank $s$ --- in
particular, this imposes well-known topological constraints on $S$. A
similar argument shows that there exists a (topologically trivial)
$\K$-vector sub-bundle $S_{D,Q}\subset S_D\subset S$ such that $\rk_\K
S_{D,Q}=\dim_\K \cK(D,Q)$ and such that
$\Gamma(M,S_{D,Q})=\cK(D,Q)\otimes_\K \cinf$.

As mentioned in the introduction, a basic problem in the analysis of
flux compactifications (which is also of mathematical interest in its
own right) is to find efficient methods for translating constrained
generalized Killing (s)pinor equations for some collection
$\xi_1,\ldots, \xi_s$ of sections of $S$ into a system of algebraic
and differential conditions for differential forms which are
constructed as bilinears in $\xi_1,\ldots,\xi_s$. In Section
\ref{sec:fierz}, we show how the geometric algebra formalism can be
used to provide an efficient and conceptually clear solution to this
problem. Before doing so, however, we have to recall the basics of the
geometric algebra approach to spin geometry, which we proceed to do
next.  

\section{The \KA bundle of  a pseudo-Riemannian manifold}
\label{sec:KA}

This section lays out the basics of the geometric algebra formalism
and develops some specialized aspects which will be needed later on.
In Subsections \ref{sec:KAprep} and \ref{sec:KAdef}, we start with the
Clifford bundle of the cotangent bundle of a pseudo-Riemannian
manifold $(M,g)$, viewed as a bundle of unital and associative --- but
non-commutative --- algebras which is naturally associated to
$(M,g)$. The basic idea of `geometric algebra' is to use a certain
isomorphic realization of the Clifford bundle in which the underlying
vector bundle is identified with the exterior bundle of $M$. In this
realization, the multiplication of the Clifford bundle
transports to a fiberwise multiplication of the exterior bundle; when
endowed with this associative but non-commutative multiplication, the
exterior bundle becomes a bundle of associative algebras known as the
{\em \KA bundle}.  In turn, the non-commutative multiplication of the \KA
bundle induces an associative but non-commutative multiplication
(which we denote by $\diamond$ and call the {\em geometric product})
on inhomogeneous differential
forms. The resulting associative algebra is known as the {\em \KA algebra}
of $(M,g)$ and can be viewed as a certain deformation of the exterior
algebra which is parameterized by the metric $g$ of $M$.  The \KA algebra
is an associative and unital algebra over the commutative and unital ring $\cC^\infty(M,\K)$ of
smooth $\K$-valued functions defined on $M$ --- so in particular
it is a $\K$-algebra upon considering the embedding $\K\subset \cC^\infty(M,\K)$ which is 
defined by associating to each element of $\K$ the corresponding constant function. The geometric
product has an expansion in terms of so-called
{\em `generalized products'}, which form a collection of binary operations
acting on inhomogeneous forms. In turn, the generalized products can
be described as certain combinations of contractions and wedge
products. The expansion of the geometric product into
generalized products can be interpreted (under certain global
conditions on $(M,g)$) as a form of `partial quantization' of a spin
system --- the role of the Planck constant being played by the inverse
of the overall scale of the metric. In this interpretation, the \KA
algebra is the quantum algebra of observables while the geometric product
is the non-commutative composition of quantum observables;
the classical limit corresponds to taking the scale of the metric to
infinity while the expansion of the geometric product into generalized
products can be viewed as a semiclassical expansion. In the classical
limit, the geometric product
reduces to the wedge product and the \KA algebra reduces to the
exterior algebra of $M$, which plays the role of the classical algebra
of observables.  Subsection \ref{sec:KAaut} discusses certain
(anti-)automorphisms of the \KA algebra which will be used intensively
later on while subsection \ref{sec:KALR} gives some properties of the
left and right multiplication operators in this algebra. In
Subsection \ref{sec:KAparallelism}, we give a brief discussion of the
decomposition of an inhomogeneous form into parts parallel and
perpendicular to a normalized one-form and of the interplay of this
decomposition with the geometric product.  Subsection
\ref{sec:KAvoltwisted} explains the role played by the volume form and
introduces the {\em `twisted Hodge operator'}, a certain variant of the
ordinary Hodge operator which is natural from the point of view of the
\KA algebra.  Subsection \ref{sec:twistedasd} discusses the
eigenvectors of the twisted Hodge operator, which we call {\em `twisted
(anti-)selfdual forms'} --- these will play a crucial role in later
considerations.  In Subsection \ref{sec:KAAlgClass}, we recall the
algebraic classification of the fiber type of the Clifford/\KA bundle,
which is an obvious application of the well-known classification of
Clifford algebras. We pay particular attention to the `non-simple
case' --- the case when the fibers of the \KA bundle fail to be simple
as associative algebras over the base field. In Subsection
\ref{sec:KAnonsimple}, we discuss the spaces of twisted
(anti-)selfdual forms in the non-simple case, showing that --- in this
case --- they form two-sided ideals of the \KA algebra. We also give a
description of such forms in terms of rank truncations, which is
convenient in certain computations even though it is not well-behaved
with respect to the geometric product. In Subsection
\ref{sec:KAreducedHodge}, we show that, in the presence of a
globally-defined one-form $\theta$ of unit norm, the spaces of twisted
selfdual and twisted anti-selfdual forms are isomorphic (as unital
associative algebras !) with the space of those inhomogeneous forms
which are orthogonal to $\theta$ --- a space which always forms a
subalgebra of the \KA algebra. We also show that the components of an
inhomogeneous form which are orthogonal and parallel to $\theta$
determine each other when the form is twisted (anti-)selfdual and give
explicit formulas for the relation between these components in terms
of what we call the `reduced twisted Hodge operator'.  Some of the material of this section is
`well-known' at least in certain circles, though the literature tends
to be limited in its treatment of general dimensions and signatures
and of certain other aspects.  The reader who is familiar with
geometric algebra may wish to concentrate on Subsections
\ref{sec:KAparallelism}, \ref{sec:KAvoltwisted}, \ref{sec:twistedasd},
\ref{sec:KAnonsimple} and \ref{sec:KAreducedHodge} and especially on
our treatment of parallelism and orthogonality for twisted
(anti-)selfdual forms, which is important for applications.

\subsection{Preparations: wedge and generalized contraction operators}
\label{sec:KAprep}

\paragraph{The grading automorphism.}
Let $\pi$ be that involutive $\cC^\infty(M,\K)$-linear automorphism of the exterior algebra $(\Omega_\K(M),\wedge)$
which is uniquely determined by the property that it acts as minus the identity on all one-forms. Thus:
\ben
\label{piDef}
\pi(\omega)\eqdef \sum_{k=0}^{d} (-1)^k\omega^{(k)} ~~,
~~\forall \omega=\sum_{k=0}^d\omega^{(k)}\in \Omega_\K(M)~~,~~{\rm where}~~\omega^{(k)}\in \Omega_\K^k(M)~~.
\een
Taking wedge products from the left and from the right with some inhomogeneous form $\omega\in \Omega_\K(M)$
defines $\cinf$-linear operators $\wedge^L_\omega$ and $\wedge^R_\omega$:
\ben
\label{lrwedgeDef}
\wedge^L_\omega(\eta) =\omega\wedge \eta~~,~~\wedge^R_\omega(\eta) =\eta\wedge \omega~~,~~
\forall \omega,\eta\in \Omega_\K(M)~~
\een
which satisfy the following identities by virtue of the fact that the wedge product is associative:
\ben
\label{lrwedgeAssoc}
\wedge^L_{\omega_1}\circ \wedge^L_{\omega_2}=\wedge^L_{\omega_1\wedge
\omega_2}~~,~~ \wedge^R_{\omega_1}\circ
\wedge^R_{\omega_2}=\wedge^R_{\omega_2\wedge \omega_1}~~,~~
\wedge^L_{\omega_1}\circ \wedge^R_{\omega_2}=\wedge^R_{\omega_2}\circ
\wedge^L_{\omega_1}~~,~~\forall \omega_1,\omega_2\in \Omega_\K(M)~~
\een
as well as the following relation, which encodes graded-commutativity of the wedge product:
\ben
\label{lrwedgeRel}
\wedge^L_{\omega}=\wedge^R_{\omega} \circ \pi^k\Longleftrightarrow
\wedge^R_{\omega}=\wedge^L_\omega \circ \pi^k~~,~~\forall \omega\in
\Omega^k_\K(M)~~.
\een
\paragraph{The inner product.}
Let $\langle~,~\rangle$ denote the symmetric non-degenerate $\cinf$-bilinear pairing
(known as the {\em inner product} of inhomogeneous forms) induced by the metric $g$ on the exterior bundle. 
To be precise, this pairing is defined through:
\be
\langle \alpha_1\wedge \ldots \wedge \alpha_k, \beta_1\wedge \ldots
\wedge \beta_l\rangle=\delta_{k l}\det ({\hat
g}(\alpha_i,\beta_j)_{i,j=1\ldots k})~~,~~\forall \alpha_i, \beta_j\in
\Omega^1_\K(M)~~,
\ee
a relation which fixes the convention used later in our computations
(cf. Section \ref{sec:application}) via the normalization property:
\be
\langle 1_M,1_M\rangle=1~~.
\ee
Here, ${\hat g}$ is the metric induced by $g$ on $T^\ast_\K M$, which
gives the following pairing on one-forms
$\alpha=\alpha_a e^a,\beta=\beta_be^b\in \Omega^1_\K(M)$:
\ben
\label{hatg}
{\hat g}(\alpha,\beta)=g^{ab}\alpha_a\beta_b~~{\rm
for}~~\alpha=\alpha_a e^a~~{\rm and}~~\beta=\beta_b e^b~~.
\een
The fixed rank components of $\Omega_\K(M)$ are mutually orthogonal
with respect to the pairing $\langle~,~\rangle$:
\be
\langle \omega,\eta\rangle=0~~,~~\forall \omega\in \Omega^k_\K(M)~~,~~\forall \eta\in \Omega^l_\K(M)~~,~~\forall k\neq l~~,
\ee
so the rank decomposition $\Omega_\K(M)=\oplus_{k=0}^d\Omega^k_\K(M)$
is an orthogonal direct sum decomposition with respect to this
pairing. Notice that the restriction of $\langle~,~\rangle$ to
$\Omega^1_\K(M)$ coincides with \eqref{hatg}. Also notice that $\pi$
is self-adjoint with respect to the pairing $\langle~,~\rangle$:
\be
\langle \pi(\omega),\eta\rangle=\langle \omega,\pi(\eta)\rangle~~,
~~\forall \omega,\eta\in \Omega_\K(M)~~.
\ee
\paragraph{Interior products.}
The $\langle~,~\rangle$-adjoints of the left and right wedge product
operators \eqref{lrwedgeDef} are denoted by $\iota^R_\omega$ and
$\iota^L_\omega$ and are called the right and left {\em generalized
contraction} (or {\em interior product}) operators, respectively:
\ben
\label{lrconDef}
\langle \wedge^L_\omega(\eta),\rho\rangle=\langle \eta,
\iota^R_\omega(\rho)\rangle ~~,~~ \langle
\wedge^R_\omega(\eta),\rho\rangle=\langle \eta,
\iota^L_\omega(\rho)\rangle ~~,~~\forall \omega, \eta,\rho\in
\Omega_\K(M)~~.
\een
Properties \eqref{lrwedgeAssoc} translate into:
\ben
\label{lrconAssoc}
\iota^L_{\omega_1}\circ \iota^L_{\omega_2}=\iota^L_{\omega_1\wedge
\omega_2}~~,~~ \iota^R_{\omega_1}\circ
\iota^R_{\omega_2}=\iota^R_{\omega_2\wedge \omega_1}~~,~~
\iota^L_{\omega_1}\circ \iota^R_{\omega_2}=\iota^R_{\omega_2}\circ
\iota^L_{\omega_1}~~,~~\forall \omega_1,\omega_2\in \Omega_\K(M)~~,
\een
while relation \eqref{lrwedgeRel} is equivalent with:
\ben
\label{lrconRel}
\iota^L_{\omega}=\pi^k\circ \iota^R_\omega \Longleftrightarrow
\iota^R_{\omega} =\pi^k\circ \iota^L_\omega ~~,~~\forall \omega\in
\Omega^k_\K(M)~~.
\een
We also have $\wedge^{L,R}_{1_M}=\id_{\Omega_\K(M)}$ and
$\iota^{L,R}_{1_M}=\id_{\Omega_\K(M)}$. Together with
\eqref{lrwedgeAssoc} and \eqref{lrconAssoc}, this shows that
$\wedge^L$ and $\wedge^R$ define a structure of
$(\Omega_\K(M),\wedge)$-bimodule on $\Omega_\K(M)$ while $\iota^L$ and
$\iota^R$ define another $(\Omega_\K(M),\wedge)$-bimodule structure on
the same space. These two bimodule structures are adjoint to each
other with respect to the pairing $\langle~,~\rangle$.

Identities \eqref{lrwedgeRel} and \eqref{lrconRel} show that
$\wedge^L_\omega$ and $\wedge^R_\omega$ determine each other while
$\iota^L_\omega$ and $\iota^R_\omega$ also determine each other. 
From now on we choose to work with {\em left}
wedge-multiplication:
\be
\wedge_\omega\eqdef \wedge^L_\omega~~
\ee
and with the following generalized contraction operator: 
\be
\iota_\omega=\iota^R_{\tau(\omega)}~~\Longleftrightarrow 
\langle \omega\wedge \eta,\rho\rangle=\langle \eta,\iota_{\tau(\omega)}(\rho)\rangle~~,
\ee
which satisfy:
\ben
\label{wedgeiotaProps} \wedge_{\omega_1}\circ
\wedge_{\omega_2}=\wedge_{\omega_1\wedge \omega_2}~~,~~
\iota_{\omega_1}\circ \iota_{\omega_2}=\iota_{\omega_1\wedge
\omega_2}~~,~~\forall \omega_1,\omega_2\in \Omega_\K(M)~~
\een
as well as:
\be
\wedge_{1_M}=\iota_{1_M}=\id_{\Omega_\K(M)}
\ee
and thus define two different structures of
left module on the space $\Omega_\K(M)$ over the ring $(\Omega_\K(M),\wedge)$.
\paragraph{Wedge and interior product with a one-form.}
For later reference, let us consider the case when $\theta\in
\Omega^1_\K(M)$ is a one-form. Recall that the metric $g$ induces
mutually-inverse `musical isomorphisms' $~_\sharp:\Gamma(M,T_\K
M)\rightarrow \Omega^1_\K(M)$ and $~^\sharp:\Omega^1_\K(M)\rightarrow
\Gamma(M,T_\K M)$ defined by raising and lowering of indices,
respectively:
\beqa
&&X=X^a e_a\Longrightarrow X_\sharp=X_a e^a~~\mbox{where}~~X_a\eqdef
g_{ab}X^b~~,\nn\\ &&\theta=\theta_a e^a\Longrightarrow
\theta^\sharp=\theta^a e_a~~\mbox{where}~~\theta^a\eqdef
g^{ab}\theta_b~~.\nn
\eeqa
These isomorphisms satisfy:
\beqa
&&g(X,Y)={\hat g}(X_\sharp, Y_\sharp)=\langle X_\sharp,
Y_\sharp\rangle~~,~~\forall X,Y\in \Gamma(M,T_\K M)~~,\nn\\
&&g(\theta_1^\sharp, \theta_2^\sharp)={\hat
g}(\theta_1,\theta_2)=\langle \theta_1,\theta_2\rangle~~,~~\forall
\theta_1,\theta_2\in \Omega^1_\K(M)~~.\nn
\eeqa
We have:
\be
X_\sharp=X\lrcorner g~~,~~\theta=\theta^\sharp\lrcorner g~~,
\ee
where $X\lrcorner $ denotes the ordinary left contraction of a tensor
with a vector field.  It is not hard to see that the left contraction
$\iota_\theta$ with a one-form coincides with the ordinary left
contraction $\theta^\sharp\lrcorner$ with the vector field
$\theta^\sharp$:
\be
\iota_\theta\omega=\theta^\sharp\lrcorner \omega~~,~~\forall \theta\in \Omega^1_\K(M)~~,~~\forall \omega\in \Omega_\K(M)~~.
\ee
Since $\theta\wedge\theta=0$, properties \eqref{wedgeiotaProps}
imply\footnote{In general, we have $\iota_\omega\circ
\iota_\omega=\wedge_\omega\circ \wedge_\omega=0$ for any $\omega\in
\Omega^\odd_\K(M)$.}
\ben
\label{wedgeiotanilp}
\wedge_\theta\circ \wedge_\theta=\iota_\theta\circ \iota_\theta=0~~.
\een
Furthermore, the similar property of $\theta^\sharp\lrcorner$ implies
that $\iota_\theta$ is an odd derivation of the exterior algebra:
\ben
\label{iotawedgeDer}
\iota_\theta(\omega\wedge \eta)=(\iota_\theta\omega)\wedge \eta+\pi(\omega)\wedge \iota_\theta\eta~~,
~~\forall \omega, \eta\in \Omega_\K(M)~~.
\een
\paragraph{Local expressions.}
If $e_a$ is an arbitrary local frame of $M$ with dual coframe $e^a$
(thus $e^a(e_b)=\delta^a_b$), we let $g_{ab}=g(e_a,e_b)$ and
$g^{ab}={\hat g}(e^a,e^b)$, so we have $g^{ab}g_{bc}=\delta^a_c$. The
vector fields $(e^a)^\sharp$ satisfy $(e^a)^\sharp \lrcorner g=e^a$ and
are given explicitly by:
\be
(e^a)^\sharp=g^{ab}e_b~~;
\ee
they form the contragradient local frame defined by $(e_a)$.
We have $e_a=g_{ab}(e^b)^\sharp$ and $g((e^a)^\sharp, (e^b)^\sharp)=g^{ab}$.
Thus:
\ben
\label{iotavcon}
\iota_{e^a}=(e^a)^\sharp\lrcorner =
g^{ab}e_b\lrcorner\Longleftrightarrow e_a\lrcorner=g_{ab}\iota_{e^b}~~.
\een

\subsection{Definition and first properties of the \KA algebra}
\label{sec:KAdef}

\paragraph{The geometric product.}
Following an idea originally due to Chevalley and Riesz
\cite{Chevalley, Riesz}, we identify $\Cl(T_\K^{\ast}M)$ with the
exterior bundle $\wedge T^{\ast}_\K M$, thus realizing the Clifford
product as the {\em geometric product}, which is the
unique fiberwise associative, unital and bilinear
binary composition\footnote{Of course, we can view $\diamond$ as a section of the vector bundle $\Hom(\wedge T^{\ast}_\K
M \otimes \wedge T^{\ast}_\K M, \wedge T^{\ast}_\K M)$. } $\diamond: \wedge T^{\ast}_\K M \times_{M}\wedge
T^{\ast}_\K M \rightarrow \wedge T^{\ast}_\K M$ whose induced action
on sections (which we again denote by $\diamond$) satisfies the
following relations for all $\theta \in \Omega^1_\K(M)$ and all
$\omega\in \Omega_\K(M)$:
\ben
\label{basicid}
\theta\diamond \omega=\theta \wedge \omega+\iota_\theta\omega~~,~~\pi(\omega)\diamond \theta=
\theta\wedge \omega - \iota_\theta \omega~~.
\een
Equations \eqref{basicid} determine the geometric composition of any
two inhomogeneous forms via the requirement that the geometric product is
associative and $\cC^\infty(M,\K)$-bilinear. 

The unit of the fiber $\Cl(T^{\ast}_{\K,x} M)$ at a point $x\in M$ corresponds to
the element $1\in \K=\wedge^0 T_{\K,x}^{\ast} M$, which is the unit of
the associative algebra $(\wedge T^{\ast}_{\K,x} M,
\diamond_{x})\approx \Cl(T^{\ast}_{\K,x} M)$. Hence the unit section
of the Clifford bundle $\Cl(T^{\ast}_{\K} M)$ is identified with the
constant function $1_M:M\rightarrow \K$ given by $1_M(x)=1$ for all
$x\in M$. Through this construction, the Clifford bundle is identified
with the bundle of algebras $(\wedge T^{\ast}_\K M, \diamond)$, which
is known \cite{Graf} as the {\em \KA bundle} of $(M,g)$. 
When endowed with the geometric product, the space $\Omega_\K(M)$ of all inhomogeneous $\K$-valued smooth forms
on $M$ becomes a unital and associative (but non-commutative) algebra
$(\Omega_\K(M),\diamond)$ over the ring $\cinf$, known as the {\em \KA algebra} of $(M,g)$.
The unit of the \KA algebra is the constant function $1_M$. We have a
unital isomorphism of associative algebras over $\cinf$ between
$(\Omega_\K(M),\diamond)$ and the $\cinf$-algebra $\Gamma(M,\Cl(T^\ast_\K M))$
of all smooth sections of the Clifford bundle. The \KA algebra can be
viewed as a $\Z_2$-graded associative algebra with even and odd parts
given by:
\be
\Omega^{\rm ev}_\K(M)\eqdef \oplus_{k={\rm even}}\Omega^k_\K(M)~~,
~~\Omega^{\rm odd}_\K(M)\eqdef \oplus_{k={\rm odd}}\Omega^k_\K(M)~~,
\ee
since it is easy to check the inclusions:
\beqa
&&\Omega^\ev_\K(M)\diamond \Omega^\ev_\K(M)\subset \Omega^\ev_\K(M)~~,~~
\Omega^\odd_\K(M)\diamond \Omega^\odd_\K(M)\subset \Omega^\ev_\K(M)~~,\\
&&\Omega^\ev_\K(M)\diamond \Omega^\odd_\K(M)\subset \Omega^\odd_\K(M)~~,~~
\Omega^\odd_\K(M)\diamond \Omega^\ev_\K(M)\subset \Omega^\odd_\K(M)~~.
\eeqa
However, it is not a $\Z$-graded algebra since the geometric product
of two forms of definite rank need not be a form of definite rank. We
let $P_\ev=\frac{1}{2}(1+\pi)$ and $P_\odd=\frac{1}{2}(1-\pi)$ be the
complementary idempotents associated with the decomposition into even
and odd parts:
\ben
\label{PevDef}
P_\ev(\omega)=\omega_\ev~~,~~P_\odd(\omega)=\omega_\odd~~,
\een
where $\omega=\omega_\ev+\omega_\odd\in \Omega_\K(M)$, with
$\omega_\ev\in \Omega^\ev_\K(M)$ and $\omega_\odd\in \Omega^\odd_\K(M)$.

\paragraph{Generalized products. Connection with quantization of spin systems.}
The geometric product $\diamond $ can be viewed as a deformation of the
wedge product (parameterized by the metric $g$) and reduces to the
latter in the limit $g\rightarrow \infty$; in this limit, the \KA
algebra reduces to the exterior algebra $(\Omega_\K(M),\wedge)$. Under
some mild assumptions, the geometric product can be
described quite elegantly in the language of supermanifolds, as the
star product induced by fiberwise Weyl quantization of a pure spin
system \cite{Hirshfeld1, Hirshfeld2, Henselder1, Henselder2}. For
this, consider the parity-changed tangent bundle $\Pi T M$ of $M$ (a
supermanifold with body $M$) and introduce odd coordinates $\zeta^a$
on the fibers of $\Pi T U$, corresponding to a coframe $e^a$ of $M$
defined on a small enough open subset $U\subset M$.  Inhomogeneous
differential forms \eqref{FormExpansion} correspond to functions
defined on $\Pi T M$ having the following local expansion:
\ben
\label{sprod}
f_\omega(x,\zeta)=_U
\sum_{k=0}^{d}\frac{1}{k!}\omega^{(k)}_{a_1\ldots a_k}(x)\zeta^{a_1}\ldots \zeta^{a_k}~~.
\een
This allows us to represent the geometric product through the fermionic analogue $\star$ of the Moyal
product, using a certain `vertical' \cite{Henselder2} quantization
procedure\footnote{This is most easily explained in the case when
$d=\dim M$ is even, namely $d=2r$. Then  `vertical'
  fermionic Weyl quantization (in which one quantizes only along the
  odd directions of $\Pi T M$, while treating the body $M$ as
  classical) can be performed by choosing an almost complex structure
  on $M$, which induces a decomposition $T_\C M=W\oplus W^\ast$
  of the complexified tangent bundle of $M$, with $W$ a complex vector bundle. This allows us to define
  fermionic Fock representations at each point $x\in M$ given by
  annihilation and creation operators
  ${\bf a}_k\eqdef \frac{1}{\sqrt{2}}(\gamma_k+i\gamma_{r+k})$,
  ${\bf a}_k^\dagger\eqdef \frac{1}{\sqrt{2}}(\gamma_k-i\gamma_{r+k})$ (where
  $i=1\ldots r$) defined at $x$, with coherent states given by $|z\rangle_{x}\eqdef e^{-\sum_{k=1}^{d/2}z^k
    {\bf a}_k^\dagger }|0\rangle_{x}$, where $|0\rangle_{x}$ is the
  vacuum at $x$ and $z^k=\frac{1}{2}(\zeta^k+i\zeta^{r+k})$ are odd
  complex coordinates along the fibers of $\Pi T_\C M$. Identifying
  $\Omega_\C (M)$ with the algebra of complex functions on $\Pi T
  M$, we have ${\bf a}_k=\frac{\partial}{\partial z^k}$ and
  ${\bf a}_k^\dagger={\bar z}^k$, so the bundle of spin Fock spaces can be
  identified with the sub-bundle $\wedge W^\ast $ of
  $\wedge T^{\ast}_\C M$. The star product $\star$ takes the form: \be
f_{\omega\diamond \eta}=f_\omega\star f_\eta=f_\omega \exp\left(g^{ k {\bar
        l}}\overleftarrow{\frac{\partial}{\partial z^k}}\overrightarrow{\frac{\partial}{\partial {\bar
          z}^l}} + g^{{\bar k}
    l}\overleftarrow{\frac{\partial}{\partial{{\bar
         z}^k}}}\overrightarrow{\frac{\partial}{\partial z^l}}\right)f_\eta~~, \ee which agrees with (\ref{fstar}).}:
\ben
\label{fstar}
f_{\omega}\star f_{\eta} =
f_{\omega\diamond \eta}=f_\omega \exp
\left(g^{ab}\frac{\overleftarrow{\partial}}{\partial{\zeta^a}}
\frac{\overrightarrow{\partial}}{\partial{\zeta^b}}\right)f_\eta~~.
\een
Expanding the exponential in \eqref{fstar} gives the following
expressions for two general inhomogeneous forms $\omega,\eta\in
\Omega_\K(M)$(cf. \cite{Houri, Kubiznak, Cariglia}):
\beqan
\label{starprod}
\omega\diamond \eta=\sum_{k=0}^{\left[\frac{d}{2}\right]}
\frac{(-1)^k}{(2k)!}\omega\wedge_{2k}\eta + \sum_{k=0}^{\left[\frac{d-1}{2}\right]} \frac{(-1)^{k+1}}{(2k+1)!}
\pi(\omega)\wedge_{2k+1}\eta~~,
\eeqan
where the binary $\cinf$-bilinear operations $\wedge_k$ are
the {\em contracted wedge products} \cite{Houri, Kubiznak, Cariglia,
BennTucker}, defined iteratively through:
\be
\omega\wedge_0\eta=\omega\wedge \eta~~,~~\omega\wedge_{k+1}\eta
=g^{ab}(e_a \lrcorner \omega)\wedge_{k}(e_b\lrcorner \eta)=g_{ab}
(\iota_{e^a}\omega) \wedge_k (\iota_{e^b}\eta) ~~.
\ee
We also have the following expansions for the {\em graded} $\diamond
$-commutator and {\em graded} $\diamond $-anticommutator of $\omega$ with $\eta$:
\ben
\label{StarGrCom}
[[\omega,\eta]]_{-,\diamond }=2\sum_{k=0}^{\left[\frac{d-1}{2}\right]} \frac{(-1)^{k+1}}{(2k+1)!} 
\pi(\omega)\wedge_{2k+1}\eta~~,~~
[[\omega,\eta]]_{+,\diamond }=2\sum_{k=0}^{\left[\frac{d}{2}\right]} \frac{(-1)^k}{(2k)!}\omega\wedge_{2k}\eta~~.
\een
For forms of definite ranks, the graded $\diamond$-commutator and graded $\diamond$-anticommutator are of course defined through:
\be
[[\omega,\eta]]_{-,\diamond }\eqdef \omega\diamond \eta-(-1)^{pq}\eta\diamond \omega~~~,~~
[[\omega,\eta]]_{+,\diamond }\eqdef \omega\diamond \eta+(-1)^{pq}\eta\diamond \omega~~,
~\forall \omega\in \Omega^p_\K(M)~,~
\forall \eta\in \Omega^q_\K(M)~,~~~~~~~~
\ee
being extended by linearity to the entire space $\Omega_\K(M)$. 
Using \eqref{starprod} one can easily deduce the following relation for homogeneous forms:
\be
\eta\diamond\omega=(-1)^{pq}\sum_{k=0}^p\frac{(-1)^{k(p-k+1)+[\frac{k}{2}]}}{k!}\omega\wedge_k\eta~~~,
~~~\forall \omega\in \Omega^p_\K(M)~~,~~\forall \eta\in \Omega^q_\K(M)~~,~~p\leq q~~.
\ee
We will mostly use, instead of $\wedge_k$, the so-called {\em generalized products}
$\bigtriangleup_k$, which are defined by rescaling the contracted wedge products:
\ben
\label{GenProd}
\bigtriangleup_k=\frac{1}{k!}\wedge_k~~.
\een
These have the advantage that the various factorial prefactors in the
expansions above disappear when those expansions are re-expressed in
terms of generalized products.

Expansions \eqref{starprod} and \eqref{StarGrCom} can also
be obtained directly from the definition of the geometric product using
\eqref{basicid}, which shows that the purely mathematical identities
given above also hold irrespective of any interpretation through the theory of
quantization of spin systems.

\subsection{(Anti-)automorphisms of the \KA algebra}
\label{sec:KAaut}

Direct computation shows that $\pi$ is an involutive automorphism
(known as {\em the main} or {\em grading} automorphism) of the \KA
bundle (a property which, in the limit $g\rightarrow \infty$, recovers
the well-known fact that $\pi$ is also an automorphism of the exterior
bundle). The \KA bundle also admits an involutive anti-automorphism
$\tau$ (known as {\em the main anti-automorphism} or as {\em
reversion}), which is given by:
\ben
\label{taudef}
\tau(\omega)\eqdef (-1)^{\frac{k(k-1)}{2}}\omega~~,~~\forall \omega\in \Omega^k_\K(M)~~.
\een
It is the unique anti-automorphism of $(\wedge T^\ast_\K M,\diamond)$
which acts trivially on all one-forms (i.e. which satisfies
$\tau(\theta)=\theta$ for any form $\theta$ of rank one).  Direct
computation (or the fact that the exterior product is recovered from
the diamond product in the limit of infinite metric) shows that $\tau$
is also an anti-automorphism of the exterior bundle $(T^\ast_\K
M,\wedge)$. We also notice that $\pi$ and $\tau$ commute.  All in all,
we have the relations:
\be
\pi \circ \tau=\tau \circ \pi~~,~~\pi\circ \pi=\tau\circ\tau=\id_{\Omega_\K(M)}~~.
\ee
Note that $\Cl^\ev(T^\ast_\K M)$ identifies with the sub-bundle of
unital subalgebras $\wedge^\ev T^\ast_\K M=\oplus_{k={\rm
even}}{\wedge^k T^\ast_\K M}$ of the \KA bundle, whose space of
smooth sections $\Omega^\ev_\K(M)$ can be described as the eigenspace of
$\pi$ corresponding to the eigenvalue $+1$:
\be
\Omega^\ev_\K(M)=\cK(1-\pi)=\{\omega\in \Omega_\K(M)~|~\pi(\omega)=\omega\}~~.
\ee

\subsection{The left and right geometric multiplication operators}
\label{sec:KALR}

Let $L_\omega$, $R_\omega$ be the $\cinf$-linear operators of left and right
multiplication with $\omega\in \Omega_\K(M)$ in the \KA algebra:
\be
L_\omega(\eta)\eqdef \omega\diamond \eta~~,~~R_\omega(\eta)\eqdef \eta\diamond \omega~~,
~~\forall \omega,\eta\in \Omega_\K(M)~~.
\ee
These satisfy:
\be
L_{\omega_1}\circ R_{\omega_2}=R_{\omega_2}\circ L_{\omega_1}~~,~~L_{\omega_1}\circ L_{\omega_2}=
L_{\omega_1\diamond \omega_2}~~,~~R_{\omega_1}\circ R_{\omega_2}=R_{\omega_2\diamond \omega_1}~~
\ee
as a consequence of associativity of the geometric product. We also have:
\be
L_\omega\circ \pi=\pi\circ L_{\pi(\omega)}~~,~~R_\omega\circ \pi=\pi\circ R_{\pi(\omega)}~~,~~
L_\omega\circ \tau=\tau\circ R_{\tau(\omega)}~~,~~\tau\circ L_{\omega}=R_{\tau(\omega)}\circ \tau~~,
~~\forall \omega\in \Omega_\K(M)~~,
\ee
since $\pi$ is an involutive algebra automorphism while $\tau$ is an
involutive anti-automorphism. Identity \eqref{basicid} can be written
as:
\ben
\label{basicop1}
L_\theta=\wedge_\theta+\iota_\theta~~,~~R_\theta\circ \pi=\wedge_\theta-\iota_\theta~~,
~~\forall \theta\in \Omega^1_\K(M)~~,
\een
being equivalent with:
\ben
\label{basicop2}
\wedge_\theta=\frac{1}{2}(L_\theta+R_\theta\circ \pi)~~,
~~\iota_\theta= \frac{1}{2}(L_\theta-R_\theta\circ \pi)~~,
~~\forall \theta\in \Omega^1_\K(M)~~.
\een
This shows that the operators $\wedge_\theta$ and $\iota_\theta$ (and
thus --- given properties \eqref{wedgeiotaProps} --- also the
operators $\wedge_\omega$ and $\iota_\omega$ for any $\omega\in
\Omega_\K(M)$) are determined by the geometric product.

\paragraph{Remark.} Equation \eqref{basicid} implies:
\ben
\label{contractionwedge}
\iota_\theta \omega=\frac{1}{2}[[\theta, \omega]]_{-,\diamond}~~,
~~\theta\wedge \omega=\frac{1}{2}[[\theta,\omega]]_{+,\diamond}~~,
~~\forall \theta\in \Omega^1_\K(M)~~,~~\forall \omega\in \Omega_\K(M)~~.
\een
The first identity in \eqref{contractionwedge} shows that the operator
$\iota_\theta$ is an odd $\cinf$-linear derivation (in fact, an odd differential ---
since $\iota_\theta\circ \iota_\theta=0$) of the \KA algebra:
\be
\iota_\theta(\omega\diamond\eta)=\iota_\theta(\omega)\diamond\eta+
\pi(\omega)\diamond \iota_\theta(\eta)~~,~~\forall \omega,\eta\in \Omega_\K(M)~~,~~\forall \theta\in \Omega^1(M)~~.
\ee
In the limit $g\rightarrow \infty$, this property recovers
\eqref{iotawedgeDer}. Notice that $\wedge_\theta$ is {\em not} a
derivation of the \KA algebra --- however, it satisfies
$\wedge_\theta\circ \wedge_\theta=0$.

\subsection{Orthogonality and parallelism}
\label{sec:KAparallelism}

Let $\theta\in \Omega^1_\K(M)$ be a fixed one-form which satisfies the
normalization condition:
\ben
\label{normalization}
{\hat g}(\theta,\theta)=1~~{\rm i.e.}~~\iota_\theta\theta=1~~.
\een
This condition is equivalent with:
\ben
\label{normalization2}
\theta\diamond\theta=1~~,
\een
a fact which follows from \eqref{basicid} and from the identity $\theta\wedge \theta=0$
(which, together,  imply $\theta\diamond \theta=\iota_\theta\theta$).

We say that an inhomogeneous form $\omega\in \Omega_\K(M)$ is {\em
parallel} to $\theta$ (we write $\theta \parallel \omega$) if
$\theta\wedge \omega=0$ and {\em orthogonal} to $\theta$ (we write
$\theta\perp\omega$) if $\iota_\theta\omega=0$. Thus:
\be
\theta\parallel\omega\stackrel{{\rm def.}}{\Longleftrightarrow} \omega\in \cK(\wedge_\theta)~~,
~~\theta\perp\omega\stackrel{{\rm def.}}{\Longleftrightarrow} \omega\in \cK(\iota_\theta)~~,
\ee
where we remind the reader that $\cK(A)$ denotes the kernel of any $\K$-linear operator $A:\Omega_\K(M)\rightarrow \Omega_\K(M)$.
Properties \eqref{wedgeiotanilp} imply $\cI(\iota_\theta) \subset
\cK(\iota_\theta)$ and $\cI(\wedge_\theta)\subset
\cK(\wedge_\theta)$,  where $\cI(A)$ denotes the image of any $\K$-linear operator $A:\Omega_\K(M)\rightarrow \Omega_\K(M)$.
These inclusions are in fact equalities, as we shall see in a moment.

\paragraph{Proposition.} Any inhomogeneous differential form
$\omega\in \Omega_\K(M)$ decomposes uniquely as:
\be
\omega=\omega_{\parallel}+\omega_\perp~~,
\ee
where $\theta\parallel \omega_{\parallel}$ and $\theta \perp
\omega_\perp$. Moreover, the parallel and orthogonal parts of $\omega$
are given by:
\be
\omega_{\parallel}=\theta\wedge (\iota_\theta\omega)~~,
~~\omega_\perp=\iota_\theta (\theta\wedge \omega)~~.
\ee
In fact, the $\cinf$-linear operators $P_{\parallel}\eqdef \wedge_\theta\circ
\iota_\theta$ and $P_\perp\eqdef \iota_\theta\circ \wedge_\theta$ are
complementary idempotents:
\be
P_{\parallel}+P_{\perp}=\id_{\Omega_\K(M)}~~,~~P_{\parallel}\circ P_{\parallel}=P_{\parallel}~~,
~~P_\perp\circ P_\perp=P_\perp~~,~~P_{\parallel}\circ P_\perp=P_\perp\circ P_{\parallel}=0~~.
\ee
\noindent {\bf Proof.} The statements of the proposition follow
immediately from the fact that $\wedge_\theta$ and $\iota_\theta$ are
nilpotent and because $\iota_\theta$ is an odd derivation of the wedge
product, which implies:
\ben
\label{thetaid}
\iota_\theta(\theta\wedge \omega)=\omega-\theta\wedge (\iota_\theta\omega)~~,
\een
where we used the normalization condition \eqref{normalization}.

\

\noindent As an immediate corollary of the proposition, we find the well-known
equalities:
\be
\cK(\iota_\theta)=\cI(\iota_\theta)~~,~~\cK(\wedge_\theta)=\cI(\wedge_\theta)~~
\ee
as well as the characterizations:
\beqan
\label{parperpchar}
\!\!\!\!\!\!\!&&\theta\parallel\omega\Longleftrightarrow \omega
= \theta\wedge \alpha ~{\rm~with~\alpha\in \Omega_\K(M)}\Longleftrightarrow \theta\diamond \omega=-\pi(\omega)\diamond\theta\Longleftrightarrow
\omega\in \cK(L_\theta+R_\theta\circ \pi)\nn~,~~~~~~\\
\!\!\!\!\!\!\!&& \theta\perp \omega\Longleftrightarrow \omega =\iota_\theta \beta ~{\rm~with~\beta\in \Omega_\K(M)}
\Longleftrightarrow \theta\diamond \omega=\pi(\omega)\diamond \theta\Longleftrightarrow
\omega \in \cK(L_\theta-R_\theta\circ \pi),~~~~~~~
\eeqan
where we used relations \eqref{basicop2}. Thus $\theta\parallel \omega$ iff. $\omega$ graded anticommutes with $\theta$ and
$\theta\perp \omega$ iff. $\omega$ graded commutes with $\theta$ in the \KA algebra. 

\paragraph{Behavior with respect to the geometric product.} 
Consider the following $\cinf$-submodules of $\Omega_\K(M)$:
\be
\Omega_\K^{\parallel}(M)\eqdef \{\omega\in \Omega_\K(M)~|~\theta \parallel\omega\}~~,~~\Omega_\K^\perp(M)\eqdef \{\omega\in \Omega_\K(M)~|~\theta\perp \omega\}~~.
\ee
Using the characterizations in \eqref{parperpchar}, we find:
\beqan
\label{parperptrans}
\theta\parallel\omega ~~{\rm and}~~\theta \perp \eta ~&\Longrightarrow&~ \theta \parallel (\omega\diamond \eta) ~~{\rm and}~~  \theta \parallel (\eta\diamond \omega)~~, \nn\\
\theta\parallel\omega,\eta ~&\Longrightarrow&~ \theta \perp (\omega\diamond \eta) ~~,\nn\\
\theta\perp \omega,\eta ~&\Longrightarrow&~ \theta\perp (\omega\diamond \eta)~~,
\eeqan
which translate into:
\ben
\label{ometaparts}
(\omega\diamond \eta)_\parallel=\omega_\parallel \diamond \eta_\perp+\omega_\perp\diamond \eta_\parallel~~,~~
(\omega\diamond \eta)_\perp=\omega_\parallel \diamond \eta_\parallel+\omega_\perp\diamond \eta_\perp~~.
\een
We thus have the inclusions:
\beqa
&&\Omega^\parallel_\K(M)\diamond \Omega^\parallel_\K(M)\subset \Omega^\perp_\K(M)~~,~~
\Omega^\perp_\K(M)\diamond \Omega^\perp_\K(M)\subset \Omega^\perp_\K(M)~~,\\
&&\Omega^\parallel_\K(M)\diamond \Omega^\perp_\K(M)\subset \Omega^\parallel_\K(M)~~,~~
\Omega^\perp_\K(M)\diamond \Omega^\parallel_\K(M)\subset \Omega^\parallel_\K(M)~~.
\eeqa
Together with the identity $\iota_\theta(1_M)=0$ (which shows that
$\theta\perp 1_M$), the last property in \eqref{parperptrans} shows
that $\Omega^\perp_\K(M)$ is a unital subalgebra of the \KA algebra.

Notice that characterizations \eqref{parperpchar} imply
that the involutions $\pi$ and $\tau$ preserve parallelism and
orthogonality to $\theta$:
\beqan
\label{piparperp}
\theta\parallel\omega\Longrightarrow \theta\parallel\pi(\omega)~~&,&~~\theta\perp \omega\Longrightarrow \theta\perp \pi(\omega)~~,\nn\\
\theta\parallel\omega\Longrightarrow \theta\parallel\tau(\omega)~~&,&~~\theta\perp \omega\Longrightarrow \theta\perp \tau(\omega)~~.
\eeqan
\paragraph{The top component of an inhomogeneous form.}
The parallel part of $\omega\in \Omega_\K(M)$ can be written as:
\ben
\label{omegatopDef}
\omega_\parallel=\theta\wedge \omega_\top~~,
\een
where:
\ben
\label{TopDef}
\omega_\top\eqdef \iota_\theta\omega\in \Omega^\perp_\K(M)~~.
\een
This shows that $\omega$ determines and is determined by the two
inhomogeneous forms $\omega_\perp$ and $\omega_\top$, both of which belong to
$\Omega^\perp_\K(M)$. In fact, any $\omega\in \Omega_\K(M)$ can be written {\em uniquely} in the form:
\ben
\label{Omdperp}
\omega=\theta\wedge \alpha+\beta~~{\rm with}~~\alpha,\beta\in \Omega^\perp_\K(M)~~,
\een
namely we have $\alpha=\omega_\top$ and $\beta=\omega_\perp$. This gives a $\cinf$-linear isomorphism:
\ben
\label{Omdpisom}
\Omega_\K(M)\stackrel{\iota_\theta+P_\perp}{\longrightarrow} \Omega^\perp(M)\oplus\Omega^\perp(M)~~.
\een
which sends $\omega\in \Omega_\K(M)$ into the pair
$(\omega_\top,\omega_\perp)$ and whose inverse sends a pair
$(\alpha,\beta)$ with $\alpha,\beta\in \Omega_\K^\perp(M)$ into the
form \eqref{Omdperp}. Since $\omega_\top$ is orthogonal to $\theta$,
we have $\theta\wedge \omega_\top=\theta\diamond
\omega_\top=\pi(\omega_\top)\diamond \theta$ and thus
$\omega_\parallel=\theta\diamond \omega_\top$. It follows that the
decomposition of $\omega$ can be written entirely in terms of the
geometric product:
\ben
\label{omegadecKA}
\omega=\theta\diamond \omega_\top+\omega_\perp~~.
\een
An easy computation using this formula gives:
\ben
\label{omegaetatop}
(\omega\diamond\eta)_\perp=\omega_\perp\diamond \eta_\perp+\pi(\omega_\top)\diamond \eta_\top~~,~~
(\omega\diamond\eta)_\top=\omega_\top\diamond \eta_\perp+\pi(\omega_\perp)\diamond \eta_\top~~.
\een
\subsection{The volume form and the twisted Hodge duality operator}
\label{sec:KAvoltwisted}

\paragraph{The ordinary volume form.}
From now on, we shall assume that $M$ is oriented (in particular, the
$\K$-line bundles $\Lambda^d T^\ast_\K M$ are trivial for
$\K=\R,\C$). Consider the volume form determined on $M$ by the metric
and by this orientation, which has the following expression in a local
frame defined on $U\subset M$:
\be
\vol_M=_U\frac{1}{d!}\sqrt{|\det g|}\epsilon_{a_1\ldots a_d}e^{a_1\ldots a_d}~~.
\ee
Here, $\det g $ is the determinant of the matrix
$(g(e_a,e_b))_{a,b=1\ldots d}$ while $\epsilon_{a_1\ldots a_d}$ are
the local coefficients of the Ricci density --- defined as the
signature of the permutation $\left(\begin{array}{cc}1~ 2~\ldots~
d~\\a_1~a_2~\ldots~ a_d\end{array}\right)$. The volume form satisfies:
\ben
\label{volSquared}
\vol_M\diamond \vol_M=(-1)^{q+\frac{d(d-1)}{2}}=(-1)^{q+\left[\frac{d}{2}\right]}=
\twopartdef{+1~,~}{p-q\equiv_4 0,1\Longleftrightarrow p-q\equiv_8 0,1,4,5}{-1~,~}{p-q\equiv_4 2,3
\Longleftrightarrow p-q\equiv_8 2, 3, 6, 7}~~,
\een
where we used the congruences:
\be
\frac{d(d-1)}{2}\equiv_2\left[\frac{d}{2}\right]~~~,~~~~~
q+\frac{d(d-1)}{2}\equiv_2\twopartdef{\frac{p-q}{2}~~~,~}{d=\mbox{even}}
{\frac{p-q-1}{2}~,~}{d=\mbox{odd}}~.
\ee
We remind the reader that $p$ and $q$ denote the number of positive
and negative eigenvalues of the metric tensor, respectively.
\paragraph{The ordinary Hodge operator.}
Recall that the ordinary $\cinf$-linear Hodge operator $\ast$ is defined through:
\ben
\label{HodgeDef}
\omega\wedge (\ast \eta)=\langle \omega,\eta\rangle \vol_M~~,
~~\forall \omega,\eta\in \Omega^k_\K (M)~~,~~\forall k=0\ldots d
\een
and satisfies the following properties, which we list for convenience of the reader:
\beqa
&~&\omega\wedge \eta=(-1)^q \langle \eta, \ast \omega\rangle \vol_M ~~,
~~\forall \omega\in \Omega^k(M)~~,~~\forall \eta\in
\Omega^{d-k}(M)~~,~~\forall k=0\ldots d~~,\nn\\ &~& \langle \ast
\omega, \ast \eta\rangle=(-1)^q \langle \omega,
\eta\rangle~~,~~\forall \omega,\eta\in \Omega(M)~~,\nn\\ &~& \vol_M=\ast
1_M \Longleftrightarrow \ast \vol_M=(-1)^q 1_M~~,\nn\\ &~&\ast
\omega=\iota_\omega\vol_M~~,~~\forall \omega\in \Omega(M)~~,\\ &~&\ast
\circ \ast =(-1)^q\pi^{d-1}~~,\nn\\ &~& \ast \circ \pi=(-1)^d\pi\circ
\ast~~\nn.
\eeqa
We also note the identity:
\be
\tau\circ \ast= (-1)^{\left[\frac{d}{2}\right]}\ast\circ ~\tau\circ \pi^{d-1}~~,
\ee
which follows by direct computation upon using the congruence:
\be
\frac{k(k-1)}{2}+\frac{(d-k)(d-k-1)}{2}\equiv_2 \frac{d(d-1)}{2}+k(d-1)~~.
\ee
\paragraph{The modified volume form.}
Consider the following $\K$-valued top form on $M$:
\ben
\label{nudef}
\nu\eqdef c_{p,q}(\K)\vol_M~~\mbox{where}~~c_{p,q}(\K)\eqdef
\twopartdef{1~,~}{\K=\R}{i^{q+\left[\frac{d}{2}\right]}~,~}{\K=\C}~~,
\een
which satisfies:
\ben
\label{NuSquared}
\nu\diamond \nu=\twopartdef{(-1)^{q+\left[\frac{d}{2}\right]}1_M~,~}{\K=\R}{+1_M~,~}{\K=\C}~~.
\een
We have the normalization property:
\be
\langle \nu, \nu\rangle= \twopartdef{(-1)^q 1_M~,~}{\K=\R}{(-1)^{\left[\frac{d}{2}\right]}1_M~,~}{\K=\C}~~
\ee
and the identity:
\ben
\label{nucomm}
L_\nu=R_\nu\circ \pi^{d-1}\Longleftrightarrow \nu \diamond \omega=\pi^{d-1}(\omega)\diamond \nu ~~,
~~\forall \omega\in \Omega_\K(M)~~,
\een
where $\pi^{d-1}$ represents the composition of $d-1$ copies of the main automorphism $\pi$:
\be
\pi^{d-1}=\twopartdef{\id_{\Omega_\K(M)}~,~}{d=\mbox{odd}}{\pi~,~}{d=\mbox{even}}~~.
\ee
In particular, $\nu$ is a central element of the \KA algebra iff. $d$ is odd.

\paragraph{The twisted Hodge operator.}
Let us define the ($\cinf$-linear) {\em twisted Hodge operator} ${\tilde \ast}:\Omega_\K(M)\rightarrow \Omega_\K(M)$ through the formula:
\be
{\tilde \ast} \omega\eqdef \omega\diamond \nu~~,~~\forall \omega\in \Omega_\K(M)~~.
\ee
Identity \eqref{NuSquared} shows that (unlike
what happens for the ordinary Hodge operator) the square of the
twisted Hodge operator is {\em always} a scalar multiple of the
identity:
\be
{\tilde \ast}\circ {\tilde \ast}=\twopartdef{(-1)^{q+\left[\frac{d}{2}\right]}\id_{\Omega_\K(M)}~,~}{\K=\R}{\id_{\Omega_\K(M)}~,~}{\K=\C}~~.
\ee
A simple computation shows that the twisted and ordinary Hodge operators are related through:
\be
{\tilde \ast}=c_{p,q}(\K) \ast \circ ~\tau~~.
\ee
In particular, the ordinary Hodge operator admits the representation:
\be
\ast \omega=\tau(\omega)\diamond \vol_M
=\frac{1}{c_{p,q}(\K)}\tau(\omega)\diamond \nu~~,~~\forall \omega\in
\Omega_\K(M)~~.
\ee

\subsection{Twisted (anti-)selfdual forms}
\label{sec:twistedasd}

Let us assume that $\K=\C$ or that $\K=\R$ and $p-q\equiv_4 0, 1$, so
that the twisted Hodge operator ${\tilde \ast}$ squares to the identity. In this case, the
twisted Hodge operator has real eigenvalues equal to $\pm 1$ and we
can consider {\em inhomogeneous} real forms belonging to the
corresponding eigenspaces.  A form $\omega\in \Omega_\K(M)$ will be
called {\em twisted selfdual} if ${\tilde \ast}\omega=+\omega$ and
{\em twisted anti-selfdual} if ${\tilde \ast}\omega=-\omega$.  We let
$\Omega^\pm_\K(M)\eqdef \{\omega\in \Omega_\K(M)|\omega\diamond \nu=\pm
\omega\}\subset \Omega_\K(M)$ be the $\cinf$-submodules of twisted selfdual and
twisted anti-selfdual forms on $M$.

\paragraph{The ideals $\Omega^\pm_\K(M)$.}
The elements $p_\pm\eqdef \frac{1}{2}(1\pm \nu)$ are complementary idempotents of the
\KA algebra:
\be
p_\pm\circ p_\pm=p_\pm~~,~~p_++p_-=1_M~~,~~p_\pm\circ p_\mp=0~~.
\ee
Notice that these idempotents are central only when $\nu$ is central,
i.e. only when $d$ is odd.  The operators $P_\pm\eqdef R_{p_\pm}$ defined
through right $\diamond$-multiplication with these elements:
\be
P_\pm(\omega)\eqdef \omega \diamond p_\pm =\frac{1}{2}(\omega\pm \omega\diamond \nu)~~
~~(\omega\in \Omega_\K(M))\Longleftrightarrow
P_\pm=\frac{1}{2}(1\pm {\tilde \ast})
\ee
are complementary idempotents in the algebra of endomorphisms of the $\cinf$-module $\Omega_\K(M)$:
\be
P_\pm^2=P_\pm ~~,~~P_+\circ P_-=P_-\circ P_+=0~~,~~P_++P_-=\id_{\Omega_\K(M)}~~.
\ee
Therefore, the images
$\Omega^\pm_\K(M)=P_\pm(\Omega_\K(M))=\Omega_\K(M) p_\pm$ are
complementary {\em left} ideals of the \KA algebra, giving the direct
sum decomposition:
\be
\Omega_\K(M)=\Omega_\K^+(M)\oplus \Omega_\K^-(M)~~.
\ee
In particular, $(\Omega_\K^\pm(M),\diamond)$ are associative
subalgebras of the \KA algebra. These subalgebras have units (given by
$p_\pm$) iff. $d$ is odd, in which case they are {\em two-sided} ideals
of $(\Omega_\K(M),\diamond)$.

\paragraph{Local characterization.}
With respect to a local coframe $e^a$ above an open subset
$U\subset M$, we have the expansions:
\beqan
\label{HodgeComponents}
\ast (e^{a_1\ldots a_k})&=&\frac{1}{(d-k)!}\sqrt{|\det g|}\epsilon^{a_1\ldots
a_k}_{~~~~~~a_{k+1}\ldots a_d}e^{a_{k+1}\ldots a_d}~~,\nn\\ {\tilde
\ast} (e^{a_1\ldots a_k})&=&\frac{1}{(d-k)!}c_{p,q}(\K)\sqrt{|\det g|}\epsilon^{a_k\ldots
a_1}_{~~~~~~a_{k+1}\ldots a_d}e^{a_{k+1}\ldots a_d}~~,
\eeqan
where indices are raised with $g^{ab}$.
Using \eqref{HodgeComponents}, one easily checks that an inhomogeneous
form $\omega\in \Omega_\K(M)$ with expansion \eqref{FormExpansion} satisfies
${\tilde \ast} \omega=\pm \omega$ iff. its non-strict coefficients
satisfy the conditions:
\ben
\label{TwistedHodgeCoeffs}
\omega^{(k)}_{a_1\ldots a_k}=\pm \frac{(-1)^{k(d-k)}}{(d-k)!}c_{p,q}(\K)\sqrt{|\det g|}~
\epsilon_{a_1\ldots a_k}^{~~~~~~a_d\ldots a_{k+1}}\omega^{(d-k)}_{a_{k+1}\ldots a_d}~~~,
~~~\forall k=0,\ldots, d~~.
\een
We note here for future reference the expansions for the Hodge dual and
the twisted Hodge dual of any $k$-form $\omega$ :
\beqa
&&(\ast\omega)_{a_{k+1}...a_d}=\frac{(-1)^{k(d-k)}}{(d-k)!}\sqrt{|\det g|}~\epsilon_{a_{k+1}\ldots a_d}{}^{a_1\ldots a_{k}}\omega_{a_1\ldots a_{k}}~~,\\
&&(\tilde\ast\omega)_{a_{k+1}...a_d}=\frac{(-1)^{k(d-k)}}{(d-k)!}c_{p,q}(\K)\sqrt{|\det g|}~\epsilon_{a_d\ldots a_{k+1}}{}^{a_1\ldots a_{k}}\omega_{a_1\ldots a_{k}}~~.
\eeqa

\subsection{Algebraic classification of fiber types}
\label{sec:KAAlgClass}

The fibers of the \KA bundle are isomorphic with the Clifford algebra
$\Cl_\K(p,q)=\Cl(p,q)\otimes_\R \K$, whose classification is
well-known. For $\K=\C$, we have an isomorphism of algebras
$\Cl_\C(p,q)\approx \Cl_\C(d,0)\eqdef \Cl_\C(d)$ and the
classification depends only on the mod $2$ reduction of $d$; for
$\K=\R$, it depends on the mod $8$ reduction of $p-q$. The {\em Schur
algebra} $\S_\K(p,q)$ is the largest division algebra contained in the
center of $\Cl_\K(p,q)$; it is determined up to isomorphism of
algebras, being isomorphic with $\R$, $\C$ or $\H$. The Clifford
algebra is either simple (in which case it is isomorphic with a matrix
algebra $\Mat(\Delta_\K(d),\S_\K(p,q))$) or a direct sum of two
central simple algebras (namely, the direct sum
$\Mat(\Delta_\K(d),\S_\K(p,q))\oplus \Mat(\Delta_\K(d),\S_\K(p,q))$),
where the positive integers $\Delta_\K(d)$ are given by well-known
formulas recalled below.  We say that the Clifford algebra is {\em
normal} if its Schur algebra is isomorphic to the base field. It is
convenient for our purpose to organize the various cases according to
the isomorphism type of the Schur algebra and to whether the Clifford
algebra is simple or not:

\paragraph{When $\K=\C$.} In this case, the Schur algebra is always
isomorphic with $\C$ (so $\Cl_\C(p,q)\approx \Cl_\C(d)$ is always
normal) and we always have $\nu\diamond\nu=+1$ and
$\Delta_\C(d)=2^{\left[\frac{d}{2}\right]}$. Moreover:

\begin{itemize}
\item The algebra is simple iff. $d={\rm even}$, in which case
$\Cl_\C(d)\approx \Mat(\Delta_\C(d),\C)$ and $\nu$ is non-central.

\item The algebra is non-simple iff. $d={\rm odd}$, in which case
$\Cl_\C(d)\approx \Mat(\Delta_\C(d),\C)\oplus \Mat(\Delta_\C(d),\C)$
and $\nu$ is central.
\end{itemize}

\paragraph{When $\K=\R$.}

\begin{enumerate}

\item The Schur algebras and the numbers $\Delta_\R(d)$ are as follows
(see, for example, \cite{Okubo}):

\begin{itemize}
\item $\S\approx \R$ ({\em normal case}). Occurs iff. $p-q\equiv_8
0,1,2$ and we have $\Delta_\R(d)=2^{\left[\frac{d}{2}\right]}$.
\item $\S \approx \C$ ({\em almost complex case}). Occurs
iff. $p-q\equiv_8 3,7$ and we have
$\Delta_\R(d)=2^{\left[\frac{d}{2}\right]}$.
\item $\S\approx \H$ ({\em quaternionic case}). Occurs
iff. $p-q\equiv_8 4,5,6$ and we have
$\Delta_\R(d)=2^{\left[\frac{d}{2}\right]-1}$.
\end{itemize}

\item The simple and non-simple cases occur as follows:

\begin{itemize}
\item $\Cl(p,q)$ is simple iff. $p-q\equiv_8 0,2, 3, 4, 6, 7$.
\item $\Cl(p,q)$ is non-simple iff. $p-q\equiv_8 1,5$. In this case,
we always have $\nu\diamond \nu=+1$ and $\nu$ is central.
\end{itemize}
\end{enumerate} The situation when $\K=\R$ is summarized in Table
\ref{table:AlgClassif}. For both $\K=\R$ and $\K=\C$, the Clifford
algebra is non-simple iff. $\nu$ is central and satisfies $\nu\diamond
\nu=1$. In this case --- for both $\K=\R$ and $\K=\C$ --- the Clifford
algebra admits two inequivalent irreducible representations by
$\K$-linear operators, which are related by the main automorphism of
the Clifford algebra and both of which are non-faithful; their Schur
algebra equals $\C$ when $\K=\C$ but may equal either $\R$ or $\H$
when $\K=\R$.

\begin{table}
\centering
\begin{tabular}{|c|c|c|}
\hline
$\K=\R$ & $\nu\diamond\nu=+1$ & $\nu\diamond\nu=-1$ \\
\hline\hline
$\nu$ is central     & $1(\R),5(\H)$ & $3(\C), 7(\C)$ \\
\hline
$\nu$ is not central & $0(\R),4(\H)$ & $2(\R), 6(\H)$ \\
\hline
\end{tabular}
\caption{Properties of $\nu$ according to the mod 8 reduction of $p-q$
for the case $\K=\R$.  At the intersection of each row and column, we
indicate the values of $p-q~({\rm mod}~8)$ for which the modified
volume form $\nu$ has the corresponding properties. In parentheses, we
also indicate the isomorphism type of the Schur algebra for that value
of $p-q~({\rm mod}~8)$. The real Clifford algebra $\Cl(p,q)$ is
non-simple iff. $p-q\equiv_8 1,5$, which corresponds to the upper left
corner of the table. Notice that $\nu$ is central iff. $d$ is odd.}
\label{table:AlgClassif}
\end{table}

\subsection{Twisted (anti-)selfdual forms in the non-simple case}
\label{sec:KAnonsimple}

In this Subsection, let us assume that we are in the non-simple
case. Then $\nu\diamond \nu=+1$ and (since $d$ is odd in the
non-simple case) $\nu$ is a central element of the \KA algebra:
\be
\nu\diamond \omega=\omega\diamond \nu~~,~~\forall \omega\in \Omega_\K(M)~~.
\ee
Using the fact that $\nu$ is central, an easy computation shows that
$P_+$ and $P_-$ are (non-unital) {\em algebra} endomorphisms of the
\KA algebra, in fact:
\be
P_\pm(\omega\diamond \eta)=P_\pm(\omega)\diamond P_\pm(\eta)=
P_\pm(\omega)\diamond \eta=\omega\diamond P_\pm(\eta)~~,~~\forall \omega, \eta\in \Omega_\K(M)~~
\ee
and:
\be
P_\pm(1_M)=p_\pm~~.
\ee
In this case, $\Omega^\pm_\K(M)$ are complementary {\em two-sided}
ideals of the \KA algebra (indeed, $p_\pm$ are central) --- in
particular, $(\Omega^\pm_\K(M),\diamond)$ are unital algebras, their
units being given by $p_\pm$.

\paragraph{Truncation and prolongation.} Since $d$ is odd in the non-simple case, we have the decomposition:
\be
\Omega_\K(M)=\Omega_\K^<(M)\oplus\Omega_\K^>(M)~~,
\ee
where:
\be
\Omega_\K^<(M)\eqdef \oplus_{k=0}^{\left[\frac{d}{2}\right]} \Omega^k_\K(M)~~,
~~\Omega_\K^>(M)\eqdef \oplus_{k=\left[\frac{d}{2}\right]+1}^d \Omega^k_\K(M)~~.
\ee
The corresponding complementary $\cinf$-linear idempotent operators $P_<, P_>:\Omega_\K(M)\rightarrow \Omega_\K(M)$ are given by:
\be
P_<(\omega)\eqdef \omega^<~~,~~P_>(\omega)\eqdef \omega^>~~,
\ee
where, for any
$\omega=\sum_{k=0}^d\omega^{(k)}\in \Omega_\K(M)$ (with $\omega^{(k)}\in \Omega^k_\K(M)$), we define $\omega^<$ (the {\em lower truncation} of
$\omega$) and $\omega^>$ (the {\em upper truncation} of $\omega$) through:
\be
\omega^<\eqdef \sum_{k=0}^{\left[\frac{d}{2}\right]}\omega^{(k)}~~~,~~~\omega^>\eqdef \sum_{k=\left[\frac{d}{2}\right]+1}^d\omega^{(k)}~~.
\ee
We have:
\be
P_>+P_<=\id_{\Omega_\K(M)}~~,~~P_>\circ P_<=P_<\circ P_>=0~~,~~P_>\circ P_>=P_>~~,~~P_<\circ P_<=P_<~~.
\ee
When $\omega$ is twisted (anti-)selfdual (i.e. $\omega\in \Omega^\epsilon_\K(M)$ with $\epsilon=\pm 1$),
we have: ${\tilde \ast}\omega=\epsilon \omega$, which implies:
\be
\omega^>=\epsilon {\tilde \ast}(\omega^<)~~,~~\forall \omega\in \Omega^\epsilon_\K(M)~~.
\ee
Hence in this case $\omega$ can be re-constructed from its lower truncation as:
\be
\omega=\omega^< +\epsilon {\tilde \ast}(\omega^<)=2 P_\epsilon (\omega^<)=P_\epsilon(2P_<(\omega))
~~,~~\forall \omega\in \Omega^\epsilon_\K(M)~~.
\ee
It follows that the restriction of $2 P_<$ to the subspace
$\Omega^\epsilon_\K(M)$ gives a $\cinf$-linear bijection between this subspace and
the subspace $\Omega^<_\K(M)$, with inverse given by the restriction
of $P_\epsilon$ to $\Omega^<_\K(M)$. We define the {\em twisted
(anti-)selfdual prolongation} of a form $\omega\in \Omega^<_\K(M)$
through:
\ben
\label{asdprol}
\omega_\pm\eqdef P_\pm(\omega)~~,~~\forall \omega\in \Omega^<_\K(M)~~.
\een
Of course, the form $\omega\in \Omega^<_\K(M)$ can be recovered from
its two prolongations as $\omega=\omega_++\omega_-$.

\paragraph{The truncated algebra $(\Omega^<_\K(M),\bdiamond_\pm)$.}

We stress that $P_<$ does not preserve the geometric product
on its entire domain of definition $\Omega_\K(M)$ --- in fact, its
image $\Omega^<_\K(M)$ is {\em not} a subalgebra of the \KA algebra
since it is not stable with respect to $\diamond$-multiplication. To
cure this problem, we use the linear isomorphisms mentioned above to
transfer the multiplication $\diamond$ of the unital subalgebra
$\Omega^\epsilon_\K(M)$ to an associative and unital multiplication
$\bdiamond_\epsilon$ defined on $\Omega^<_\K(M)$ through:
\ben
\label{bdDef}
\omega\bdiamond_\epsilon \eta=2P_<(P_\epsilon(\omega)\diamond
P_\epsilon(\eta))\in \Omega^<_\K(M)~\Longleftrightarrow
P_\epsilon(\omega\bdiamond_\epsilon \eta)=P_\epsilon(\omega)\diamond
P_\epsilon(\eta)~~,~~\forall \omega,\eta\in \Omega^<_\K(M)~.
\een

Since $P_\epsilon$ {\em is} a morphism of algebras on its entire domain of definition $\Omega_\K(M)$,
we have $P_\epsilon(\omega)\diamond P_\epsilon(\eta)=P_\epsilon(\omega \diamond \eta)$, so \eqref{bdDef} gives:
\ben
\label{bdiamondDef}
\omega\bdiamond_\epsilon\eta =2P_<(P_\epsilon(\omega\diamond\eta))~~,~~\forall \omega,\eta\in \Omega^<_\K(M)~~.
\een
Since $2P_\epsilon (\omega\diamond \eta) = (1+\epsilon {\tilde \ast}) (P_<(\omega\diamond \eta)+P_>(\omega\diamond\eta))$ and
${\tilde \ast}\circ P_<=P_>\circ {\tilde \ast}$, this implies:
\ben
\label{bdiamond}
\omega\bdiamond_\epsilon \eta=P_<(\omega\diamond \eta)+\epsilon {\tilde \ast} P_>(\omega\diamond \eta)
=(\omega\diamond \eta)^<+
\epsilon {\tilde \ast} [(\omega\diamond \eta)^>]~~,
\een
a formula which can be used to implement the product $\bdiamond_\epsilon$ in a
symbolic computation system. Combining everything shows that we have
mutually inverse isomorphisms of algebras:
\ben
\label{diagram:bdiamond}
\scalebox{1}{
\xymatrix@1{
(\Omega^{<}_\K(M),\bdiamond_\epsilon) \ar@<0.5ex>[r]^{P_\epsilon|_{\Omega^{<}_\K(M)}} {~~~}
& {~~~}(\Omega_\K^\epsilon (M),\diamond) \ar@<0.5ex>[l]^{2 P_<|_{\Omega_\K^\epsilon (M)}}~~.
}}
\een
Thus $(\Omega^<_\K(M),\bdiamond_\epsilon)$ provides a model for the unital associative algebra
$(\Omega^\epsilon_\K(M),\diamond)$.

\paragraph{Local expansions.} Let us further assume that $d\geq 3$. Then the covector fields $e^a\in
\Omega^1_\K(M)$ defined by a local pseudo-orthonormal frame above $U\subset M$ belong to the
subspace $\Omega^<_\K(M)$ and we consider their twisted (anti-)selfdual
prolongations:
\ben
\label{eaprol}
e^a_\pm\eqdef P_\pm(e^a)\in \Omega^\pm_\K(M)~~.
\een
Since $e^{a_1\ldots a_k}=e^{a_1}\wedge \ldots \wedge
e^{a_k}=e^{a_1}\diamond \ldots \diamond e^{a_k}$ and since $P_\pm$ are
endomorphisms of the \KA algebra, we find that the prolongations of
$e^{a_1\ldots a_k}$ are given by $\diamond$-monomials in the
prolongations of $e^a$:
\ben
\label{eAprol}
e^{a_1\ldots a_k}_\pm\eqdef P_\pm(e^{a_1\ldots a_k})=e^{a_1}_\pm\diamond \ldots \diamond e^{a_k}_\pm~~.
\een
In particular, the twisted (anti-)selfdual forms $\{e^{a_1\ldots
a_k}_\pm|1\leq a_1<\ldots < a_k\leq d, k=0,\ldots,
\left[\frac{d}{2}\right]\}$ constitute a basis of the free
$\cC^\infty(U,\K)$-module $\Omega^\pm_\K(U)$ (since $\{e^{a_1\ldots
a_k} |1\leq a_1<\ldots < a_k\leq d, k=0,\ldots,
\left[\frac{d}{2}\right]\}$ form a basis of the module
$\Omega^<_\K(U)$ and since the operation of taking the prolongation is
an isomorphism of $\cC^\infty(U,\K)$-modules). In fact, any twisted
(anti-)selfdual form $\omega\in \Omega^\pm_\K(M)$
(cf. \eqref{FormExpansion},\eqref{HomFormExpansion}) expands as:
\be
\omega=_U 2\sum_{k=0}^{\left[\frac{d}{2}\right]}\frac{1}{k!}
\omega^{(k)}_{a_1\ldots a_k}e^{a_1\ldots a_k}_\pm~~,~~\forall \omega\in \Omega^\pm_\K(M)~~,
\ee
where $\omega^{(k)}_{a_1\ldots a_k}\in \cC^\infty(U,\K)$ are as in
\eqref{HomFormExpansion}. The coefficients with $k\geq
\left[\frac{d}{2}\right]+1$ are determined by those with $k\leq
\left[\frac{d}{2}\right]$ through relations
\eqref{TwistedHodgeCoeffs}.  The lower truncation of such $\omega$ has
the local expansion:
\be
\omega^<=_U\sum_{k=0}^{\left[\frac{d}{2}\right]}\frac{1}{k!}\omega^{(k)}_{a_1\ldots a_k}e^{a_1\ldots a_k}~~.
\ee
We also note the explicit expressions:
\be
e^{a_1\ldots a_k}_\pm=\frac{1}{2}(e^{a_1\ldots a_k}\pm c_{p,q}(\K)
\epsilon^{a_k\ldots a_1}_{~~~~~~a_{k+1}\ldots a_d}e^{a_{k+1}\ldots a_d})~~,
\ee
where $c_{p,q}(\K)$ was defined in \eqref{nudef}.

\subsection{Orthogonality in the non-simple case}
\label{sec:KAreducedHodge}

Assuming that we are in the non-simple case, let us consider the
situation when we have a distinguished one-form $\theta\in
\Omega^1_\K(M)$ which satisfies the normalization condition
$\iota_\theta\theta=1$.

\paragraph{The isomorphism of algebras between $\Omega^\perp_\K(M)$
and $\Omega_\K^\epsilon (M)$.} For any $\omega\in \Omega_\K(M)$, consider the decomposition
$\omega=\omega_{\parallel}+\omega_\perp$ into parts
$\omega_{\parallel}=\theta\wedge (\iota_\theta\omega) $ and
$\omega_\perp=\iota_\theta(\theta \wedge
\omega)=\omega-\omega_{\parallel}$ parallel and perpendicular to
$\theta$. Since $\theta\parallel\nu$ (indeed, we have $\theta\wedge \nu=0$) and ${\tilde \ast}
\omega=\omega\diamond \nu$, properties \eqref{parperptrans} imply:
\be
\theta \parallel {\tilde \ast}(\omega_{\perp})~~,~~\theta \perp {\tilde \ast}(\omega_\parallel)~~,
\ee
which gives:
\ben
\label{HodgeParPerp}
({\tilde \ast}\omega)_{\parallel}={\tilde \ast}(\omega_\perp)~~, ~~({\tilde \ast}\omega)_\perp= {\tilde \ast}(\omega_{\parallel})~~.
\een
The subalgebras $\Omega^\epsilon_\K(M)$ and $\Omega_\K^\perp(M)$ of
the \KA algebra can be identified with each other using the operator $2P_\perp$, which
takes a twisted (anti-)selfdual form $\omega$ into $2\omega_\perp$. Indeed,
if ${\tilde \ast \omega}=\epsilon \omega$ (with $\epsilon=\pm 1$),
then ${\tilde \ast}(\omega_\parallel) =\epsilon\omega_\perp$ and ${\tilde \ast}(\omega_\perp) =\epsilon\omega_\parallel$. 
Hence the last of
equations \eqref{ometaparts} implies:
\be
(\omega\diamond\eta)_\perp={\tilde \ast}(\omega_\perp)\diamond{\tilde \ast}(\eta_\perp)+\omega_\perp\diamond\eta_\perp=2\omega_\perp \diamond \eta_\perp~~,
\ee
since ${\tilde \ast}(\omega_\perp)\diamond{\tilde \ast}(\eta_\perp)=\omega_\perp\diamond \nu\diamond \eta_\perp\diamond \nu=\omega_\perp\diamond \eta_\perp\diamond \nu \diamond \nu=
\omega_\perp\diamond \eta_\perp$, where we used the fact that $\nu$ is central in the \KA algebra and that it squares to $1_M$. Thus:
\be
(\omega\diamond \eta)_\perp=2\omega_\perp \diamond \eta_\perp \Longleftrightarrow
2 P_\perp(\omega\diamond \eta)=2(2P_\perp\omega)\diamond (2P_\perp\eta)~~,~~\forall \omega,\eta\in \Omega^\epsilon_\K(M)~~.
\ee
We also have:
\be
2P_\perp(p_\pm)=P_\perp(1\pm \nu)=P_\perp(1_M)=1_M~~,
\ee
where we used the fact that $\theta||\nu$. These properties show that
the restriction $2P_\perp|_{\Omega^\epsilon_\K(M)}$ is a unital morphism of
algebras from $(\Omega^\epsilon_\K(M),\diamond)$ to
$(\Omega^\perp_\K(M),\diamond)$. An easy computation shows that it is
an isomorphism whose inverse equals the restriction of $P_\epsilon$ to
$\Omega^\perp_\K(M)$. It follows that we have mutually inverse
unital isomorphisms of algebras:
\ben
\label{DiagramASDPerp}
\scalebox{1}{
\xymatrix@1{
(\Omega_\K^\epsilon (M), \diamond) \ar@<0.7ex>[r]^{2P_\perp|_{\Omega_\K^\epsilon (M)}} {~~~} &
{~~~}(\Omega^\perp_\K (M),\diamond) \ar@<0.7ex>[l]^{P_\epsilon|_{\Omega_\K^\perp (M)}} ~~.
}}
\een
Combining with the results of Subsection \ref{sec:KAnonsimple}, we have thus found two isomorphic models for the unital
subalgebra $(\Omega^\epsilon_\K(M),\diamond)$:
\ben
\label{diagram:twomodels}
\scalebox{1}{
\xymatrix@1{(\Omega^{<}_\K(M),\bdiamond_\epsilon) ~~~~\ar@<0.5ex>[r]^{~P_\epsilon|_{\Omega^{<}_\K(M)}~} &
{~~~}(\Omega_\K^\epsilon (M),\diamond) \ar@<0.5ex>[l]^{~2P_{<}|_{\Omega_\K^\epsilon (M)}~~} ~~~~\ar@<0.5ex>[r]^{~~2P_\perp|_{\Omega_\K^\epsilon (M)}}  &
{~~~~}(\Omega^\perp_\K (M),\diamond) \ar@<0.5ex>[l]^{~~P_\epsilon|_{\Omega_\K^\perp (M)}~}}}~~.
\een

\paragraph{The reduced twisted Hodge operator.} Since
$\theta \parallel \nu$, we can write:
\be
\nu=\theta\wedge \nu_\top~~,~~
\ee
where the {\em reduced volume form} $\nu_\top$ is defined through:
\ben
\label{nu0def}
\nu_\top\eqdef \iota_\theta\nu = \theta\diamond \nu =\nu\diamond \theta ~~.
\een
The last two equalities in \eqref{nu0def} follow from \eqref{basicid}
and from the fact that (in the non-simple case) $\nu$ is central in the \KA algebra (since $d$ is odd in
this case).  Multiplying the last equation with $\theta$ in the \KA
algebra and using the fact that $\theta\diamond \theta={\hat
g}(\theta,\theta)=1$ gives:
\ben
\label{thetanu0comm}
\nu=\nu_\top\diamond\theta=\theta\diamond \nu_\top~~.
\een
Notice the identity:
\ben
\label{nu0squared}
\nu_\top^2=+1_M~~,
\een
which follows from \eqref{nu0def} using the fact that $\nu$ is
central, the normalization condition for $\theta$ and the property
$\nu\diamond \nu=+1_M$, which always holds in the
non-simple case.  Defining the {\em reduced twisted Hodge operator}
${\tilde \ast}_0$ through:
\ben
\label{ReducedHodgeDef}
{\tilde \ast}_0\omega\eqdef \pi(\omega)\diamond \nu_\top~~,~~\forall \omega\in
\Omega_\K(M)\Longleftrightarrow {\tilde \ast}_0=R_{\nu_\top}\circ \pi~~,
\een
we have $\pi(\nu_\top)=\nu_\top$, so equation \eqref{nu0squared} implies:
\ben
\label{RedHodgeSquared}
{\tilde \ast}_0\circ {\tilde \ast}_0=+\id_{\Omega_\K (M)}~~.
\een
For later reference, we note the identities (where we use \eqref{thetanu0comm} and
the fact that $\pi(\nu_\top)=\nu_\top$):
\be
[\pi, R_{\nu_\top}]_{-,\circ}=[\pi,{\tilde \ast}_0]_{-,\circ}=0~~
\ee
as well as:
\be
[L_\theta,{\tilde \ast}_0]_{+,\circ}=[R_\theta,{\tilde \ast}_0]_{+,\circ}=0~~,
\ee
which follow by easy computation. Using \eqref{basicop2}, the last
identities imply the following anticommutation relations, which will be
important below:
\ben
\label{TwistedHodgeAcomms}
[\wedge_\theta, {\tilde \ast}_0]_{+,\circ}=[\iota_\theta,{\tilde \ast}_0]_{+,\circ}=0~~.
\een
To find explicit expressions for the parallel and perpendicular parts
of ${\tilde \ast}\omega$, notice that ${\tilde \ast}\omega=\omega
\diamond \nu=\omega\diamond \nu_\top\diamond \theta = \theta \wedge \pi
(\omega\diamond \nu_\top)-\iota_\theta \pi(\omega\diamond \nu_\top)=\theta
\wedge (\pi(\omega)\diamond \nu_\top) -\iota_\theta (\pi(\omega)\diamond
\nu_\top)$, where we used \eqref{basicid} and the fact that
$\pi(\nu_\top)=+\nu_\top$.
Thus:
\ben
\label{tildeastdec}
\!\!\!\!\! ({\tilde \ast}\omega)_{\parallel}=\theta \wedge \left[\pi(\omega)\diamond \nu_\top\right]
=\theta \wedge \left[(\pi(\omega)\diamond \nu_\top)_\perp\right]~,
~({\tilde \ast}\omega)_\perp=-\iota_\theta \left[\pi(\omega)\diamond \nu_\top\right]=
-\iota_\theta \left[(\pi(\omega)\diamond \nu_\top)_{\parallel}\right].
\een
The decomposition $\omega=\omega_{\parallel}+\omega_\perp$ and
the fact that $\iota_\theta \nu_\top=0$ (thus $\theta\perp \nu_\top$) imply (using \eqref{parperptrans}):
\be
\theta \parallel (\omega_{\parallel}\diamond \nu_\top)~~,~~\theta \perp (\omega_\perp \diamond \nu_\top)~~,
\ee
and (using \eqref{piparperp} and the fact that $\pi(\nu_\top)=+1$):
\be
\theta \parallel (\pi(\omega_{\parallel})\diamond \nu_\top)~~,~~\theta \perp (\pi(\omega_\perp) \diamond \nu_\top)~~.
\ee
These relations show that:
\be
(\omega\diamond \nu_\top)_{\parallel}=\omega_{\parallel}\diamond \nu_\top~~{\rm and}~~(\omega\diamond \nu_\top)_\perp=\omega_\perp\diamond \nu_\top~~
\ee
as well as:
\ben
\label{ta0props1}
(\pi(\omega)\diamond \nu_\top)_{\parallel}=\pi(\omega_{\parallel})\diamond \nu_\top~~{\rm and}~
~(\pi(\omega)\diamond \nu_\top)_\perp=
\pi(\omega_\perp)\diamond \nu_\top~~.
\een
Combining the last relation with \eqref{tildeastdec} gives:
\ben
\label{ta0props2}
({\tilde \ast}\omega)_{\parallel}=\theta \wedge \left[\pi(\omega_\perp)\diamond \nu_\top\right]~~,
~~({\tilde \ast}\omega)_\perp=
-\iota_\theta \left[\pi(\omega_{\parallel})\diamond \nu_\top\right]~.
\een
Equations \eqref{ta0props2} and \eqref{HodgeParPerp} read:
\ben
\label{reducedasd}
({\tilde \ast}\omega)_{\parallel}={\tilde \ast}(\omega_\perp)=\theta\wedge {\tilde \ast}_0(\omega_\perp)~~,
~~~({\tilde \ast}\omega)_\perp={\tilde \ast}(\omega_{\parallel})=-
\iota_\theta {\tilde \ast}_0 (\omega_{\parallel})~~,
\een
while \eqref{ta0props1} gives:
\be
({\tilde \ast}_0\omega)_{\parallel}={\tilde \ast}_0(\omega_{\parallel})~~,~~({\tilde \ast}_0\omega)_\perp={\tilde \ast}_0(\omega_\perp)~~.
\ee
In particular, we have:
\be
[{\tilde \ast}_0,P_\parallel]_{-,\circ}=[{\tilde \ast}_0,P_\perp]_{-,\circ}=0~~
\ee
and ${\tilde \ast}=\wedge_\theta \circ {\tilde \ast}_0\circ P_\perp -\iota_\theta\circ {\tilde \ast}_0\circ P_\parallel=
\wedge_\theta \circ P_\perp \circ {\tilde \ast}_0 -\iota_\theta\circ P_\parallel \circ {\tilde \ast}_0$, which gives:
\ben
\label{tildedec}
{\tilde \ast}=\wedge_\theta\circ {\tilde \ast}_0 -\iota_\theta \circ {\tilde \ast}_0
\een
upon using $\wedge_\theta \circ P_\parallel=\iota_\theta \circ
P_\perp=0\Leftrightarrow \wedge_\theta \circ
P_\perp=\wedge_\theta~~{\rm and}~~\iota_\theta \circ P_\parallel
=\iota_\theta$.  The relations above imply the following:

\paragraph{Lemma.} Consider the operators $\alpha_\theta\eqdef \wedge_\theta \circ{\tilde \ast}_0$ and
$\beta_\theta=-\iota_\theta \circ{\tilde \ast}_0$.
Then:
\ben
\label{alphabeta}
\alpha_\theta\circ \alpha_\theta=\beta_\theta\circ \beta_\theta=0~~,
~~\alpha_\theta\circ \beta_\theta=P_{\parallel}~~,~~\beta_\theta\circ \alpha_\theta=P_\perp~~.
\een
\noindent {\bf Proof.} The statement follows by direct computation
using properties \eqref{TwistedHodgeAcomms} and
\eqref{RedHodgeSquared}.

\

\noindent Notice that \eqref{reducedasd} takes the form:
\ben
\label{rasd}
({\tilde \ast}\omega)_{\parallel}={\tilde \ast}(\omega_\perp)=\alpha_\theta (\omega_\perp)~~,
~~~({\tilde \ast}\omega)_\perp={\tilde \ast}(\omega_{\parallel})=\beta_\theta (\omega_{\parallel})~~,
\een
For reader's convenience, we also list a few other properties which follow immediately from the above:
\beqa
\alpha_\theta \circ P_\parallel=0 \Longrightarrow \alpha_\theta \circ P_\perp=\alpha_\theta~~&,&~~
P_\perp \circ \alpha_\theta =0 \Longrightarrow P_\parallel \circ \alpha_\theta=\alpha_\theta~~,\nn\\
\beta_\theta \circ P_\perp=0 \Longrightarrow \beta_\theta \circ P_\parallel=\beta_\theta~~&,&~~
P_\parallel \circ \beta_\theta =0 \Longrightarrow P_\perp \circ \beta_\theta=\beta_\theta~~,\nn\\
P_\parallel \circ {\tilde \ast}=\alpha_\theta~~&,&~~P_\perp\circ {\tilde \ast}=\beta_\theta~~.\nn
\eeqa

\paragraph{Proposition.} Let $\omega\in \Omega_\K(M)$. Then the following statements are equivalent:

\

\noindent (a) $\omega$ is twisted (anti-)selfdual, i.e. ${\tilde \ast} \omega=\epsilon\omega$ for $\epsilon=\pm 1$~~.

\

\noindent (b) The components $\omega_{\parallel}$ and $\omega_\perp$ satisfy the following equivalent relations:
\ben
\label{redasd}
\omega_{\parallel}=\epsilon \theta\wedge {\tilde \ast}_0(\omega_\perp)~~,~~\omega_\perp=
-\epsilon \iota_\theta {\tilde \ast}_0 (\omega_{\parallel})~~.
\een
In this case, $\omega_{\parallel}$ and $\omega_\perp$ determine each
other and thus any of them determines $\omega$.

\

\noindent {\bf Proof.} The fact that the two relations listed in
\eqref{redasd} are equivalent to each other is an immediate
consequence of the Lemma.  The rest of the proposition follows from
\eqref{tildedec}.

\

\noindent Recalling the definition \eqref{TopDef}, we have $\theta \perp \omega_\top$ and
$\omega_{\parallel}=\wedge_\theta \omega_\top$, so the decomposition of $\omega$ reads $\omega=\theta\wedge\omega_\top+\omega_\perp$.
Using \eqref{TwistedHodgeAcomms}, we find:
\be
\beta_\theta\circ\wedge_\theta=P_\perp\circ {\tilde \ast}_0={\tilde \ast}_0 \circ P_\perp~~,
\ee
which implies:
\ben
\label{betaexpl}
\beta_\theta(\omega_{\parallel})={\tilde \ast}_0 (\omega_\top)\Longrightarrow ({\tilde \ast}\omega)_\perp=
{\tilde \ast}(\omega_\parallel)={\tilde \ast}_0 (\omega_\top)~~,
\een
where in the last equality we used \eqref{rasd}. Hence the previous proposition has the following:

\paragraph{Corollary.} The following statements are equivalent for any $\omega\in \Omega_\K(M)$:

\

\noindent (a)  $\omega$ is twisted (anti-)selfdual, i.e. ${\tilde \ast} \omega=\epsilon\omega$ for $\epsilon=\pm 1$.

\

\noindent (b) The inhomogeneous forms $\omega_\top=\iota_\theta
\omega$ and $\omega_\perp=\iota_\theta (\theta\wedge \omega)$ satisfy
the equation:
\ben
\label{TopPerp}
\omega_\perp=\epsilon {\tilde \ast}_0 \omega_\top\Longleftrightarrow
\omega_\top=\epsilon {\tilde \ast}_0 \omega_\perp~~.
\een
In this case, $\omega_\perp$ and $\omega_\top$ determine each other
and thus any of them determines $\omega$.  Explicitly, $\omega_\top$
determines $\omega$ through the formula:
\ben
\label{OmegaFromTop}
\omega=(\wedge_\theta +\epsilon {\tilde \ast}_0) (\omega_\top)~~,
\een
while $\omega_\perp$ determines $\omega$ through:
\ben
\label{OmegaFromPerp}
\omega=(\id_{\Omega_\K(M)} +\epsilon \wedge_\theta \circ {\tilde \ast}_0) (\omega_\perp)~~.
\een

\noindent The corollary shows that the maps
$\wedge_\theta +\epsilon {\tilde \ast}_0
:\Omega_\K^\perp (M)\stackrel{\sim}{\rightarrow}
\Omega^\epsilon_\K(M)$ are isomorphisms of $\cC^\infty(M,\K)$-modules,
whose inverses are given by $\omega\rightarrow \omega_\top$. We stress that these maps
are {\em not} isomorphisms of algebras.

\paragraph{The morphism $\varphi_\epsilon$.} For later reference,
consider the $\cinf$-linear operator:
\be
\varphi_\epsilon\eqdef 2P_\perp\circ P_\epsilon: \Omega_\K(M)\rightarrow \Omega^\perp_\K(M)
\ee
which acts as follows on $\omega=\theta\diamond\omega_\top+\omega_\perp=\theta\wedge \omega_\top+\omega_\perp\in \Omega_\K(M)$:
\ben
\label{phiDef}
\varphi_\epsilon(\omega)=\epsilon {\tilde \ast}_0(\omega_\top)+\omega_\perp=
\epsilon \nu_\top\diamond \omega_\top+\omega_\perp~~,
\een
where we used equation \eqref{betaexpl} and we noticed that $ \nu_\top\diamond \omega_\top=\pi(\omega_\top)
\diamond \nu_\top={\tilde \ast}_0(\omega_\top)$ (since $\omega_\top$ and $\nu_\top$ are
orthogonal to $\theta$ and since $\rk\nu_\top=d-1$ is even).  We have $\varphi_\epsilon(\theta)=\epsilon \nu_\top$
(since $\theta_\top=1$ and $\theta_\perp=0$) and
$\varphi_\epsilon(\omega)=\omega$ for all $\omega\in \Omega^\perp_\K(M)$; in
fact, these properties determine $\varphi_\epsilon$. One has the
following:

\paragraph{Proposition.} The map $\varphi_\epsilon$ is idempotent, i.e. $\varphi_\epsilon\circ \varphi_\epsilon=\varphi_\epsilon$.
Moreover, it is a (unital) morphism of algebras from $(\Omega_\K(M),\diamond)$ to $(\Omega^\perp(M),\diamond)$.

\

\noindent {\bf Proof.} Idempotency follows by noticing that $\varphi_\epsilon|_{\Omega^\perp_\K(M)}=\id_{\Omega^\perp_\K(M)}$.
The fact that $\varphi_\epsilon$ is a morphism of algebras follows since both $P_\perp$ and $P_\epsilon$ are such. 
Finally, unitality of $\varphi_\epsilon$ follows by computing: 
\be
\varphi_\epsilon(1_M)=P_\perp(1_M+\epsilon \nu)=1_M~~,
\ee
where we used $P_\perp(1_M)=1_M$ and $P_\perp(\nu)=0$. 

\

\noindent It is clear that $\varphi_\epsilon$ is surjective. Moreover, the last
proposition of the previous paragraph implies: 
\be
\cK(\varphi_\epsilon)=\Omega^{-\epsilon}_\K(M)~~.  
\ee It follows that
$\varphi_\epsilon$ restricts to an isomorphism from
$\Omega^\epsilon_\K(M)$ to $\Omega^\perp_\K(M)$ --- which, of course, equals the isomorphism
$2P_\perp|_{\Omega^\epsilon_\K(M)}$ of diagram \eqref{DiagramASDPerp}. Notice the relations:
\ben
\label{varphiP}
P_\epsilon\circ \varphi_\epsilon=P_\epsilon~~,~~\varphi_\epsilon\circ P_\epsilon=\varphi_\epsilon~~,
~~P_\perp\circ \varphi_\epsilon=\varphi_\epsilon~~,~~\varphi_\epsilon\circ  P_\perp=P_\perp~~,
\een
where the first equality follows from the fact that $2P_\epsilon\circ P_\perp$ restricts to the identity on $\Omega^\epsilon_\K(M)$ (see diagram \eqref{DiagramASDPerp}). Also notice the property:
\ben
\label{varphinu}
\varphi_\epsilon(\nu)=\epsilon 1_M~~,
\een
which follows by direct computation upon using $P_\perp(1_M)=1_M$, $P_\perp(\nu)=0$ and the fact that $\nu\diamond \nu=1_M$. 

\section{Describing bundles of pinors}
\label{sec:pin}

In this section, we discuss the realization of pin bundles within the
geometric algebra formalism --- focusing especially on the non-simple
case, when the irreducible pin representations are non-faithful.
Subsection \ref{sec:pinbasic} discusses an approach to pinor bundles
which is particularly well adapted to the geometric algebra
formalism. In this approach (which, in some ways, goes back to Dirac --- see
\cite{Trautman, FriedrichTrautman} for a beautiful treatment), one
defines pinors as sections of a bundle $S$ of modules over the \KA algebra, the fiberwise
module structure being described by a morphism
$\gamma:(\Omega_\K(M),\diamond)\rightarrow (\End(S),\circ)$ of bundles of algebras.
For the case when $\gamma$ is irreducible on the fibers, the well-known representation theory relevant for this construction
and its relation with the fiber type classification of the \KA bundle
is recalled in Subsection \ref{sec:pinrep}, paying attention to
characterizing the kernel and image of $\gamma$.  In Subsection
\ref{sec:pininverse}, we introduce a certain `partial inverse'
$\gamma^{-1}$ of $\gamma$, which provides a sort of `vertical
dequantization' map.  Subsection \ref{sec:pintrace} discusses a trace
on the subalgebra $\Omega^\gamma_\K(M)$, which is related by $\gamma$
to the natural fiberwise trace on the pin bundle.

\subsection{Basic considerations}
\label{sec:pinbasic}

We define a bundle of $\K$-{\em pinors} to be a
bundle $S$ of modules over the Clifford bundle $\Cl(T^\ast_\K M)$,
i.e. a $\K$-vector bundle all of whose fibers $S_{x}$ ($x\in M$) are
modules over the corresponding Clifford algebras $\Cl(T^{\ast}_{\K,x}
M)$.  In our language, such a bundle is simply a bundle of modules
over the \KA bundle $(\wedge T^{\ast}_\K M,\diamond )$. In the
particular case when the morphism $\gamma:(\wedge T^{\ast}_\K
M,\diamond )\rightarrow \End(S)$ induces an {\em irreducible}
representation of the Clifford algebra $\Cl(T_{\K,x}^{\ast} M)$ on
each fiber $\End(S_{x})$, a bundle of pinors will be called a {\em pin
  bundle}.  Bundles of $\K$-{\em spinors} and $\K$-{\em spin} bundles
are defined similarly, but replacing $\Cl(T^\ast_\K M)$ with
$\Cl^\ev(T^\ast_\K M)$, i.e. replacing the \KA bundle with its even sub-bundle. 
Physically, smooth sections of a (s)pin bundle describe $\K$-valued (s)pinors of spin $1/2$ defined over the manifold
$M$.  As explained in \cite{FriedrichTrautman, Trautman}, a pin bundle
$S$ in our sense exists on $M$ iff. $M$ admits a so-called ${\rm
  Clifford}^c$-structure, i.e. a `reduction' of the structure group of
the bundle of pseudo-orthonormal frames of $T_\K M$ to a certain
extension of the Clifford (a.k.a. Lipschitz) group over $\K$. In this
case, the choices globally available for $S$ depend --- up to
isomorphism --- on the choices of a ${\rm Clifford}^c$ structure and
can be obtained from such a structure by applying the associated
bundle construction.  The map induced on sections satisfies:
\ben
\label{gammamf}
\gamma(\omega\diamond  \eta)=\gamma(\omega)\circ \gamma(\eta)~~,
~~\forall \omega,\eta\in \Omega_\K(M)
\een
as well as:
\be
\gamma(1_M)=\id_S~~,
\ee
where $\id_S\in \Gamma(M, \End(S))$ denotes the identity section of
the bundle $\End(S)$.

In the language of `vertical quantization' of spin systems, the (${\rm
L}^2$-completion of the) space $\Gamma(M,S)$ of smooth sections of $S$
plays the role of the Hilbert space (when $\K=\R$, one of course has
to consider the complexification of $S$ instead).  In this
interpretation, $\gamma$ plays the role of quantization map, giving a
morphism from the algebra of quantum observables of the system (which
is the \KA algebra) to the (${\rm L}^2$-completion of the) algebra
$(\Gamma(M,\End(S)),\circ)$, which plays the role of algebra of
`vertical' operators acting in the Hilbert space. These statements can be made
quite precise provided that certain global conditions are imposed on
$(M,g)$, but the rigorous treatment of this issue falls outside of the
scope of this paper.

\paragraph{Notation.} If $(e_a)_{a=1\ldots d}$ is a local
frame of $TM$ above an open subset $U\subset M$ (with dual coframe
$(e^a)_{a=1\ldots d}$), then any inhomogeneous differential form
$\omega$ on $M$ expands locally as in \eqref{FormExpansion}. We define
$\gamma^a\stackrel{{\rm def.}}{=}\gamma(e^a)\in \Gamma(U,\End(S))$,
so $\gamma(e^{a_1\ldots a_k})=\gamma^{a_1\ldots
a_k}\eqdef \frac{1}{k!}\epsilon_{a_1\ldots a_k}\gamma^{a_1}\circ \ldots
\circ \gamma^{a_k}\in \Gamma(U,\End(S))$ (the complete
antisymmetrization of the composition $\gamma^{a_1}\circ \ldots \circ
\gamma^{a_k}$). We have:
\be
\gamma(\omega)\eqdef _U\sum_{k=0}^d{\frac{1}{k!}\omega^{(k)}_{a_1\ldots
a_k}\gamma^{a_1\ldots a_k}}~~.
\ee
The locally-defined sections $\gamma^a\in
\Gamma(U,\End(S))$ correspond to physicists' `gamma matrices'.

\subsection{Representation theory}
\label{sec:pinrep}

Let $S$ be a pin bundle with underlying morphism $\gamma: (\Omega_\K(M),\diamond)\rightarrow (\End(S),\circ)$. 

\paragraph{Injectivity and surjectivity of $\gamma$.}
It is important to note that $\gamma$ need not be fiberwise injective
or surjective, i.e. the morphisms of algebras
$\gamma_{x}:(\Lambda T_{\K,x}^\ast M,\diamond_{x})\approx
\Cl(T_{\K,x}^\ast M)\rightarrow \End(S_{x})$ need not be injective
or surjective. The following characterization is convenient in this regard:

\begin{itemize}
\item $\gamma$ is fiberwise injective iff. the fiber of the \KA bundle
is simple as an associative algebra.
\item $\gamma$ is fiberwise surjective iff. the Schur algebra of the
fiber of the \KA bundle is isomorphic with the base field $\K$.
\end{itemize}

\noindent This gives the following classification:

\paragraph{The case $\K=\C$.} Then $\gamma$ is always fiberwise
surjective, being fiberwise injective iff. $d$ is even.

\paragraph{The case $\K=\R$.} Then $\gamma$ is fiberwise surjective
iff. $p-q\equiv_8 0,1,2$. It is fiberwise injective iff. $p-q\equiv_8
0,2,3,4,6,7$. This is summarized in table \ref{table:InjSurj}.

\begin{table}
\centering
\begin{tabular}{|c|c|c|}
\hline
$\K=\R$ & injective & non-injective \\
\hline\hline
surjective     & $ 0(\R), 2(\R)$ & $1(\R)$ \\
\hline
non-surjective & $3(\C), 7(\C), 4(\H), 6(\H)$ & $5(\H)$ \\
\hline
\end{tabular}
\caption{Fiberwise character of $\gamma$ for the case $\K=\R$.  At the
intersection of each row and column, we indicate the values of
$p-q~({\rm mod}~8)$ for which the map induced by $\gamma$ on each
fiber of the \KA algebra has the corresponding properties. In
parentheses, we also indicate the isomorphism type of the Schur
algebra for that value of $p-q~({\rm mod}~8)$. Note that $\gamma$ is
fiberwise surjective exactly for the normal case, i.e. when the Schur
algebra is isomorphic with $\R$.}
\label{table:InjSurj}
\end{table}

\paragraph{The Schur bundle and the image of $\gamma$.}
The Schur algebra $\S_\K(p,q)$ of Subsection \ref{sec:KAAlgClass} is realized naturally in the
representation space.  Picking a point $x$ on $M$, let $\Sigma_x$ be
the subalgebra of $(\End(S_x),\circ)$ consisting of those
endomorphisms $T_{x}\in \End(S_{x})$ which commute with any
operator lying in the image of $\gamma_{x}$:
\be
\Sigma_{x}\eqdef \{T_x\in \End(S_{x})|[\gamma_{x}(\omega_x),T_x ]_{-,\circ}=0~,
~\forall \omega_x\in \wedge T^\ast_{x} M\}~~.
\ee
Then $\Sigma_x$ is isomorphic with $\S_\K(p,q)$ for all
$x\in M$. The bundle determined by $\Sigma_x$ when $x$ varies on $M$ will be
called {\em the Schur bundle} of $\gamma$; it is a bundle of
sub-algebras of $(\End(S),\circ)$. Of course, the space $S_{x}$ can be
viewed as a left $\Sigma_{x}$-module via the obvious action of the
elements of $\Sigma_{x}$ --- whereby $S$ can be viewed as a bundle of
modules over the Schur bundle. The image $\gamma(\Lambda T^\ast_\K M)$
of $\gamma$ coincides with the sub-bundle $\End_\Sigma(S)\subset \End(S)$
whose fiber at $x\in M$ is given by:
\be
\End_\Sigma(S)_{x}=\{T_x\in \End(S)~|~[T_x, U_x]_{-,\circ}=0~,~\forall U\in \Sigma_x\}~~,
\ee
while its space of globally-defined smooth sections is:
\be
\Gamma(M,\End_\Sigma(S))=\{T\in \Gamma(M,\End(S))|[T,V]_{-,\circ}=0~,~\forall V \in \Gamma(M,\Sigma)\}~~.
\ee
We also note that the image of the map induced by $\gamma$ on sections is given by:
\be
\gamma(\Omega_\K(M))=\Gamma(M,\End_\Sigma(S))~~.
\ee
\paragraph{Irreducible algebra representations of the fiber of the \KA bundle.}
We end by recalling some well-known facts from the representation theory of Clifford algebras:
\begin{itemize}
\item A simple Clifford algebra admits (up to $\K$-linear equivalence)
a single non-trivial irreducible representation by $\K$-linear
operators, whose dimension equals $\Delta_\K(d)\dim_\K \S_\K(p,q)$.
\item A non-simple Clifford algebra admits (up to $\K$-linear
equivalence) two non-trivial irreducible representations by
$\K$-linear operators, whose real dimensions are both equal to $\Delta_\K(d)\dim_\R
\S_\K(p,q)$.  The two representations map the Clifford volume
element determined by some given orientation into a sign factor times
the identity operator of the representation space, and are
distinguished one from another by the value of that signed factor.
\end{itemize}

\subsection{A partial inverse of $\gamma$ in the non-simple case}
\label{sec:pininverse}

In the non-simple case, the bundle morphism $\gamma$ has the
property:
\ben
\label{epsilondef}
\gamma(\nu)=\epsilon_\gamma\id_S~~,
\een
where $\epsilon_\gamma\in \{-1,1\}$ is a sign factor which we shall
call the {\em signature} of $\gamma$. Direct computation using \eqref{epsilondef} gives:
\be
\gamma\circ P_{\epsilon_\gamma}=\gamma~~,~~\gamma\circ P_{-\epsilon_\gamma}=0~~,
\ee
which implies that $\gamma$ vanishes when restricted to the sub-bundle
$\wedge^{-\epsilon_\gamma} T^\ast_\K M$ and that its restriction to
$\wedge^{\epsilon_\gamma} T^\ast_\K M$ gives an isomorphism between
this latter sub-bundle of algebras and the sub-bundle of algebras
$(\End_\Sigma(S),\circ)$ of $\End(S)$. We have $\ker(\gamma)=\wedge^{-\epsilon_\gamma}
T^\ast_\K M$.  Hence the corresponding map on
sections (which we denote again by $\gamma$) has kernel
$\cK(\gamma)=\Omega^{-\epsilon_\gamma}_\K(M)$ while its restriction to
$\Omega^{\epsilon_\gamma}_\K(M)$ gives an isomorphism between this
latter subalgebra of the \KA algebra and the subalgebra
$\Gamma(M,\End_\Sigma(S))$ of $(\Gamma(M,\End(S)),\circ)$.

\paragraph{The sub-bundle $\wedge^\gamma T^\ast_\K M$ and the subalgebra $\Omega^\gamma_\K(M)$.} Let
us introduce notation which will allow us to treat all cases uniformly.

\begin{itemize}
\item In the non-simple case (when the signature $\epsilon_\gamma$ is defined), we
let $\wedge^\gamma T^\ast_\K M\eqdef \wedge^{\epsilon_\gamma} T^\ast_\K M$ and
$\Omega^\gamma_\K(M)\eqdef  \Omega^{\epsilon_\gamma}_\K(M)$.
\item In the simple case, we let $\wedge^\gamma T^\ast_\K M\eqdef \wedge T^\ast_\K M$ and
$\Omega^\gamma_\K(M)\eqdef  \Omega_\K(M)$.
\end{itemize}
In both cases, we have $\Omega^\gamma_\K(M)=\Gamma(M,\wedge^\gamma T^\ast_\K M)$. Notice that
$\wedge^\gamma T^\ast_\K M$ is always a sub-bundle of unital algebras of the \KA bundle
while  $\Omega^\gamma_\K(M)$ is always a unital subalgebra of the \KA algebra.

\paragraph{The partial inverse of $\gamma$. } Consider the bundle isomorphism:
\ben
\label{gammarc}
\gamma|_{\wedge^\gamma T^\ast_\K M}^{\End_\Sigma(S)}:\wedge^\gamma T^\ast_\K M\stackrel{\sim}{\longrightarrow} \End_\Sigma(S)
\een
obtained by restricting the domain of definition of $\gamma$ to the
sub-bundle $\wedge^\gamma T^\ast_\K M$ of the exterior bundle and the
co-domain of definition to the sub-bundle $ \End_\Sigma(S)$ of
$\End(S)$.  We let:
\be
\gamma^{-1}\eqdef \left(\gamma|_{\wedge^\gamma T^\ast_\K M}^{\End_\Sigma(S)}\right)^{-1}:\End_\Sigma(S)
\stackrel{\sim}{\longrightarrow} \wedge^\gamma T^\ast_\K M
\ee
be the inverse of \eqref{gammarc}. The corresponding maps on sections give the mutually-inverse isomorphisms of algebras displayed in diagram \eqref{diagram:gammainv}.
\ben
\label{diagram:gammainv}
\scalebox{1}{
\xymatrix@1{
(\Omega_\K^\gamma (M), \diamond) \ar@<0.7ex>[r]^{{\tiny {\gamma|^{{\tiny \Gamma(M,\End_\Sigma(S))}}_{{\tiny \Omega^\gamma_\K(M)}}}}~~~} {~~~~~} &
{~~~~~}(\Gamma(M,\End_\Sigma(S)),\circ) \ar@<0.7ex>[l]^{\gamma^{-1}~~~~~} ~~.
}}
\een

\paragraph{Notation.} We define:
\ben
\label{checkTsecDef}
\check{T}\eqdef \gamma^{-1}(T)\in \Omega^\gamma_\K(M)~~,~~\forall T\in \Gamma(M,\End_\Sigma(S))~~.
\een
In the context of `vertical quantization' of spin systems,
$\gamma^{-1}$ plays the role of a (partial) `vertical dequantization
map', so $\check{T}$ in equation \eqref{checkTsecDef} is the dequantization of a `vertical'
operator $T$ acting in the Hilbert space. The partial inverse of
$\gamma$ allows us to transfer statements about operators acting on
pinors to statements about differential forms --- an observation which
will be used intensively in what follows. Notice the relations:
\be
\gamma^{-1}\circ \gamma=P_{\epsilon_\gamma}~~,~~\gamma\circ \gamma^{-1}=\id_{\End_\Sigma(S)}~~,
~~\gamma\circ P_{\epsilon_\gamma}=\gamma~~,~~\gamma\circ P_{-\epsilon_\gamma}=0~~.
\ee

\paragraph{Local expression for $\gamma^{-1}$.} Considering a local pseudo-orthonormal
coframe $e^a$ and recalling that $\gamma(e^a)=\gamma^a$, we find:
\ben
\label{ginve}
\gamma^{-1}(\gamma^a)=e^a_\gamma~~,
\een
where we have set:
\be
e^a_\gamma=\twopartdef{e^a_{\epsilon_\gamma}~,~}{\mathrm{we~are~in~the~non-simple~case}}{e^a~,~}
{\mathrm{we~are~in~the~simple~case}}~~,
\ee
with $e^a_{\epsilon_\gamma}$ defined as in \eqref{eaprol}. Relation \eqref{ginve} implies:
\be
\gamma^{-1}(\gamma^{a_1\ldots a_k})=e^{a_1\ldots a_k}_{\gamma}
\twopartdef{e^{a_1\ldots a_k}_{\epsilon_\gamma}~,~}{\mathrm{we~are~in~the~non-simple~case}}
{e^{a_1\ldots a_k}~,~}{\mathrm{we~are~in~the~simple~case}}~~,
\ee
where $e^{a_1\ldots a_k}_\pm$ are defined in \eqref{eAprol} and
we used the fact that $\gamma^{a_1\ldots a_k}=\gamma^{a_1}\circ
\ldots \circ \gamma^{a_k}$ for all mutually-distinct $a_1\ldots a_k$,
the fact that $\gamma^{-1}$ is an isomorphism of algebras when
corestricted to its image and (for the non-simple case) identity
\eqref{eAprol}.

\subsection{Trace on $\Omega^\gamma_\K(M)$}
\label{sec:pintrace}

The subalgebra $\Omega_\K^\gamma(M)$ admits a $\cinf$-linear map 
$\cS:\Omega^\epsilon_\K(M)\rightarrow \cC^\infty(M,\K)$ given by:
\be
\cS(\omega)=\omega^{(0)} N_{p,q} \rk_\K(S)~~,
\ee
where $\omega^{(0)}$ is the rank $0$ component of $\omega$ (see expansion \eqref{FormExpansion}),
$S$ is any of the $\K$-pinor bundles and $N_{p,q}$ equals $1$ or $2$ according to
whether the corresponding fiberwise representation is faithful or not (notice that
$N_{p,q}\rk_\K S$ is the dimension of the smallest {\em faithful}
representation of the fiberwise Clifford algebra). One has:
\be
\cS(\omega)=\tr(\gamma(\omega))~~,~~\forall \omega\in \Omega^\gamma_\K(M)~~
\ee
as well as:
\be
\cS(1_M)=N_{p,q}\dim_\K(S)~~,~~\cS(\omega\diamond \eta)=
\cS(\eta\diamond \omega)~~,~~\forall \omega,\eta\in \Omega^\gamma_\K(M)~~.
\ee

\section{The Fierz isomorphism and generalized Killing forms}
\label{sec:fierz}

In this section, we take up the issue of translating constrained
generalized Killing pinor equations into conditions on differential
forms.  To simplify presentation, we shall assume from the outset that
the Schur algebra is isomorphic with the base field $\K$, so that
either $\K=\C$ or we are in the normal case with $\K=\R$.  We start in
Subsection \ref{sec:fierzbisp} with a discussion of the bundle of
bipinors.  Subsection \ref{sec:fierzdef} considers a certain
isomorphism of bundles of algebras (which we shall call {\em the Fierz
isomorphism}) that provides an identification of the bundle of bipinors with the bundle
$(\wedge^\gamma T^\ast_\K M,\diamond)$ and allows for a concise description of those Fierz identities which involve
four pinors.  This construction makes essential use of a choice of
bilinear and non-degenerate `admissible' form $\cB$ on the pin bundle
(such inner products were classified in \cite{AC0}, see also
\cite{AC1}). Subsection \ref{sec:fierzprop} extracts some basic
properties of this isomorphism which will be useful later on. In
Subsection \ref{sec:fierzcomp}, we give a brief discussion of
completeness relations for the endomorphism algebra of the pin
bundle. Subsection \ref{sec:fierzexp} gives an explicit local
expansion of the Fierz isomorphism which depends on the choice of a
local pseudo-orthonormal coframe. In Subsection \ref{sec:fierzalg}, we show
how algebraic constraints on pinors translate very directly into
constraints on differential forms if one uses the basic properties of
the Fierz isomorphism. Subsection \ref{sec:fierzdiff} takes up the
problem of translating generalized Killing pinor equations into conditions
on differential forms.  Using the Fierz isomorphism, we show that any
connection on the pin bundle which is compatible with $\cB$ defines
a certain algebra connection on the \KA bundle (i.e. a linear
connection which is a fiberwise derivation of the geometric product)
such that the Fierz isomorphism is flat with respect
to the connections induced on its domain and codomain. Using this
property, we show how one can easily translate generalized Killing
conditions on pinors into differential constraints on forms
defined on $M$. Subsection \ref{sec:fierzaltdiff} gives another form
of such differential constraints, which is used in Appendix
\ref{sec:direct} for comparison with the component approach outlined
in \cite{MartelliSparks}. In Subsection \ref{sec:fierzCGKA}, we
discuss some basic aspects of the algebro-differential system of
constraints on inhomogeneous forms which results from our analysis. As
expected, our formulation allows one to extract basic
structural properties of this system, thereby providing a starting
point for a natural generalization of the classical theory of Killing
forms. Finally, Subsection \ref{sec:fierz2} considers the particular
cases of one and two independent constrained generalized Killing pinors with a
definite and symmetric $\Spin(d)$-invariant metric, the
first of which is relevant to the application discussed in Section
\ref{sec:application}.

\subsection{Bipinor algebras}
\label{sec:fierzbisp}

Let $S$ be a pin bundle over $(M,g)$ with underlying morphism 
$\gamma:(\wedge T^\ast_\K M, \diamond)\rightarrow (\End(S),\circ)$. 

\paragraph{Admissible bilinear pairings on the pin bundle.} It is
well-known that $S$ carries so-called {\em admissible} non-degenerate
bilinear pairings $\cB$ whose action on sections of $S$ satisfies:
\ben
\label{Bproperties}
\cB(\gamma(\omega)\xi,\xi')=\cB(\xi,\gamma(\tau_\cB (\omega))\xi')~~,~~
\cB(\xi',\xi)=\sigma_\cB \cB(\xi,\xi')~~,~~\forall \omega\in
\Omega_\K(M)~~,~~\forall \xi,\xi'\in \Gamma(M,S)~~
\een
as well as another property which can be found in \cite{AC1}
but will not be relevant for what follows.  In the formulas above, we
used the following anti-automorphism of the \KA algebra:
\be
\tau_\cB=\pi^{\frac{1-\epsilon_\cB}{2}}\circ \tau=
\twopartdef{\tau~,~}{\epsilon_\cB=+1~}{\pi\circ \tau~,~}{\epsilon_\cB=-1}~,
\ee
where $\tau$ is the reversion anti-automorphism defined in
\eqref{taudef}.  The numbers $\epsilon_\cB$ (the {\em type} of $\cB$)
and $\sigma_\cB$ (the {\em symmetry} of $\cB$) equal $+1$ or $-1$,
depending on $p,q$ and the precise choice of $\cB$; such bilinear
pairings were classified in \cite{AC0, AC1}.  Notice that the first
equation in \eqref{Bproperties} implies:
\be
\gamma(\omega)^t=\gamma(\tau_\cB(\omega))~~,
~~\forall \omega\in \Omega_\K(M)~~,
\ee
i.e.:
\ben
\label{optranspose}
(~)^t\circ \gamma=\gamma\circ \tau_\cB~~,
\een
where $T^t$ denotes the transpose of $T\in \Gamma(M,\End(S))$
with respect to $\cB$, which is defined through:
\ben
\label{transpose}
\cB(T\xi,\xi')=\cB(\xi,T^t\xi')~~,~~\forall \xi,\xi'\in \Gamma(M,S)~~.
\een
This operation satisfies $(T^t)^t=T$ and $(\id_S)^t=\id_S$, the first
identity being a consequence of the signed symmetry property of $\cB$
(the second identity listed in \eqref{Bproperties}). The operation
$T\rightarrow T^t$ of taking the $\cB$-transpose defines a $\cinf$-linear 
anti-automorphism of the algebra $(\Gamma(M,\End(S)),\circ)$.

\paragraph{Local expressions.} Given a local pseudo-orthonormal coframe $e^a$
above $U\subset M$, the first of properties \eqref{Bproperties}
amounts to:
\be
\cB(\gamma^a\xi,\xi')=\epsilon_\cB \cB(\xi,\gamma^a\xi')~~,~~\forall \xi,\xi'\in \Gamma(U,S)~~,
\ee
which means that $\gamma^a=\gamma(e^a)~$ satisfy $~(\gamma^a)^t=\epsilon_\cB \gamma^a$~,
a relation which implies:
\be
(\gamma^{a_1\ldots a_k})^t=\epsilon_\cB^k \gamma^{a_k\ldots a_1}~~,
\ee
where $\epsilon_\cB^k=(\epsilon_\cB)^k$.
Since $[\gamma^a,\gamma^b]_{+,\circ}=2\eta^{ab}$, we also have:
\be
(\gamma^a)^{-1}=\gamma_a~~\mbox{where}~~\gamma_a\eqdef \eta_{ab}\gamma^b~~,
\ee
which in turn gives:
\be
(\gamma^{a_1\ldots a_k})^{-1}=\gamma_{a_k\ldots a_1}~~.
\ee
Combining the above, we find:
\be
((\gamma^{a_1\ldots a_k})^{-1})^t=(\gamma_{a_k\ldots a_1})^t=\epsilon_\cB^k\gamma_{a_1\ldots a_k}~~,
\ee
which implies:
\ben
\label{relB}
\cB((\gamma^{a_1\ldots a_k})^{-1}\xi,\xi')=\epsilon_\cB^k\cB(\xi,\gamma_{a_1\ldots a_k}\xi')~~.
\een
These relations will be useful later.

\paragraph{The isomorphism $E$.} The non-degenerate pairing $\cB$ induces a
bundle isomorphism $\rho:S\stackrel{\sim}{\rightarrow}S^{\ast}$,
whose action on sections is given by:
\be
\rho(\xi)(\xi')\eqdef \cB(\xi',\xi)~~,~~\forall \xi,\xi'\in \Gamma(M,S)~~.
\ee
On the other hand, we have a natural bundle isomorphism $q:S\otimes
S^{\ast} \stackrel{\sim}{\rightarrow} \End(S)$, given on sections
as follows:
\be
q(\xi\otimes \eta )(\xi')\eqdef \eta (\xi')\xi~~,
~~\forall \xi,\xi'\in \Gamma(M,S)~~,~~\forall \eta \in \Gamma(M, S^\ast)~~.
\ee
The two maps above combine to give a bundle isomorphism $E\eqdef
q\circ (\id_S\otimes \rho):S\otimes
S\stackrel{\sim}{\rightarrow}\End(S)$. Setting $E_{\xi,\xi'}\eqdef
E(\xi\otimes \xi')\in \Gamma(M,\End(S))$ for all $\xi,\xi'\in
\Gamma(M,S)$, we have:
\ben
\label{Eeq}
E_{\xi_1,\xi_2}\circ E_{\xi_3,\xi_4}=
\cB(\xi_3,\xi_2)E_{\xi_1,\xi_4}~~,~~\forall \xi_1,\xi_2,\xi_3,\xi_4\in \Gamma(M,S)~~,
\een
an identity which follows from the explicit form:
\ben
\label{Edef}
E_{\xi,\xi'}(\xi'')=\cB(\xi'',\xi')\xi~~
\een
{\em without} making use of the signed symmetry property of
$\cB$. Note that $E$ depends on the choice of $\cB$ (since $\rho$ does).

\paragraph{The bundle of bipinors and the bipinor algebra of $S$.} The
bundle isomorphism $E$ allows us to transfer the fiberwise composition
of operators from $\End(S)$ to an associative and bilinear
fiberwise composition $\bullet$ defined on the {\em bundle of
bipinors} $S\otimes S$, whose action on sections takes the form:
\ben
\label{bispinorcomp}
u \bullet v\eqdef E^{-1}(E(u)\circ E(v))~~,~~\forall u,v \in \Gamma(M,S\otimes S)~~.
\een
This operation satisfies:
\ben
\label{4tcomp}
(\xi_1\otimes \xi_2)\bullet (\xi_3\otimes
\xi_4)=\cB(\xi_3,\xi_2)\xi_1\otimes \xi_4~~,~~\forall
\xi_1,\xi_2,\xi_3,\xi_4\in \Gamma(M,S)~~.
\een
The composition $\bullet$ makes the bundle of bipinors into a bundle
of unital associative algebras which is isomorphic with the bundle of
algebras $(\End(S),\circ)$; of course, the unit section $\id_S$ of
$\End(S)$ maps to the unit section of $S\otimes S$, which we denote by
$\cI\eqdef E^{-1}(\id_S)$. The unital associative algebra
$\Gamma(M,S\otimes S)=\Gamma(M,S)\otimes_{\cinf}\Gamma(M,S)$
consisting of smooth sections of the bipinor bundle will be called
the {\em bipinor algebra} of $S$; it is an algebra over the ring
$\cinf$.

\paragraph{The bipinor $\cinf$-algebra of a submodule of sections.}
If $\cK$ is any $\K$-linear subspace of $\Gamma(M,S)$, then the set: 
\be
\cK\otimes_\cinf \cK\eqdef \{\xi\otimes\xi'|\xi,\xi'\in \cK\}\subset \Gamma(M,S\otimes S)\approx \Gamma(M,S)\otimes_\cinf \Gamma(M,S)~~
\ee
is a $\K$-linear subspace of $\Gamma(M,S\otimes S)$. 

When $\cK\subset \Gamma(M,S)$ is a {\em submodule} of the
$\cinf$-module $\Gamma(M,S)$, then the subspace $\cK\otimes_\cinf
\cK\subset \Gamma(M,S\otimes S)$ is a (generally non-unital)
subalgebra of the bipinor algebra of $S$, which we shall call the {\em
bipinor algebra of $\cK$}. This associative algebra defined over
$\cinf$ depends on the choice of $\cB$.  In particular, the space of
smooth global sections $\Gamma(M,K)$ of any vector sub-bundle $K$ of
$S$ is a $\cinf$-submodule of $\Gamma(M,S)$ and the corresponding
bipinor algebra $\Gamma(M,K)\otimes_\cinf
\Gamma(M,K)=\Gamma(M,K\otimes K)$ will be called the {\rm bipinor
algebra of $K$.}

\paragraph{The bipinor $\K$-algebra of a $\cB$-flat subspace of sections.}
Another interesting case arises when the $\K$-linear subspace
$\cK\subset \Gamma(M,S)$ is {\em $\cB$-flat}, by which we mean that
$\cK$ satisfies the condition:
\be
\cB(\xi,\xi'){\rm ~is~a~constant~function~on~}M~~,~~\forall \xi,\xi'\in \cK~~.
\ee
In this case, $\cK\otimes_\cinf \cK$ is a $\K$-subalgebra of the
bipinor algebra $(\Gamma(M,S\otimes S),\bullet)$, which we shall call
the {\em (flat) bipinor $\K$-algebra} determined by $\cK$.

\subsection{The Fierz isomorphism and Fierz identities. Fierz algebras. }
\label{sec:fierzdef}

\paragraph{The Fierz isomorphism $\check{E}$.} Let us now assume that
we are in one of the cases when the Schur algebra is isomorphic with
the base field.  Then one can also transport to the bundle of
bipinors the isomorphism
$\gamma^{-1}: \End(S)\stackrel{\sim}{\rightarrow} (\wedge
T^{\ast}_\K M)^\gamma$ to get an isomorphism of bundles of algebras:
\be
\check{E}\eqdef \gamma^{-1}\circ E:(S\otimes S,\bullet)
\stackrel{\sim}{\rightarrow} (\wedge^\gamma T^\ast_\K M, \diamond)~~,
\ee
which we shall call {\em the Fierz isomorphism}. On sections, this
induces a $\cinf$-linear isomorphism of algebras (denoted, as usual, by the same symbol):
\be
\check{E}\eqdef \gamma^{-1}\circ E:(\Gamma(M,S\otimes S),\bullet)
\stackrel{\sim}{\rightarrow} (\Omega^\gamma_\K(M), \diamond)~~,
\ee
which identifies the bipinor algebra with the subalgebra
$\Omega^\gamma_\K(M)$ of the \KA algebra. Note that $\check{E}$
depends on the choice of admissible form $\cB$.

\paragraph{Fierz identities involving four pinors.} Setting
$\check{E}_{\xi,\xi'}\eqdef \check{E}(\xi\otimes
\xi')=\gamma^{-1}(E_{\xi, \xi'})\in \Omega_\K^\gamma (M)$ (for
$\xi,\xi'\in \Gamma(M,S)$), equation \eqref{Eeq} implies the following
identity in the subalgebra $\Omega^\gamma_\K(M)$ of the \KA algebra:
\ben
\label{FierzGeom}
\check{E}_{\xi_1,\xi_2}\diamond \check{E}_{\xi_3,\xi_4}=
\cB(\xi_3,\xi_2)\check{E}_{\xi_1,\xi_4}~~,~~\forall \xi_1,\xi_2,\xi_3,\xi_4\in \Gamma(M,S)~~.
\een
Equation \eqref{FierzGeom} is the condensed expression of Fierz
identities involving four pinors. These identities simply express the
fact that $\gamma$ (and thus $\check{E}$) is an isomorphism of bundles of
algebras, rather than simply an isomorphism of vector bundles --- and
are, in fact, equivalent with this property once fiberwise linearity of $\gamma$
is assumed. The construction of $\check{E}$ is summarized in the
commutative diagram \eqref{diagram:checkE}, which applies provided that $\S_\K(p,q)\approx \K$. In the diagram, we 
show the action of the various morphisms on sections. 
\ben
\label{diagram:checkE}
\scalebox{1}{\xymatrix{
\Gamma(M,S\otimes S) \ar[d]^\wr_{\check{E}} \ar[r]^{\id_S\otimes \rho}_{\sim}
\ar[dr]^E_{\sim} & \Gamma(M,S\otimes_{\K} S^{\ast}) \ar[d]^q_\wr \\
\Omega^\gamma_\K(M) \ar[r]^\sim_{\gamma} & \Gamma(M,\End_\K(S))
}}
\een

\paragraph{Notation.} Let $\cK\subset \Gamma(M,S)$ be any $\K$-linear subspace of $\Gamma(M,S)$. 
The image of the $\K$-linear subspace $\cK\otimes_\cinf \cK\subset \Gamma(M,S\otimes S)$ 
through the Fierz isomorphism will be denoted by: 
\ben
\label{CK}
\check{\cK}\eqdef \check{E}(\cK\otimes_\cinf\cK)\subset \Omega^\gamma_\K(M)~~
\een
and is a $\K$-linear subspace of $\Omega_\K^\gamma(M)$. 

\paragraph{The Fierz $\cinf$-algebra of a submodule of sections.}
When the subspace $\cK\subset \Gamma(M,S)$ is a {\em submodule} of the $\cinf$-module $\Gamma(M,S)$, then the subspace
$\cK\otimes_{\cinf} \cK\subset \Gamma(M,S\otimes S)$ is a (generally non-unital) 
subalgebra (over $\cinf$) of the bipinor algebra of $S$. 
Its image \eqref{CK} through the Fierz isomorphism
is a (generally non-unital) subalgebra of the $\cinf$-algebra
$(\Omega^\gamma_\K(M),\diamond)$, which we shall call the {\em Fierz
subalgebra} determined by $\cK$. This algebra over $\cinf$ encodes the
Fierz identities between bilinears constructed from pinors which
belong to $\cK$. A particular case arises when $\cK=\Gamma(M,K)$ where
$K\subset S$ is some vector sub-bundle of $S$ --- in which situation
$\cK\otimes_\cinf \cK$ is the bipinor algebra of $K$. The
corresponding Fierz subalgebra $\check{\cK}$ will then be called the
Fierz subalgebra determined by the sub-bundle $K$. With the further
assumption that $\cB$ is a scalar product (as happens in the
application of Subsection \ref{sec:application}), the morphism $E$ can
then be used to identify the Fierz algebra of $K$ with
the $\cC^\infty(M,\K)$-algebra $\Gamma(M,\End(K))$ of globally-defined
endomorphisms of the bundle $K$; in particular, the Fierz
subalgebra is unital in such cases.

\paragraph{The Fierz $\K$-algebra of a $\cB$-flat subspace of sections.} 

When $\cK$ is a $\cB$-flat $\K$-linear subspace of $\Gamma(M,\K)$, the vector 
space $\cK\otimes_\cinf \cK$ is a $\K$-subalgebra of the bipinor algebra 
$(\Gamma(M,S\otimes S),\bullet)$. It follows that its image \eqref{CK} through the 
Fierz isomorphism is a $\K$-subalgebra of the algebra $(\Omega^\gamma_\K(M),\diamond)$, 
which will be called the {\em (flat) Fierz $\K$-algebra} determined by $\cK$.

\subsection{Some properties of the Fierz isomorphism}
\label{sec:fierzprop}

A simple computation using \eqref{Edef} shows that the following
identities hold for any $T\in \Gamma(M,\End(S))$:
\ben
\label{Teq}
T \circ E_{\xi,\xi'}=E_{T\xi, \xi'}~~,
~~E_{\xi,\xi'}\circ T=E_{\xi, T^t\xi'}~~,~~\forall \xi,\xi'\in \Gamma(M,S)~~,
\een
i.e.
\ben
\label{Teq2}
L_T\circ E=E\circ(T\otimes \id_S)~~,~~R_T\circ E=E\circ (\id_S\otimes T^t)~~,
\een
where $L_T$ and $R_T$ are the operators of left and right
multiplication with $T$ in the algebra $(\Gamma(M,\End(S)),\circ)$.
Applying $\gamma^{-1}$ to identities \eqref{Teq} and setting $\check{T}\eqdef \gamma^{-1}(T)\in \Omega^\gamma_\K(M)$ gives:
\ben
\label{Teqstar}
\check{T} \diamond  \check{E}_{\xi, \xi'}=\check{E}_{T \xi, \xi'}~~,
~~\check{E}_{\xi, \xi'}\diamond  \check{T} = \check{E}_{\xi, T^t \xi'}~~,
\een
i.e. (substituting $T\rightarrow T^t$ into the second equation):
\ben
\label{TEtranspose}
\check{T} \diamond  \check{E}_{\xi, \xi'}
= \check{E}_{T \xi, \xi'}~~,
~~\check{E}_{\xi, \xi'}\diamond  (T^t)^{\mathbf{\check{~}}} = \check{E}_{\xi, T\xi'}~~.
\een
This also reads:
\ben
\label{TEtransposeop}
L_{\check{T}}\circ \check{E}=\check{E}\circ(T\otimes \id_S)~~,
~~R_{\check{T}}\circ \check{E}=\check{E}\circ (\id_S\otimes T^t)~~,
\een
where $L_{\check{T}}$ and $R_{\check{T}}$ are the operators of left and
right multiplication with $\check{T}$ in the \KA algebra. Equation
\eqref{optranspose} implies $\gamma^{-1}\circ (~)^t=\tau_\cB\circ
\gamma^{-1}$, i.e.: \be \gamma^{-1}(T^t)=\tau_\cB
(\gamma^{-1}(T))\Longleftrightarrow (T^t)^{\mathbf{\check{~}}}
=\tau_\cB(\check{T})~~,~~\forall T\in \Gamma(M,\End(S))~~.
\ee
We can thus write the second relation of \eqref{Teqstar} in the form:
\ben
\label{TeqstarReal}
\check{E}_{\xi, \xi'}\diamond  \tau_\cB(\check{T}) = \check{E}_{\xi, T\xi'}~~~,
~~~\forall T\in \Gamma(M,\End(S))~~,~~\forall \xi,\xi'\in \Gamma(M,S)~~.
\een
We also notice the relation:
\ben
\label{ETranspose}
(E_{\xi,\xi'})^t=\sigma_\cB E_{\xi',\xi}\Longleftrightarrow
\tau_\cB(\check{E}_{\xi,\xi'})=\sigma_\cB \check{E}_{\xi',\xi}~~,
~~\forall \xi, \xi'\in \Gamma(M,S)~~,
\een
which follows from the signed symmetry of $\cB$ together with
definition \eqref{Edef}. The last identity can also be written as:
\ben
\label{Et}
(~)^t\circ E=E\circ \transp_\cB~\Longleftrightarrow ~\tau_\cB\circ \check{E}=\check{E}\circ \transp_\cB~~,
\een
where $\transp_\cB\eqdef\sigma_\cB \transp$ and
$\transp:\Gamma(M,S\otimes S)\rightarrow \Gamma(M,S\otimes S)$ is the
$\cinf$-linear operator which is defined as follows on decomposable
elements:
\ben
\label{transpdef}
\transp(\xi\otimes \xi')=\xi'\otimes \xi~~,~~\forall \xi,\xi'\in \Gamma(M,S)~~.
\een
One easily checks that $\transp_\cB$ is an anti-automorphism of the
bipinor algebra.  For later reference, note the identity:
\ben
\label{Etr}
\tr\circ L_T\circ E=\cB \circ (T\otimes \id_S)\Longleftrightarrow
\tr(T \circ E_{\xi,\xi'})=\cB(T\xi,\xi')~~,~~\forall \xi,\xi'\in \Gamma(M,S)~~,
\een
which will be useful below.

\subsection{Local completeness relations for the endomorphism algebra of the pin bundle}
\label{sec:fierzcomp}

Let $\tr:\End(S)\rightarrow \cO_\K$ be the natural\footnote{This
map can be defined naturally on the bundle of endomorphisms of any
vector bundle, making no reference whatsoever to any bilinear pairing on
the bundle.} trace on $\End(S)$ (which is a morphism of $\K$-vector
bundles). Recall that we assume the Schur algebra to be isomorphic with the base field $\K$. Then a convenient
generating set of local sections for the vector bundle $\End(S)$
above a sufficiently small open subset $U\subset M$ is given (see, for
example, \cite{Okubo}) by the operators $\{\gamma^{a_1\ldots
a_k}|k=0,\ldots d, 1\leq a_1< \ldots < a_k\leq d\}$, where $d=p+q$ and
$\gamma^a=\gamma(e^a)$ with $(e^a)$ a pseudo-orthonormal local
coframe of $M$ above $U$. The following identity (the
`completeness relation') holds:
\ben
\label{completeness}
T=_U\frac{\Delta_\R(d)}{2^d}
\sum_{k=0}^{d}\frac{1}{k!}\tr(\gamma^{a_k\ldots a_1}\circ T)
\gamma_{a_1\ldots a_k}~~,~~\forall T\in \Gamma(U,\End(S))~~,
\een
where $\Delta_\R(d)=2^{[\frac{d}{2}]}$ for $\S \approx \R$
or $\S \approx \C$ (see Section \ref{sec:KAAlgClass}).

\paragraph{Remark.} The $\K$-vector bundles
$\End(S)\otimes \End(S)$ and $\End(\End(S))$ can be
identified through the bundle isomorphism ${\cal
W}:\End(S)\otimes \End(S)\rightarrow \End(\End(S))$,
which acts as follows on sections:
\be
{\cal W}(A\otimes  B)(T)=A~\tr(B\circ T)~~,
~~\forall A, B, T\in \Gamma(M,\End(S))~~.
\ee
Using this isomorphism, we transport the composition $\circ$ of
$\End(\End(S))$ to an associative composition $\bullet$ defined on
$\End(S)\otimes \End(S)$, whose action on sections is given by:
\be
(A\otimes  B)\bullet (A'\otimes B')=
{\cal W}^{-1}({\cal W}(A\otimes B)\circ {\cal W}(A'\otimes B'))~~,
~~\forall A, A', B, B'\in \Gamma(M,\End(S))~~.
\ee
An easy computation shows that $\bullet$ has the explicit form:
\ben
\label{biendcomp}
(A\otimes B)\bullet (A'\otimes B')=\tr(A'\circ B) A\otimes B'~~,
~~\forall A, A', B, B'\in \Gamma(M,\End(S))~~.
\een
The unit section $\id_{\End(S)}$ of the bundle
$\End(\End(S))$ corresponds via ${\cal W}$ to the unit section
$\cI\eqdef {\cal W}^{-1}(\id_{\End(S)})$ of
$(\End(S)\otimes \End(S),\bullet)$. Expression
\eqref{biendcomp} shows that the completeness relation is equivalent
with the following decomposition of the unit $\cI|_U$ of the algebra
$(\Gamma(U,\End(S)\otimes \End(S)),\bullet)$:
\ben
\label{CompBi}
\cI=_U\sum_{k=0}^{d}\frac{1}{k!}\gamma_{a_1\ldots a_k}\otimes \gamma^{a_k\ldots a_1}~~.
\een

\subsection{Explicit expansion of the Fierz isomorphism when the Schur algebra equals the base field}
\label{sec:fierzexp}

Given a local pseudo-orthonormal coframe of $M$, explicit expansions for the
isomorphism $\check{E}$ can be derived using the results of
\cite{Okubo} (see also \cite{Rand}) for any choices of the base field
$\K$ and of the signature type $(p,q)$. A complete discussion is quite
involved given the different behavior in various cases and will be
taken up in detail in a different publication. Below, we shall
consider only the case when the Schur algebra is
isomorphic with the base field $\K$, i.e. the case when $\K=\C$ and
the case when $\K=\R$ with $p-q\equiv_8 0,1,2$.

In this case, the local completeness relation \eqref{completeness}
holds. Applying it to the endomorphism $T=E_{\xi, \xi'}$ and using
relation \eqref{Etr}, we find the local expansion:
\ben
\label{Eexpansion}
E_{\xi,\xi'}=_U\frac{1}{2^{\left[\frac{d+1}{2}\right]}}
\sum_{k=0}^{d}\frac{1}{k!}\cB(\gamma_{a_k\ldots a_1}\xi, \xi')
\gamma^{a_1\ldots a_k}~~,~~\forall \xi,\xi'\in \Gamma(U,S)~~,
\een
which implies the following local expansion of the Fierz
isomorphism upon applying $\gamma^{-1}$ to both sides:
\ben
\label{checkEexpansion}
\check{E}_{\xi,\xi'}=\frac{1}{2^{\left[\frac{d+1}{2}\right]}}\bcE_{\xi,\xi'}~~,
\een
(an identity holding in $\Omega^\gamma_\K(M)$), where:
\ben
\label{checkomega}
\bcE_{\xi,\xi'}=\sum_{k=0}^d\bcE^{(k)}_{\xi,\xi'}
\een
(as in \eqref{FormExpansion}), with:
\ben
\label{checkomegaexp}
\bcE^{(k)}_{\xi,\xi'}=_U\frac{1}{k!}\bcE^{(k)}_{a_1\ldots a_k}(\xi,\xi')e^{a_1\ldots a_k}_\gamma~~,
\een
where, using \eqref{relB}, we have:
\ben
\label{checkomegacomp}
\bcE^{(k)}_{a_1\ldots a_k}(\xi,\xi')=
\cB(\gamma_{a_k\ldots a_1}\xi, \xi')=\epsilon_\cB^k\cB(\xi, \gamma_{a_1\ldots a_k}\xi')~~.
\een

\subsection{Expressing the algebraic constraints through differential forms}
\label{sec:fierzalg}

Consider the case when we have a single algebraic constraint $Q\xi=0$,
where $Q \in \Gamma(M,\End(S))$. As in Section \ref{sec:CGK}, we let $\cK(Q)$ denote the space of solutions of the algebraic
constraint, which is a sub-module of the $\cC^\infty(M,\K)$-module $\Gamma(M,S)$. Recall that there generally exists no
sub-bundle of $S$ whose space of smooth sections equals $\cK(Q)$.

\paragraph{The dequantized constraint and the $\cinf$-subalgebra of constrained inhomogeneous forms.}
The inhomogeneous form:
\be
\check{Q}=\gamma^{-1}(Q)\in \Omega^\gamma_\K(M)~~
\ee
will be called the {\em dequantization} of $Q$. Relation \eqref{Teq2} gives:
\be
L_Q\circ E=E\circ (\id_S\otimes Q)~~,~~R_{Q^t}\circ E =E\circ (Q\otimes \id_S)~~,
\ee
where $L_{Q},R_{Q^t}$ are --- as above --- the operators on
$\Gamma(M,\End(S))$ given by left and right composition with $Q$ and
$Q^t$.  Using the fact that $E$ is an isomorphism as well as the
identities $\cK(Q\otimes \id_S)=\cK(Q)\otimes_{\cinf} \Gamma(M,S)$ and $\cK(\id_S\otimes Q)=\Gamma(M,S)\otimes_{\cinf} \cK(Q)$, 
we find $\cK(Q)\otimes_{\cinf} \cK(Q)= \cK(Q\otimes \id_S)\cap \cK(\id_S\otimes Q)$ and:
\be
E(\cK(Q)\otimes_{\cinf} \cK(Q))=\cK (L_Q)\cap \cK (R_{Q^t})~~.
\ee
Applying $\gamma^{-1}$ to both sides of this relation gives the
following description of the Fierz algebra of the sub-module of sections $\cK(Q)$,
which we shall call the {\em  $\cinf$-algebra of $Q$-constrained inhomogeneous forms}:
\beqan
\label{Qcons_forms}
\check{\cK}_Q\eqdef  \check{E}(\cK(Q)\otimes_\cinf \cK(Q)) &=&
\cK(L_{\check{Q}})\cap \cK(R_{\tau_\cB(\check{Q})})\cap  \Omega^\gamma_\K(M)=\nn\\
&=&\cK(L_{\check{Q}}+R_{\tau_\cB(\check{Q})})\cap
\cK(L_{\check{Q}}-R_{\tau_\cB(\check{Q})})\cap  \Omega^\gamma_\K(M)~.~~~~~~~~~
\eeqan
Here, $L_{\check{Q}},R_{\check{Q}}$ are the left and right
$\diamond$-multiplication operators of Section \ref{sec:KALR} and the
second equality above is obvious. With these definitions, we have the equivalence:
\ben
\label{AlgebraicConstraints}
\check{E}_{\xi, \xi'}\in \check{\cK}_Q ~\Longleftrightarrow~
\xi\otimes \xi'\in \cK(Q)\otimes_{\cinf} \cK(Q)~~,
\een
i.e.:
\ben
\label{AlgConstraints}
\check{Q} \diamond  \check{E}_{\xi, \xi'}=\check{E}_{\xi, \xi'}
\diamond  \tau_\cB(\check{Q}) = 0 ~\Longleftrightarrow~\xi=0~~\mbox{or}~~\xi'=0~~\mbox{or}~~Q\xi=Q\xi'=0~~.
\een

\paragraph{Behavior under $\tau_\cB$.} The equivalence \eqref{AlgConstraints} can be written
in the following form, which follows by applying $\tau_\cB$ to the
second equation in the left hand side:
\ben
\label{Alg1}
\check{Q} \diamond \check{E}_{\xi,\xi'}=\check{Q}\diamond
\tau_\cB(\check{E}_{\xi,\xi'}) = 0~\Longleftrightarrow~ \xi\otimes\xi'\in \cK(Q)\otimes_{\cinf} \cK(Q)
\een
or (using \eqref{ETranspose}) in the form:
\ben
\label{Alg2}
\check{Q} \diamond  \check{E}_{\xi,\xi'}=\check{Q}\diamond \check{E}_{\xi',\xi} = 0
~\Longleftrightarrow~ \xi\otimes\xi'\in \cK(Q)\otimes_{\cinf} \cK(Q)~~.
\een
In fact, relation \eqref{Et} and the obvious fact that $\cK(Q)\otimes_{\cinf}
\cK(Q)$ is invariant under the anti-automorphism $\transp_\cB$ imply that the
algebra of $Q$-constrained differential forms is invariant under $\tau_\cB$:
\be
\tau_\cB(\check{\cK}_Q)=\check{\cK}_Q~~.
\ee

\paragraph{Remark.} Using \eqref{TeqstarReal} and \eqref{checkEexpansion} and separating ranks
shows that \eqref{AlgConstraints} can be written as:
\be
\xi\otimes \xi'\in \cK(Q)\otimes_{\cinf} \cK(Q) \Longleftrightarrow
\bcE^{(k)}_{Q\xi,\xi'}=\bcE^{(k)}_{\xi,
Q\xi'}=0~~,~~\forall k=0\ldots d~~,
\ee
which amounts to the following description
upon using the explicit form \eqref{checkomegacomp} of $\bcE$:
\ben
\label{AlgebraicConstraintscomp}
\xi\otimes \xi'\in \cK(Q)\otimes_{\cinf} \cK(Q) \Longleftrightarrow
\cB(\xi,Q^t\circ\gamma_{a_1\ldots a_k} \xi') =\cB(\xi,\gamma_{a_1\ldots
a_k} \circ Q\xi')=0~~,~~\forall \xi, \xi'\in \cK(Q)~~.
\een
In Appendix \ref{sec:Qdirect}, we show that equations
\eqref{AlgebraicConstraintscomp} are equivalent with certain relations
which were first discussed in \cite{MartelliSparks} for the special
case $d=p=8,~q=0$.

\subsection{Conditions on differential forms implied by the generalized Killing equation}
\label{sec:fierzdiff}

In our formulation, a {\em Clifford connection} $\nabla^S=\dd
x^m\otimes \nabla^S_m$ on $S$ is a connection which satisfies:
\!\!\!\!\!\!\!\!\!\ben
\label{cliff_conn}
[\nabla^S_m,\gamma(\omega)]_{-,\circ}=\gamma(\nabla_m \omega)~,~\forall
\omega\in \Omega_\K(M)\Longleftrightarrow (\nabla_m^S)^\ad\circ
\gamma=\gamma\circ \nabla_m\Longleftrightarrow P_{\epsilon_\gamma}\circ \nabla_m
=\gamma^{-1}\circ (\nabla^S_m)^\ad\circ \gamma~~~~~~~
\een
where $\nabla_m$ is the connection induced on $\wedge T^{\ast}_\K M$
by the Levi-Civita connection of $(M,g)$ and
$(\nabla_m^S)^\ad:\Gamma(M,\End(S))\rightarrow
\Gamma(M,\End(S))$ is the connection induced by $\nabla_m^S$ on
$\End(S)$:
\be
(\nabla_m^S)^\ad(T)\eqdef [\nabla_m^S,T]_{-,\circ}=\nabla_m^S\circ T - T\circ
\nabla_m^S~~,~~\forall T\in \Gamma(M,\End(S))~~.
\ee
Notice that $P_{\epsilon_\gamma}\circ \nabla$ is the connection
induced by $\nabla$ on the sub-bundle $(\wedge T^\ast_\K
M)^{\epsilon_\gamma}$ and that this induced connection is determined
by $(\nabla^S)^\ad$ through property \eqref{cliff_conn}. In the following,
we take $\nabla^S$ to be the connection on $S$
induced by the Levi-Civita connection of $(M,g)$; it is well-known
that $\nabla^S$ is a Clifford connection in the sense discussed
above\footnote{The same is true for any metric but torsion-full
connection.}.  A discussion of this and other properties of $\nabla^S$
in index language (which also serves to fix our conventions and leads
to another derivation of certain identities extracted in this paper)
can be found in Appendix \ref{sec:identities}. Equation
\eqref{cliff_conn} is compatible with the fact that $\nabla_m$ is a derivation of the
\KA algebra $(\Omega_\K(M), \diamond )$ --- a property which can be checked
by direct computation using the fact that $\nabla_m$ is an even
derivation of the exterior algebra which is compatible with the metric.
Similarly, we consider the connection $D^\ad$ induced by $D$ on $\End(S)$:
\be
D_m^\ad(T)\eqdef D_m\circ T-T\circ D_m~~,~~\forall T\in \Gamma(M,\End(S))~~.
\ee

\paragraph{Notation.} We let $\cK(D)=\cap_{m=1}^{d}\cK(
D_m) \subset \Gamma(M,S)$ be the finite-dimensional $\K$-linear subspace of
all generalized Killing pinors with respect to $D$. 

\paragraph{The dequantized connection.}

In the following, we consider only the case when the Schur algebra is
isomorphic with the base field. Then any connection $D$ on $S$ can be written as:
\ben
\label{D}
D_m=\nabla^S_m +A_m=\nabla^S_m+ \gamma(\check{A}_m)~~,
\een
where:
\be
\check{A}_m=\gamma^{-1}(A_m)\in \Omega_\K^\gamma(M)~~
\ee
are inhomogeneous differential forms on $M$.  The Clifford connection
property \eqref{cliff_conn} of $\nabla^S$ implies:
\ben
\label{DDK}
D_m^\ad\circ \gamma=\gamma\circ \D_m\Longleftrightarrow P_{\epsilon_\gamma}\circ \D_m
=\gamma^{-1}\circ D_m^\ad\circ \gamma \Longleftrightarrow
[D_m,\gamma(\omega)]_{-,\circ}=\gamma(\D_m\omega)~~,~~\forall \omega\in \Omega_\K(M)~~,
\een
where the derivation  $\D_m$ (which we shall call the {\em
adjoint dequantized connection}) of the \KA algebra
$(\Omega_\K(M),\diamond )$ is defined through:
\ben
\label{DK}
\D_m \omega\eqdef \nabla_m \omega+[\check{A}_m,\omega]_{-, \diamond}~~,~~\forall \omega\in \Omega_\K(M)~~.
\een
Since $\nabla_m\nu=0$ and $\check{A}_m \in \Omega^\gamma_\K(M)$, we
have $[\check{A}_m,\Omega^\gamma_\K(M)]_{-,\diamond}\subset \Omega^\gamma_\K(M)$
and $\D_m(\Omega^\gamma_\K(M))\subset \Omega^\gamma_\K(M)$. Composing
equation \eqref{DDK} with $\gamma^{-1}$ from both sides gives the
following relation which will be used below:
\ben
\label{phiflat}
\D_m\circ \gamma^{-1} =\gamma^{-1}\circ D_m^\ad ~~.
\een
To arrive at \eqref{phiflat}, we noticed that $\D_m\circ \gamma^{-1}(\End_\K(S))=D_m^\ad(\Omega^\gamma_\K(M))\subset \Omega^\gamma_\K(M)$ implies
$P_{\epsilon_\gamma}\circ \D_m\circ \gamma^{-1}=\D_m\circ \gamma^{-1}$, which in turn implies \eqref{phiflat} upon using the equation
$P_{\epsilon_\gamma}\circ \D_m\circ\gamma^{-1}=\gamma^{-1}\circ D_m^\ad$, which follows upon composing the last equality in \eqref{DDK} 
with $\gamma^{-1}$ from the right and using the property $\gamma\circ\gamma^{-1}=\id_{\End_\K(S)}$. Note that $A_m$ can be combined into the object:
\be
A=e^m \otimes A_m\in \Omega^1_\K(M)\otimes_\cinf \Omega_\K^\gamma(M)~~,
\ee
while $\D_m$ can be combined into the map:
\be
\D=e^m\otimes \D_m=\nabla+e^m\otimes
(\check{A}_m)^\ad:\Omega_\K(M)\longrightarrow \Omega^1_\K(M)\otimes_{\cinf}
\Omega_\K(M)~~,
\ee
where:
\be
(\check{A}_m)^\ad(\omega):=[\check{A}_m,\omega]_{-,\diamond}
\ee
and we used $e^m\otimes \nabla_m=\nabla$.

\paragraph{Flatness of the Fierz isomorphism.} For the remainder of
this paper, we shall assume that $D$ is compatible with $\cB$ in the
usual sense that $\cB$ is $D$-flat:
\ben
\label{DBFlatness}
\dd\cB(\xi,\xi')=\cB(D\xi, \xi')+\cB(\xi,D\xi')\Leftrightarrow
\partial_m \cB(\xi,\xi')=\cB(D_m\xi, \xi')+\cB(\xi,D_m\xi')
~,~\forall \xi,\xi'\in \Gamma(M, S).
\een
In this case, the $\K$-linear subspace $\cK(D)\subset \Gamma(M,S)$ is $\cB$-flat, so it defines 
a flat bipinor $\K$-algebra $\cK(D)\otimes_\cinf\cK(D) \subset  \Gamma(M,S\otimes S)$.
Furthermore, the isomorphism $\rho:S\stackrel{\sim}{\rightarrow}
S^{\ast} $ satisfies $D_m^{\ast} \circ \rho=\rho\circ D_m$ (where
$D_m^{\ast} $ is the dual connection) while the natural isomorphism
$q:S\otimes S^{\ast} \stackrel{\sim}{\rightarrow}\End(S)$ satisfies
$D^\ad_m\circ q=q\circ (D_m\otimes \id_S+\id_S\otimes D_m^{\ast}
)$. It follows that the isomorphism $E=q\circ (\id_S\otimes \rho)$
satisfies\footnote{Note that $D_m\otimes \id_S +\id_S \otimes D_m$ is
the connection induced by $D_m$ on $S\otimes S$.}:
\ben
\label{hEflat}
D_m^\ad \circ E=E\circ (D_m\otimes \id_S +\id_S \otimes D_m)~~.
\een
In particular, we have:
\ben
\label{Einclusion}
E(\cK(D)\otimes_\cinf \cK(D))\subset \cK(D^\ad)~~,
\een
where we have introduced the following $\K$-subalgebra of $(\Gamma(M,\End(S)),\circ)$:
\be
\cK(D^\ad)\eqdef \cap_{m=1}^d\cK(D_m^\ad)~~.
\ee
On the other hand, $\gamma^{-1}$ satisfies \eqref{phiflat}. Together
with \eqref{hEflat}, this implies that the Fierz isomorphism
$\check{E}$ satisfies:
\ben
\label{Fierzflat}
\D_m \circ \check{E}=\check{E}\circ (D_m \otimes \id_S+\id_S\otimes D_m)~~.
\een
Therefore, we find:
\ben
\label{FierzFlatness}
\D_m \check{E}_{\xi,\xi'}=\check{E}_{D_m \xi ,\xi'}+\check{E}_{\xi, D_m\xi'}~~,~~
\forall \xi,\xi'\in \Gamma(M, S)~~.
\een

\paragraph{The $\K$-algebra of generalized Killing forms.}

We define the {\em $\K$-algebra of generalized Killing forms} to be the following
$\K$-subalgebra of $(\Omega^\gamma_\K(M),\diamond)$:
\be
\check{\cK}_{D}\eqdef \cK(\D) \cap \Omega^\gamma_\K(M) =\gamma^{-1}(\cK(D^\ad))~~.
\ee
The elements of $\check{\cK}_{D^\ad}$  will be called {\em generalized Killing forms}.
Relation \eqref{Einclusion} implies that the flat Fierz $\K$-algebra: 
\be
\check{\cK}(D)\eqdef \check{E}(\cK(D)\otimes_\cinf \cK(D))~~
\ee
defined by the $\cB$-flat subspace $\cK(D)\subset \Gamma(M,S)$ is a subalgebra of the 
$\K$-algebra of generalized Killing forms: 
\ben
\label{KerInclusion}
\check{\cK}(D)\subset \check{\cK}_D~~.
\een
In particular, we have: 
\ben
\label{GKA}
\D_m \check{E}_{\xi,\xi'} =0~~,~~\forall \xi,\xi' \in \cK(D)~~.
\een

\paragraph{Behavior under $\tau_\cB$.} $\Spin(d)$-invariance of $\cB$
implies that $\cB$ is flat with respect to the connection
$\nabla^S$:
\ben
\label{NablaFlatness}
\small{
\!\!\! \dd\cB(\xi,\xi')=\cB(\nabla^S\xi, \xi')+\cB(\xi,\nabla^S\xi')\Longleftrightarrow
\partial_m \cB(\xi,\xi')=\cB(\nabla^S_m\xi, \xi')+\cB(\xi,\nabla^S_m\xi')
,~\forall \xi,\xi'\in \Gamma(M,S)~,~}
\een
which is easily seen to imply the property:
\be
\left[(\nabla^S_m)^\ad(T)\right]^t=(\nabla^S_m)^\ad(T^t)~~,~~\forall T\in \Gamma(M,\End(S))~~.
\ee
Together with the assumption \eqref{DBFlatness}, identity \eqref{NablaFlatness} implies
that $A_m$ are $\cB$-antisymmetric endomorphisms of $S$:
\be
A_m^t=-A_m\Longleftrightarrow \cB(A_m\xi, \xi')=-\cB(\xi, A_m\xi')~~,
~~\forall \xi, \xi'\in \Gamma(M,S)~~.
\ee
In turn, these properties imply the relation
$[A_m,T]_{-,\circ}^t=[A_m,T^t]_{-,\circ}$, so $D_m^\ad$ satisfies:
\ben
\label{DTranspose}
D_m^\ad(T^t)=\left(D_m^\ad(T)\right)^t~~,~~\forall T\in \Gamma(M,\End(S))
~\Longleftrightarrow~ D_m^\ad \circ (~)^t =(~)^t\circ D_m^\ad ~~.
\een
Setting $T=\gamma(\omega)$ and applying $\gamma^{-1}$ to both sides gives:
\ben
\label{checkDtranspose}
\D_m(\tau_\cB(\omega))=\tau_\cB(\D_m(\omega))~~,~~\forall \omega\in \Omega^\gamma_\K(M)
~\Longleftrightarrow~ \D_m \circ \tau_\cB =\tau_\cB\circ \D_m ~~.
\een
In particular, the $\K$-algebra of generalized Killing forms is
invariant under $\tau_\cB$:
\be
\tau_\cB(\check{\cK}_D)=\check{\cK}_D~~,
\ee
a property which (by virtue of \eqref{Et}) it shares with the flat Fierz 
$\K$-algebra $\check{\cK}(D)$:
\be
\tau_\cB(\check{\cK}(D))=\check{\cK}(D)~~.
\ee
Together with \eqref{ETranspose}, identity \eqref{checkDtranspose} implies
that $\D$-flatness of $\check{E}_{\xi,\xi'}$ and $\D$-flatness of
$\check{E}_{\xi',\xi}$ are equivalent statements, so that it suffices
to require only one of the two.

\subsection{Alternate form of the differential constraints}
\label{sec:fierzaltdiff}

Consider the following local expansion, which results by applying \eqref{completeness}
to $[T,E_{\xi,\xi'}]_{-,\circ}$, where $T\in \Gamma(M,\End(S))$:
\ben
\label{ident0}
[T, E_{\xi,\xi'}]_{-,\circ}=\frac{1}{2^{\left[\frac{d}{2}\right]}}
\sum_{k=0}^{d}\frac{1}{k!} \tr(\gamma^{a_k\ldots
a_1}\circ [T,E_{\xi,\xi'}]_{-,\circ}) \gamma_{a_1\ldots a_k} ~~,~~\forall
T\in \Gamma(M,\End(S))~~,~\forall \xi, \xi' \in \Gamma(M,S)~~.
\een
An easy computation using cyclicity of $\tr$ and identity \eqref{Etr} gives:
\be
 \tr(\gamma^{a_k\ldots a_1}\circ [T,E_{\xi,\xi'}]_{-,\circ})=
-\cB([T,\gamma^{a_k\ldots a_1}]_{-,\circ}\xi, \xi')~~,
\ee
so that \eqref{ident0} becomes:
\ben
\label{ident1}
\small{
\!\! [T, E_{\xi,\xi'}]_{-,\circ}=-\frac{1}{2^{\left[\frac{d}{2}\right]}}
\sum_{k=0}^{d}\frac{1}{k!} \cB([T,\gamma^{a_k\ldots
a_1}]_{-,\circ}\xi, \xi') \gamma_{a_1\ldots a_k} ~,~~\forall T\in
\Gamma(M,\End(S))~~,~\forall \xi, \xi' \in \Gamma(M,S)~.}
\een
Setting $T=A_m=\gamma(\check{A}_m)$ in \eqref{ident1} and applying the
morphism $\gamma^{-1}$ to both sides gives:
\ben
\label{ident2}
[\check{A}_m,\bcE_{\xi,\xi'}]_{-,\diamond}
=-\sum_{k=0}^{d}\frac{1}{k!} \cB([A_m,\gamma^{a_k\ldots
a_1}]_{-,\circ}\xi,\xi') e^{a_1\ldots a_k}_\gamma~~.
\een
Using this identity in the definition \eqref{DK} of $\D_m$ gives:
\ben
\label{ident3}
\D_m \bcE_{\xi,\xi'}=\nabla_m \bcE_{\xi,\xi'}
-\sum_{k=0}^{d}\frac{1}{k!} \cB(\xi, [A_m,\gamma_{a_1\ldots
a_k}]_{-,\circ}\xi') e^{a_1\ldots a_k}_\gamma~~.
\een
Consider now relation \eqref{FierzFlatness}, written in terms of
$\bcE_{\xi,\xi'}$:
\be
\D_m \bcE_{\xi,\xi'}=\bcE_{D_m \xi,\xi'}+
\bcE_{\xi, D_m \xi'}~~,~~\forall \xi, \xi' \in \Gamma(M,S)~~.
\ee
Substituting \eqref{ident3} in the left hand side, this becomes:
\ben
\label{NablaOmega}
\nabla_m \bcE_{\xi,\xi'}=\bcE_{D_m
\xi,\xi'}+\bcE_{\xi, D_m \xi'}+\sum_{k=0}^{d}\frac{1}{k!}
\cB(\xi, [A_m,\gamma_{a_1\ldots a_k}]_{-,\circ}\xi') e^{a_1\ldots
a_k}_\gamma~~.
\een
Separating ranks, we conclude that equation \eqref{FierzFlatness} is
equivalent with the following system of identities:
\ben
\label{FierzFlatDirect}
\nabla_m \bcE_{a_1\ldots
a_k}(\xi,\xi')=\bcE_{a_1\ldots a_k}(D_m
\xi,\xi')+ \bcE_{a_1\ldots a_k}(\xi, D_m \xi')+ \cB( \xi,
[A_m,\gamma_{a_1\ldots a_k}]_{-,\circ}\xi' )~~ ~,~~\forall k=0\ldots
d~,
\een
where $ \xi, \xi' \in \Gamma(M,S)$ are arbitrary. In particular, we have:
\ben
\label{NablaE}
\nabla_m \bcE_{a_1\ldots a_k}(\xi,\xi')= \cB(\xi,
[A_m,\gamma_{a_1\ldots a_k}]_{-,\circ}\xi')~~, ~~\forall k=0\ldots d~,
~~\forall \xi, \xi' \in \cK(D)~~, 
\een
which agrees with equation \eqref{diffcps} (see appendix
\ref{sec:DiffDirect}).

\subsection{The $\K$-algebra of constrained generalized Killing forms}
\label{sec:fierzCGKA}

As before, we consider the case $\chi=1$ of the CGK equations:
\beqan
\label{fulleq}
D\xi=Q\xi=0~~,~~\mbox{with}~~ D=\nabla^S+A~~,~~A=\dd x^m\otimes A_m~~.
\eeqan
Let $\cK(D,Q)=\cK(D)\cap \cK(Q)$ denote the (finite-dimensional)
$\K$-linear subspace of $\Gamma(M,S)$ consisting of all solutions to
\eqref{fulleq}.  We define
the {\em $\K$-algebra of constrained generalized Killing (CGK) forms} determined by
$D$ and $Q$ to be the following $\K$-subalgebra of $(\Omega^\gamma_\K(M),\diamond)$:
\be
\check{\cK}_{D,Q}\eqdef \check{\cK}_D \cap \check{\cK}_Q~~.
\ee
In general,  $\check{\cK}_{D,Q}$ is a non-unital $\K$-algebra. The discussion
of the previous subsections shows that the flat Fierz $\K$-algebra determined by the $\cB$-flat subspace 
$\cK(D,Q)\subset \Gamma(M,S)$:
\ben
\label{checkKDQ}
\check{{\cK}}(D,Q)\eqdef \check{E}(\cK(D,Q)\otimes_{\cinf} \cK(D, Q))~~
\een
is a subalgebra of the $\K$-algebra of CGK forms:
\ben
\label{checkKDQinclusion}
\check{{\cK}}(D,Q) \subset \check{\cK}_{D,Q}~~.
\een
This property of the Fierz isomorphism depends essentially on the
assumption that $\cB$ is $D$-flat (an assumption which {\em is}
satisfied in the application discussed in Section
\ref{sec:application}).

\paragraph{Expression for a basis of solutions of the CGK pinor equations.}
Let $s=\dim_\K \cK(D,Q)$ denote the $\K$-dimension of the space of
solutions to the CGK equations. Choosing a basis $(\xi_i)_{i=1\ldots
s}$ of such solutions, we set:
\ben
\label{BasisE}
\check{E}_{ij}\eqdef  \check{E}_{\xi_i,\xi_j}=
\frac{1}{2^{\left[\frac{d+1}{2}\right]}}\bcE_{ij}\in \Omega^\gamma_\K(M)~~,
\een
where (cf. equation \eqref{checkomegacomp}):
\ben
\label{checkEinhom}
\bcE_{ij}=\sum_{k=0}^d\bcE^{(k)}_{ij}~~,
\een
with
\beqa
&&\bcE^{(k)}_{ij}\eqdef \bcE^{(k)}_{\xi_i,\xi_j}=_U\frac{1}{k!}
\bcE^{(k)}_{a_1\ldots a_k}(\xi_i,\xi_j)e^{a_1\ldots a_k}_\gamma~~,\\
&&\bcE^{(k)}_{a_1\ldots a_k}(\xi_i,\xi_j)=
\cB(\gamma_{a_k\ldots a_1}\xi_i, \xi_j)=
\epsilon_\cB^k\cB(\xi_i, \gamma_{a_1\ldots a_k}\xi_j)~~
\eeqa
giving the following expression for the homogeneous form-valued bilinears:
\ben
\label{checkEhom}
\bcE^{(k)}_{ij}=_U\frac{1}{k!}
\epsilon_\cB^k\cB(\xi_i, \gamma_{a_1\ldots a_k}\xi_j)e^{a_1\ldots a_k}_\gamma~~.
\een

Since $\xi_i\otimes \xi_j$ form a basis of the $\K$-vector space $\cK(D,Q)\otimes_\cinf
\cK(D,Q)$, the inhomogeneous differential forms $\check{E}_{ij}$ form
a basis of the $\K$-vector space $\check{\cK}(D,Q)$. Inclusion
\eqref{checkKDQinclusion} amounts to the following system of equations
for the inhomogeneous differential forms $\check{E}_{ij}$:
\ben
\label{fulleqij}
\D_m \check{E}_{ij}=\check{Q}\diamond  \check{E}_{ij}=
\check{E}_{ij}\diamond  \tau_\cB(\check{Q})=0~~,~~\forall i,j=1\ldots s~~.
\een
These can also be written as:
\ben
\label{DQeqs}
\D_m \check{E}_{ij}=\check{Q}\diamond  \check{E}_{ij}=0~~,
~~\forall i,j=1\ldots s~~
\een
upon applying $\tau_\cB$ to the last equation in \eqref{fulleqij} and
using the relation
$\tau_\cB(\check{E}_{ij})=\epsilon_\cB\check{E}_{ji}$ (cf.
\eqref{ETranspose}). The inhomogeneous differential forms
$\check{E}_{ij}$ also satisfy the Fierz identities:
\ben
\label{Eids}
\check{E}_{i j}\diamond  \check{E}_{k l}=
\cB_{kj}\check{E}_{i l}~~,~~\forall i,j,k,l = 1\ldots s~~,
\een
where we defined the following constants:
\be
\cB_{ij}\eqdef \cB(\xi_i,\xi_j)~~.
\ee
The algebro-differential system consisting of \eqref{DQeqs} and
\eqref{Eids} can be taken as the basis for extending the classical
theory of Killing forms, a subject which is of mathematical interest in its own
right. It provides a synthetic geometric description of the essential
conditions imposed by having a fixed number of
unbroken supersymmetries in a flux compactification,
formulated in the language of geometric algebra. When expanding the
geometric product into generalized products using \eqref{starprod},
the innocently-looking equations \eqref{DQeqs} and \eqref{Eids} take
on a form which may seem rather formidable when $s$ is sufficiently large
(see Section \ref{sec:application} for an example). Of course, the
language of geometric algebra allows one to study such systems
starting directly from the synthetic expressions \eqref{DQeqs},
\eqref{Eids}, which shows that the problem of characterizing the
solutions of such equations belongs most properly to the intersection between
K\"{a}hler-Cartan theory \cite{KahlerCartan} and the theory of {\em
noncommutative} associative algebras --- a point of view on flux
compactifications, which, in our opinion, could lead to a deeper
understanding of various problems pertaining to that subject.

\paragraph{Truncated model of the $\K$-algebra of CGK forms in the non-simple case.} Recall that, in the non-simple case, we can realize
$(\Omega^\epsilon(M),\diamond)$ as the truncated algebra
$(\Omega^<(M),\bdiamond_\epsilon)$, where $\epsilon\in\{-1,+1\}$ is
the signature of $\gamma$. An inhomogeneous form
$\omega\in\Omega^\epsilon(M)$ satisfies $\tilde
\ast\omega=\epsilon\omega$ and is called twisted selfdual if
$\epsilon=+1$ and twisted anti-selfdual if $\epsilon=-1$. Such forms can be uniquely
decomposed as $\omega=2P_\epsilon(\omega_<)=\omega_<+\epsilon\tilde\ast\omega_<$, where
$\omega_<\in\Omega^<(M)$ has rank smaller than $[\frac{d}{2}]$.  In particular, the forms 
$\check{E}_{ij}$ discussed above can be decomposed uniquely as 
$\check{E}_{ij}=\check{E}^<_{ij}+\epsilon\tilde\ast\check{E}^<_{ij}$. Similarly, we have the unique decompositions 
$\check{Q}=\check{Q}^< +\epsilon \tilde \ast \check{Q}^<$ and $\check{A}_m=\check{A}_m^<+\epsilon {\tilde \ast}\check{A}_m^<$, where $\epsilon=\epsilon_\gamma$.
Let us define a derivation $\check{D}_m^{\ad,<}$ of $(\Omega(M),\bdiamond_\epsilon)$ through:
\ben
\label{Dmadtruncated}
\check{D}_m^{\ad,<}(\omega)\eqdef \nabla_m\omega+2[\check{A}_m^<,\omega]_{-,\bdiamond_\epsilon}
=\nabla_m\omega+([\check{A}_m,\omega]_{-,\diamond})^<=(\D_m\omega)^<~~,~~\forall \omega\in \Omega^<(M)~~, 
\een
where, as usual, the (anti-)commutator with respect to $\bdiamond_\epsilon$ is defined as:
\be
[\omega,\eta]_{\pm,\bdiamond_\epsilon}=\omega\bdiamond_\epsilon\eta \pm \eta\bdiamond_\epsilon\omega~~,
~~\forall \omega,\eta\in \Omega^<(M)~~.
\ee
We have: 
\beqa
([\omega,\eta]_{\pm,\diamond})^<=P_<([\omega,\eta]_{\pm,\diamond})=2[\omega_<,\eta_<]_{-,\bdiamond_\epsilon}~~,
~~\forall \omega,\eta\in \Omega(M)~~
\eeqa
since $2P_<$ is a morphism of algebras from $(\Omega(M),\diamond)$ to $(\Omega(M),\bdiamond_\epsilon)$. 
Noticing that $[\nabla_m,P_<]_{-,\circ}=0$, we find that $\check{D}_m^{\ad,<}$ satisfies: 
\be
P_< \circ \D_m =\check{D}_m^{\ad,<} \circ  P_< ~~,
\ee
which generalizes \eqref{Dmadtruncated}. 

Using the commutation relation $[\nabla_m,P_\epsilon]_{-,\circ}=0$ and the mutually-inverse 
isomorphisms of diagram \eqref{diagram:bdiamond}, 
it is easy to see that the CGK pinor equations \eqref{DQeqs} are equivalent with the {\em truncated} 
{\em CGK pinor equations}:
\ben
\label{DQeqstruncated}
\check{D}_m^{\ad,<} \check{E}^<_{ij}=\check{Q}^<\bdiamond_\epsilon  \check{E}^<_{ij}=0~~,
~~\forall i,j=1\ldots s~~.
\een
Indeed, applying $2P_<$ to \eqref{DQeqs} gives \eqref{DQeqstruncated} while applying $P_\epsilon$ to 
\eqref{DQeqstruncated} gives \eqref{DQeqs}. 
The first of the truncated CGK pinor equations can also be written in the form 
$\nabla_m\check{E}^<_{ij}=-2[\check{A}_m,\check{E}^<_{ij}]_{-,\bdiamond_\epsilon}$. 
On the other hand, applying $2P_<$ to \eqref{Eids} gives the {\em truncated} {\em geometric Fierz identities}:
\ben
\label{FierzRel}
\check{E}^<_{ij}\bdiamond_\epsilon\check{E}^<_{kl}=\frac{1}{2} \cB_{kj}\check{E}^<_{il}~~,~~\forall i,j,k,l=1\ldots s~~.~~
\een

\subsection{A particular case: when $\K=\R$ and $\cB$ is a scalar product.}
\label{sec:fierz2}

Consider the particular case when $\K=\R$ and $\cB$ is symmetric
(thus $\sigma_\cB=+1$) and positive-definite with $\epsilon_\cB=+1$
(this happens, for example, in the application considered in Section
\ref{sec:application}). In this case, we can choose $\xi_1,\ldots, 
\xi_s$ such that $\cB_{ij}=\delta_{ij}$.  Then relations \eqref{Eids}
show that the $\K$-algebra $\check{\cK}(D,Q)$ is
isomorphic with the algebra $\Mat(s,\R)$ of square real matrices of
dimension $s$, the unit being given by $\check{C}\eqdef \sum_{i=1}^s
\check{E}_{ii}$.  An isomorphism to $\Mat(s,\R)$ is given by
$\check{E}_{ij}\rightarrow e_{ij}$, where $e_{ij}\in \Mat(s,\R)$ is
the matrix whose only non-vanishing entry equals $1$ and is found on
the $i$-th row and $j$-th column:
\be
(e_{ij})_{kl}=\delta_{ik}\delta_{jl}~~.
\ee
We have $\check{C}^{\diamond 2}=\check{C}$, $\check{C}\diamond
\check{E}_{ij}=\check{E}_{ij}\diamond \check{C}=\check{E}_{ij}$,
$\check{E}_{ii}^{\diamond 2}=\check{E}_{ii}$, $\check{E}_{ij}^{\diamond 2}=0$ for $i\neq
j$. Since $\check{E}_{ij}\diamond \check{E}_{ji}=\check{E}_{ii}$, the $\K$-algebra
$\check{\cK}(D,Q)$ is generated by the elements $(\check{E}_{ij})_{i\neq j}$, with
the relations:
\beqa
&~&\check{E}_{ij} \diamond \check{E}_{kl} = 0~~,~~\forall i\neq j,~j\neq k, ~k\neq l~~,\\
&~&\check{E}_{ij}\diamond \check{E}_{jk} = \check{E}_{ik}~~,~~\forall i, j, k~~{\rm distinct}~~,\\
&~&\check{E}_{ij}\diamond \check{E}_{ji}\diamond \check{E}_{ik} = \check{E}_{ik}~~,~~\forall i\neq j~~{\rm and}~~i\neq k~~,\\
&~&\check{E}_{ij}\diamond \check{E}_{jk}\diamond \check{E}_{kj} = \check{E}_{ij}~~,~~\forall i\neq j~~{\rm and}~~j\neq k~~.
\eeqa
This system of generators and relations can be reduced further (see
the example below) upon using the identity
$\check{E}_{ji}=\tau(\check{E}_{ij})$, where we used the fact that
$\sigma_\cB=+1$ and we noticed that $\tau_\cB=\tau$ (since
$\epsilon_\cB=+1$).

\paragraph{The case of one CGK pinor.} When $s=1$ (a single unit norm solution $\xi$ of the CGK pinor equations, which is unique up to sign), the flat Fierz $\K$-algebra
$\check{\cK}(D,Q)$ has dimension one, being isomorphic with $\R$ as a unital 
$\R$-algebra.  It is generated by the single basis element $\check{E}=\check{E}_{\xi,\xi}$ (which is
the unit of $\check{\cK}(D,Q)$), with the relation:
\ben
\label{e0}
\check{E}\diamond \check{E}=\check{E}~~.
\een
The condition $\tau(\check{E})=\check{E}$ implies that certain rank components
of $\check{E}$ vanish identically. This situation occurs in the example
considered in Section \ref{sec:application}, though in that case we prefer to
take the squared norm $\cB(\xi,\xi)$ of  $\xi$ to equal $2$, for ease of comparison
with the results of \cite{MartelliSparks}. This produces an extra factor of
two in the right hand side of \eqref{e0}. 

\paragraph{The case of two CGK pinors.} When $s=2$, the flat Fierz $\K$-algebra $\check{\cK}(D,Q)$
is generated by $\check{E}_{12}$ and $\check{E}_{21}$ with the relations:
\beqan
&~&\check{E}_{12}^{\diamond 2}= \check{E}_{21}^{\diamond 2}=0 ~~,\label{e1}\\
&~&\check{E}_{12}\diamond \check{E}_{21}\diamond
\check{E}_{12}-\check{E}_{12}=\check{E}_{21}\diamond \check{E}_{12}\diamond \check{E}_{21}-\check{E}_{21}=0~~\label{e2}.
\eeqan
We have $\check{E}_{11}=\check{E}_{12}\diamond \check{E}_{21}$,
$\check{E}_{22}=\check{E}_{21}\diamond \check{E}_{12}$ and
$\check{C}=\check{E}_{11}+\check{E}_{22}=\check{E}_{12}\diamond
\check{E}_{21}+\check{E}_{21}\diamond \check{E}_{12}$. Since
$\check{E}_{21}=\tau(\check{E}_{12})$, we can in fact generate the
entire algebra from $\check{E}_{12}$ up to applying $\tau$. Using the
fact that the first and second equation in \eqref{e1} as well as the
first and second equation in \eqref{e2} are related through
reversion, we find that the entire system is equivalent with the
following Fierz relations:
\be
\check{E}_{12}^{\diamond 2}=0 ~~,~~\check{E}_{12}\diamond \tau(\check{E}_{12})\diamond \check{E}_{12}=\check{E}_{12}~~,
\ee
together with the following $Q$- and $D$-constraints:
\be
\check{Q}\diamond \check{E}_{12}=\check{E}_{12}\diamond \tau(\check{Q})=0~~,~~\D_m \check{E}_{12}=0~~.
\ee
\paragraph{Remark.} The algebra $\Mat(2,\R)\approx \Cl(2,0)$ is also
generated by two real Pauli matrices, for example by:
\be
\sigma_1=\left[\begin{array}{cc}~0 & ~1\\~1&~0\end{array}\right]=e_{12}+e_{21}~~,~~
\sigma_3=\left[\begin{array}{cc}~1 &~0\\~0& -1\end{array}\right] =e_{11}-e_{22}~~,
\ee
subject to the relations $\sigma_1^2=\sigma_3^2=1$ and
$\sigma_1\sigma_3+\sigma_3\sigma_1=0$.  Therefore, the flat Fierz
$\K$-algebra $\check{\cK}(D,Q)$ is also generated by the inhomogeneous
differential forms:
\be
\check{\Sigma}_1=\check{E}_{12}+\check{E}_{21}~~,
~~ \check{\Sigma}_3=\check{E}_{11}-\check{E}_{22}~~,
\ee
subject to the relations:
\beqa
\check{\Sigma}_1^{\diamond 2}-\check{\Sigma}_3^{\diamond 2}=0~~&,&~\check{\Sigma}_1\diamond \check{\Sigma}_3+\check{\Sigma}_3\diamond \check{\Sigma}_1=0~~,\\
\check{\Sigma}_1^{\diamond 3}=\check{\Sigma}_1~~&,&~~\check{\Sigma}_3^{\diamond 3}=\check{\Sigma}_3~~.
\eeqa
Noticing that $\check{\Sigma}_1\diamond
\check{\Sigma}_3=\check{E}_{21}-\check{E}_{12}$, we have
$\check{C}=\check{E}_{11}+\check{E}_{22}=
\check{\Sigma}_1^{\diamond 2}=\check{\Sigma}_3^{\diamond 2}$ and
\beqa
&~&\check{E}_{11}=\frac{1}{2}(\check{C}+\check{\Sigma}_3)~~,~~\check{E}_{22}=\frac{1}{2}(\check{C}-\check{\Sigma}_3)~~,\\
&~&\check{E}_{12}=\frac{1}{2}(\check{\Sigma}_1-\check{\Sigma}_1\diamond
\check{\Sigma}_3)~~,~~
\check{E}_{21}=\frac{1}{2}(\check{\Sigma}_1+\check{\Sigma}_1\diamond
\check{\Sigma}_3)~~.
\eeqa

\section{Example: flux compactifications of 
eleven-dimensional supergravity on eight-manifolds}
\label{sec:application}

In this section, we illustrate the methods developed in the present paper by
applying them to the case of ${\cal N}=1$ warped compactifications of 
eleven-dimensional supergravity on eight-manifolds down to an $\AdS_3$ space,
a situation which was studied through direct methods in
\cite{MartelliSparks} and \cite{Tsimpis}. After some basic preparations in Subsection
\ref{sec:applprep}, Subsection \ref{sec:applCGK} gives our translation of the generalized Killing pinor equations
into a system of algebraic and differential constraints on inhomogeneous forms
defined on the compactification space and shows how our approach allows one to
recover the results of \cite{MartelliSparks}. 

\subsection{Preparations}
\label{sec:applprep}

Consider supergravity on an 11-manifold $\tilde{M}$
with Lorentzian metric ${\tilde g}$ (of `mostly plus' signature).
Besides the metric, the action of the theory contains the three-form potential with
four-form field strength $\tilde{G}\in\Omega^4(\tilde{M})$ and the
gravitino $\tilde{\Psi}_M$, which is a real pinor of spin $3/2$. We
assume that $({\tilde M},{\tilde g})$ is spinnable. For supersymmetric
bosonic backgrounds, both the gravitino VEV and its supersymmetry
variation must vanish, which requires that there exists at least one
solution ${\tilde \eta}$ to the equation:
\ben
\label{susy}
\delta_{\tilde \eta} {\tilde \Psi}_M \eqdef  \tcD_M {\tilde \eta} = 0~~,
\een
where uppercase indices run from $0$ to $10$ and $\tcD_M$ is the
supercovariant connection:
\be
\tcD_M\eqdef \nabla^{\tilde S}_M - \frac{1}{288} \left( {\tilde G}_{NPQR}
\tilde{\gamma}^{NPQR}{}_M- 8{\tilde G}_{MNPQ}\tilde{\gamma}^{NPQ} \right)~~.
\ee
Here, $\tilde{\gamma}^M$ are the gamma matrices of $\Cl(10,1)$ in that
32-dimensional real (Majorana) irreducible representation for which
${\tilde \gamma}^{(12)}\eqdef {\tilde \gamma}^1\ldots {\tilde
\gamma}^{11}=+1$ and
\be
\nabla^{\tilde S}_M=\partial_M+\frac{1}{4}{\tilde \w}_{MNP}{\tilde \gamma}^{NP}~~
\ee
is the connection on the pin bundle ${\tilde S}$ induced by the
Levi-Civita connection of ($\tilde{M}$, $\tilde{g}$).  The
eleven-dimensional supersymmetry generator ${\tilde \eta}$ (which is a
Majorana spinor field of spin $1/2$) is a smooth section of the pin
bundle ${\tilde S}$, which is a rank $32$ real vector bundle defined
on ${\tilde M}$.

As in \cite{MartelliSparks}, we consider compactification down to an
$\AdS_3$ space of cosmological constant $\Lambda=-8\kappa^2$, where
$\kappa$ is a positive real parameter --- this includes the Minkowski
case as the limit $\kappa\rightarrow 0$.  Thus ${\tilde M}=N\times M$,
where $N$ is an oriented 3-manifold diffeomorphic with $\R^3$ and carrying the
$\AdS_3$ metric while $M$ is an oriented Riemannian eight-manifold whose metric
we denote by $g$. The metric on ${\tilde M}$ is a warped product:
\beqan
\label{warpedprod}
\dd {\tilde s}^2_{11}  & = & e^{2\Delta} \dd s_{11}^2~~~{\rm where}~~~\dd s_{11}^2=\dd s^2_3+ g_{mn} \dd x^m \dd x^n~~.
\eeqan
Here, the warp factor $\Delta$ is a smooth function defined on $M$ while $\dd s_3^2$ is the
squared length element on $N$. For the field strength ${\tilde G}$, we use the ansatz:
\be
{\tilde G} = e^{3\Delta} G~~~{\rm with}~~~ G = {\rm vol}_3\wedge f+F~~,
\ee
where $f=f_m e^m\in \Omega^1(M)$, $F=\frac{1}{4!}F_{mnpq} e^{mnpq}\in \Omega^4(M)$ and
$\vol_3$ is the volume form of $N$. Small Latin indices from the middle of the alphabet run
from $1$ to $8$ and correspond to a choice of frame on $M$. For ${\tilde \eta}$, we
use the ansatz:
\be
{\tilde \eta}=e^{\frac{\Delta}{2}}\eta~~~{\rm with}~~~ \eta=\psi\otimes \xi~~,
\ee
where $\xi$ is a Majorana spinor of spin $1/2$ (a.k.a. a real pinor) on the
internal space $M$ and $\psi$ is a Majorana spinor on the $\AdS_3$
space $N$. Mathematically, $\xi$ is a section of the pinor bundle of
$M$, which is a real vector bundle of rank $16$ defined on $M$,
carrying a fiberwise representation of the Clifford algebra
$\Cl(8,0)$. Since $p-q\equiv_8 0$ for $p=8$ and $q=0$, this
corresponds to the simple normal case of Section \ref{sec:pin}. In
particular, the corresponding morphism $\gamma:(\wedge T^\ast
M,\diamond)\rightarrow (\End(S),\circ)$ of bundles of algebras is an
isomorphism, i.e. it is bijective on the fibers. We set
$\gamma^m=\gamma(e^m)$ for
some local frame of $M$. In dimension eight with Euclidean signature,
there exists an admissible \cite{AC0} (and thus $\Spin(8)$-invariant)
bilinear pairing $\cB$ on the pin bundle $S$ with $\sigma_\cB=+1$ and $\epsilon_\cB=+1$ 
which is a scalar product (i.e., it is fiberwise symmetric and positive-definite).

Assuming that $\psi$ is a Killing pinor on the $\AdS_3$ space, the
supersymmetry condition \eqref{susy} decomposes into a system
consisting of the following constraints for $\xi$:
\ben
\label{par_eq}
D_m\xi = 0~~,~~Q\xi = 0~~,
\een
where $D_m$ is a linear connection on $S$ and $Q\in \Gamma(M,\End(S))$ is
a globally-defined endomorphism of the vector bundle $S$. As in
\cite{MartelliSparks, Tsimpis},  {\em we do not require that $\xi$ has definite
chirality} (that is, $\xi$ need {\em not} satisfy the Weyl
condition).  The space of solutions of \eqref{par_eq} is a
finite-dimensional $\R$-linear subspace $\cK(D,Q)$ of the space $\Gamma(M,S)$ of
smooth sections of $S$. Of course, this subspace is trivial for
generic metrics $g$ and fluxes $F$ and $f$ on $M$, since the generic
compactification of the type we consider breaks all supersymmetry. The
interesting problem is to find those metrics and fluxes on $M$ for
which some fixed amount of supersymmetry is preserved in three
dimensions, i.e.  for which the space $\cK(D,Q)$ has some given
non-vanishing dimension, which we denote by $s$. The case $s=1$ (which
corresponds to ${\cal N}=1$ supersymmetry in three dimensions) was studied in
\cite{MartelliSparks, Tsimpis} and will be re-considered below. Direct computation using the compactification ansatz gives the
following expressions for $D$ and $Q$ (which are equivalent with those
derived in reference \cite{MartelliSparks} --- see the remark below for
ease of comparison):
\beqan
\label{Dappl}
D_m &=&\nabla^S_m+A_m~~,~~A_m= \frac{1}{4}f_p\gamma_{m}{}^{p}\circ \gamma^{(9)}
+\frac{1}{24}F_{m p q r}\gamma^{ p q r}+\kappa \gamma_m\circ \gamma^{(9)}~~,\\
\label{Qappl}
&& Q=\frac{1}{2}\gamma^m\partial_m\Delta -\frac{1}{288}F_{m p q r}\gamma^{m p q r}
-\frac{1}{6}f_p \gamma^p \circ \gamma^{(9)}
-\kappa\gamma^{(9)} ~~,
\eeqan
where $\gamma^{(9)}=\gamma^1\circ\ldots \circ\gamma^8$. Notice that the last terms in \eqref{Dappl} 
and \eqref{Qappl} depend on the cosmological constant
of the $\AdS_3$ space --- and that they vanish in the Minkowski limit $\kappa\rightarrow 0$.

Given some desired amount of supersymmetry which we want to be
preserved in three dimensions (i.e. given some desired
dimensionality $s$ of the space of solutions to \eqref{par_eq}), the general aim is to reformulate 
\eqref{par_eq} as equations on differential forms 
$\bcE^{(k)}_{\xi,\xi'}=\frac{1}{k!}\cB(\xi,\gamma_{m_1\ldots m_k}\xi') e^{a_1\ldots a_m}$ 
constructed as bilinear combinations of pinors $\xi,\xi'$ which satisfy
\eqref{par_eq}. The pinor bilinears 
will be constrained by Fierz identities. The translation to equations on the differential forms
$\bcE^{(k)}_{\xi,\xi'}$ can be achieved directly by starting from the
following equivalent reformulation of the algebraic constraints
$Q\xi=Q\xi'=0$:
\be
\cB( \xi,\left(Q^t\circ\gamma_{m_1\ldots m_k}\pm \gamma_{m_1\ldots m_k}\circ Q\right) \xi') =0~~
\ee
and treating the constraints $D_m \xi=D_m\xi'=0$
through the method outlined in \cite{MartelliSparks}. The theoretical
basis of that approach is explained in detail in Appendix
\ref{sec:direct}, where we also show how that method is equivalent
with the more general approach which we have developed in the present
paper. In the next  Subsection, we shall illustrate our approach in the simplest case $s=1$
(${\cal N}=1$ supersymmetry in three dimensions, see \cite{MartelliSparks}), so we
will require that \eqref{par_eq} admits {\em one} non-trivial solution $\xi$. 

\paragraph{Remark.} For easy comparison with \cite{MartelliSparks}, we note
that loc. cit. uses a redundant parameterization of the degrees of
freedom described by $\xi$, which is given by the following sections of $S$:
\be
\varepsilon^\pm\eqdef \frac{1}{\sqrt{2}}(\xi_{+}\pm \xi_{-})~~,
\ee
which satisfy $\gamma^{(9)} \varepsilon^\pm=\varepsilon^\mp$.  Here,
$\xi_\pm=\cP_\pm\xi$ are the positive and negative chirality
components of $\xi$, with $\cP_\pm\eqdef \frac{1}{2}(1\pm
\gamma^{(9)})$. We have $\xi=\xi_++\xi_-$ and
$\gamma^{(9)}\xi=\xi_+-\xi_-$ with $\gamma^{(9)}\xi_\pm=\pm \xi_\pm$.
The operators $\cP_\pm$ are complementary $\cB$-orthoprojectors:
\be
\cP_\pm^2=\cP_\pm~~,~~\cP_\pm \cP_\mp=0~~,~~\cP_++ \cP_-=1~~,~~(\cP_\pm)^t=\cP_\pm~~.
\ee
In particular, one has:
\ben
\label{epsxi}
\varepsilon^+=\frac{1}{\sqrt{2}}\xi~~,~~ \varepsilon^{-}=\frac{1}{\sqrt{2}}\gamma^{(9)} \xi~~,
\een
so $\varepsilon^+$ and $\varepsilon^-=\gamma^{(9)}\varepsilon^+$ are not independent. For reader's
convenience, we note the identities:
\beqa
\cP_{\epsilon_1}\gamma^{a_1\ldots
a_k}\cP_{\epsilon_2}=0&\Longrightarrow&
\cB(\xi_{\epsilon_1},\gamma^{a_1\ldots
a_k}\xi_{\epsilon_2})=0~~,~~{\rm
when}~~\epsilon_1\epsilon_2=(-1)^{k+1}~~, \nn \\
\cP_{\epsilon_1}\gamma^{a_1\ldots
a_k}\cP_{\epsilon_2}=\cP_{\epsilon_1} \gamma^{a_1\ldots a_k}
&\Longrightarrow& \cB(\xi_{\epsilon_1},\gamma^{a_1\ldots
a_k}\xi_{\epsilon_2})=\cB(\xi, \cP_{\epsilon_1}\gamma^{a_1\ldots a_k}
\xi)~~,~{\rm when}~~\epsilon_1\epsilon_2=(-1)^k ~~,\nn
\eeqa
where $\epsilon_1,\epsilon_2\in \{-1,+1\}$ and $\cP_{\pm 1}\eqdef
\cP_{\pm}$,~ $\xi_{\pm 1}\eqdef \xi_{\pm}$. Since $\epsilon_\cB=+1$, \eqref{relB} implies:
\be
(\gamma^{a_1\ldots a_k})^t=(-1)^{\frac{k(k-1)}{2}}\gamma^{a_1\ldots a_k}~~,
\ee
which shows that $\gamma^{a_1\ldots a_k}$ is $\cB$-symmetric for
$k=0,1,4,5,8$ and $\cB$-antisymmetric for $k=2,3,6,7$. In particular,
we have that $\gamma^{(9)}$ is $\cB$-symmetric.

\subsection{The case of ${\cal N}=1$ supersymmetry in 3 dimensions}
\label{sec:applCGK}

Let us consider the CGK pinor equations \eqref{par_eq} on the Riemannian
8-manifold $(M,g)$, assuming that the space of solutions has dimension $s=1$
over $\R$. Since $p-q\equiv_8 0$, we are in the normal simple case.
This case is characterized by two admissible pairings $\cB_\pm$ on $S$, which 
have the properties given in \cite{AC1}, i.e. 
$\epsilon_{\cB_\pm}=\pm 1$ and $\sigma_{\cB_\pm}=+1$.
Since any choice of admissible pairing leads to the same result, 
we choose to work with $\cB=\cB_+$ for convenience; 
this satisfies $\epsilon_\cB=+1$ and $\sigma_\cB=+1$. We can assume that 
$\cB$ is a scalar product on $S$ and we 
denote the corresponding norm by $||~||$. 

As in Subsection \ref{sec:fierzCGKA} (see equation \eqref{BasisE}), consider the inhomogeneous differential form: 
\be
\check{E}\eqdef\frac{1}{16} \sum_{k=0}^8 \bcE^{(k)}\in \Omega(M)~~
\ee
defined by a non-trivial solution $\xi$ of \eqref{par_eq}. Equation \eqref{checkEhom} gives:
\be
\bcE^{(k)}=\frac{1}{k!}\cB(\xi,\gamma_{m_1...m_k}\xi)e^{m_1...m_k} \in \Omega^k(M)~~,
~~\forall a_1,\ldots,a_k=1\ldots8~~,~~\forall k=0\ldots8~.
\ee
Using the properties of $\cB$ and relations \eqref{transpose} and \eqref{relB}, 
one easily checks that $\bcE^{(2)}=\bcE^{(3)}=\bcE^{(6)}=\bcE^{(7)}=0$ 
while the non-vanishing bilinears are:
\beqan
\label{forms8}
&&\bcE^{(0)}=\cB(\xi,\xi)~~,~~\bcE^{(1)}=\cB(\xi,\gamma_a\xi)e^a ~~
,~~ \bcE^{(4)}= \frac{1}{4!}\cB(\xi,\gamma_{a_1\ldots a_4}\xi) e^{a_1\ldots a_4}~,\nn\\
&& \bcE^{(5)}=\frac{1}{5!} \cB(\xi,\gamma_{a_1\ldots a_5}\xi) e^{a_1\ldots a_5}~~,
~~ \bcE^{(8)}=\frac{1}{8!} \cB(\xi,\gamma_{a_1\ldots a_8}\xi) e^{a_1\ldots a_8}=\cB(\xi, \gamma_{(9)}\xi)~\nu~,~~~~~~~~~~~
\eeqan
where $\nu$ is the volume form. 
Thus the inhomogeneous form which generates the Fierz algebra expands as:
\ben
\label{Basis1}
\check{E} = \frac{1}{16}\left[\bcE^{(0)}+\bcE^{(1)}+\bcE^{(4)}+\bcE^{(5)}+\bcE^{(8)}\right]~~.
\een
Note that this case also admits non-vanishing 3- and 7-form bilinears, which are proportional to the Hodge duals of 
the 5- and 1-forms respectively (but they do not appear in the generator). 
These other form-valued bilinears can also be expressed as in \eqref{checkEhom}, but upon using the other admissible pairing $\cB_{-}$, which is in fact not necessary 
for this analysis.

\paragraph{Remark.} If one wanted to write the form-valued pinor bilinears in terms of the other admissible 
pairing  $\cB_{-}=\cB_+\circ(\id_S\otimes\gamma(\nu))$, then one would be interested in constructing:
\beqa
&& \bcE_{-}^{(k)}\eqdef\frac{1}{k!}(\epsilon_{\cB_-})^{k} \cB_{-}(\xi,\gamma_{a_1...a_k}\xi)e^{a_1\ldots a_k}
=\frac{1}{k!}(-1)^{k}\cB(\xi,\gamma_{a_1...a_k}\circ\gamma(\nu)\xi)e^{a_1\ldots a_k}\in\Omega^k(M)~~,\nn\\
&&~~~~~~\forall a_1,\ldots,a_k=1\ldots8~~,~~\forall k=0\ldots8~~,
\eeqa
which, upon using $\epsilon_{\cB_{-}}=-1$ and $\sigma_{\cB_{-}}=+1$
and the well-known formula:
\be
\gamma_{a_1...a_k}=\frac{(-1)^\frac{k(k-1)}{2}}{(d-k)!}\epsilon_{a_1...a_k}{}^{a_{k+1}...a_d}
\gamma_{a_{k+1}...a_d}\circ\gamma^{(d+1)}
\ee
would lead to the homogeneous form-valued bilinears:
\beqan
\label{forms8second}
&& \bcE_{-}^{(0)}=\cB_{-}(\xi,\xi)=\lambda=\cB(\xi,\gamma(\nu)\xi)=\bcE^{(8)}\diamond\nu~,\nn\\
&& \bcE_{-}^{(3)}= -\frac{1}{3!}\cB_{-}(\xi,\gamma_{a_1\ldots a_3}\xi) e^{a_1\ldots a_3}=
-\frac{1}{3!}\cB(\xi,\gamma_{a_1\ldots a_3}\circ\gamma(\nu)\xi) e^{a_1\ldots a_3}=
 -\bcE^{(5)}\diamond\nu,\nn\\
&& \bcE_{-}^{(4)}= \frac{1}{4!}\cB_{-}(\xi,\gamma_{a_1\ldots a_4}\xi) e^{a_1\ldots a_4}=
\frac{1}{4!}\cB(\xi,\gamma_{a_1\ldots a_4}\circ\gamma(\nu)\xi) e^{a_1\ldots a_4}=
\bcE^{(4)}\diamond\nu,~~~~~~~~~~~~~~~~~~~~~~~~\\
&& \bcE_{-}^{(7)}=-\frac{1}{7!} \cB_{-}(\xi,\gamma_{a_1\ldots a_7}\xi) e^{a_1\ldots a_7}=
-\frac{1}{7!}\cB(\xi,\gamma_{a_1\ldots a_7}\circ\gamma(\nu)\xi) e^{a_1\ldots a_7}
=-\bcE^{(1)}\diamond\nu,\nn\\
&& \bcE_{-}^{(8)}=\frac{1}{8!} \cB_{-}(\xi,\gamma_{a_1\ldots a_8}\xi) e^{a_1\ldots a_8}=\beta\nu
=\frac{1}{8!} \cB(\xi,\gamma_{a_1\ldots a_8}\circ\gamma(\nu)\xi) e^{a_1\ldots a_8}=
\bcE^{(0)}\diamond\nu~.~~~~~~\nn
\eeqan
Thus, one could also write the generator:
\be
\check{E}_{-}=\frac{N}{2^d}\sum_k \bcE_{-}^{(k)}
\ee
which would lead to the same result up to twisted Hodge duality --- 
since the form-valued bilinears constructed using $\cB_{-}$ are connected through 
Hodge duality with those constructed using $\cB_{+}$. 

Following the notations and conventions of \cite{MartelliSparks}, we define:
\be
 a\eqdef\frac{1}{2}\bcE^{(0)}~~,~~ \bar{K}\eqdef\frac{1}{2}\bcE^{(1)}~~,~~
 Y\eqdef\frac{1}{2}\bcE^{(4)}~~,~~
 \bar{Z}\eqdef\frac{1}{2}\bcE^{(5)}~~,~~ W\eqdef\frac{1}{2}\bcE^{(8)}=b\nu~~,
\ee
where $2a$ is the squared norm of $\xi$ and $b$ a 
smooth function on $M$. 
With these notations, \eqref{Basis1} becomes:
\ben
\label{Basis}
\check{E} = \frac{2}{16}\left[a+\bar{K}+Y+\bar{Z}+b~\nu\right]~~.
\een
Note that we use the five-form $\bar{Z}$ instead of its sign-reversed 
 Hodge dual $\bar{\phi}\eqdef -\ast {\bar Z}$, which was used in 
\cite{MartelliSparks}. We prefer to work with $\xi=\xi_+\oplus\xi_{-}$ rather than 
with $\xi_\pm$ and $\epsilon^\pm\eqdef\frac{1}{\sqrt{2}}(\xi_+\pm\xi_{-})$, 
which were used in loc. cit. All in all, we have the following correspondence with 
the notations and conventions of \cite{MartelliSparks}:
\beqa
&& \xi_\pm=\cP_\pm\xi~~,~~\mathrm{where}~~\cP_\pm=\frac{1}{2}(1\pm\gamma^{(9)})~~,\\
&& \varepsilon^+=\frac{1}{\sqrt{2}}\xi~~,~~ \varepsilon^-=\frac{1}{\sqrt{2}}\gamma^{(9)}\xi~~,\\
&& \bar{K}_m=\xi_{+}^t\gamma_m\xi_{-}=\xi^t \cP_{+}\gamma_m\xi=\frac{1}{2}\xi^t\gamma_m\xi~~,\\
&& -(\ast\bar{Z})_{mnp}=\bar{\phi}_{mnp}=\xi_{+}^t\gamma_{mnp}\xi_{-}=\xi^t \cP_{+}\gamma_{mnp}\xi=\frac{1}{2}\xi^t\gamma^{(9)}\gamma_{mnp}\xi
~~,\\
&& Y_{mnpr}={\varepsilon^\pm}^t\gamma_{mnpr}\varepsilon^\pm=\frac{1}{2}(\xi_{+}^t\gamma_{mnpr}\xi_{+}+\xi_{-}^t\gamma_{mnpr}\xi_{-})
=\frac{1}{2}\xi^t\gamma_{mnpr}\xi~~.
\eeqa

\

\noindent The dequantizations of $A_m$ and $Q$ are given by:
\beqa
\check{A}_m &\eqdef& \gamma^{-1}(A_m) = \frac{1}{4}\iota_{(e_m)_\sharp} F + \frac{1}{4}((e_m)_\sharp \wedge f)\diamond \nu +\kappa ~(e_m)_\sharp \diamond\nu~~, \nn\\
  \check{Q} &\eqdef& \gamma^{-1}(Q) = \frac{1}{2} \dd\Delta -\frac{1}{6} f\diamond \nu -\frac{1}{12}F -\kappa~\nu~~.
\eeqa

\noindent As shown in Section \ref{sec:fierz}, the CGK pinor equations imply the following conditions for $\check{E}$:
\beqan
\label{QConstr}
 \check{Q}\diamond\check{E}&=& 0~~,\\
\label{NablaConstr}
\nabla_m\check{E} &=&- [\check{A}_m,\check{E}]_{-,\diamond}~~.
\eeqan

\noindent In turn, relations \eqref{NablaConstr} imply: 
\ben
\label{DiffConstr}
\dd\check{E} = -(e_m)_\sharp\wedge[\check{A}_m,\check{E}]_{-,\diamond}~~.
\een

\noindent In this case, \eqref{Eids} amounts to only one quadratic relation for the inhomogeneous form $\check{E}$:
\ben
\label{FierzConstr}
\check{E}\diamond\check{E}=2a\check{E}~~,
\een
which encodes the relevant Fierz identities between the form bilinears constructed from $\xi$.

\

 We remind the reader of the following  relations (see the previous sections):
\be
\ast\omega=\tau(\omega)\diamond\nu=\iota_\omega\nu~~,~~\ast\ast\omega=\pi(\omega)~~,
~~\forall\omega\in \Omega(M)~~,
\ee
where $\tau$ is the reversion defined in \eqref{taudef}. During our calculations we will 
use the following identities which hold in any dimension and signature
for any homogeneous forms $\omega\in \Omega^r(M)$ and $\eta\in
\Omega^s(M)$:
\beqan
\label{Contractions}
&& \omega\wedge\ast\eta=(-1)^{r(s-1)}\ast \iota_{\tau(\omega)}\eta ~~,~~\mathrm{when}~~r\leq s~~,\nn\\
&& \iota_{\omega}(\ast\eta)=(-1)^{r s}\ast(\tau(\omega)\wedge\eta)~~,~~\mathrm{when}~~r+s\leq d~~
\eeqan
and of the following:
\beqan
\label{IdentGeomProd}
(-1)^{[\frac{m+1}{2}]}[\pi^m(\omega)]\btu_m[\ast\tau(\omega)]
 &=& (-1)^{[\frac{m'+1}{2}]}[\ast\tau(\omega)]\btu_{m'}[\pi^{d-1}(\eta)] \nn\\
&=& (-1)^{[\frac{m''+1}{2}]}\ast\tau\left[\pi^{m''}(\omega)\btu_{m''}\eta\right]~,\\
\rm{ for}~~~\tilde\omega-m &=& \tilde\eta-m'=m''~,~~\rm{where}~~m,m',m''>0~. \nn
\eeqan
We also remind the reader that one can uniquely decompose any $k$-form $\omega\in \Omega^k(M)$ into 
parallel and orthogonal parts with respect to any fixed 1-form $\theta\in\Omega^1(M)$ 
such that $\omega=\omega_\perp+\omega_\parallel$, where
 $\omega_\parallel=\theta\wedge\omega_{\top}$ with $\omega_{\top}\in\Omega^{k-1}(M)$,
  $\omega_\perp\in\Omega^k(M)$ and 
  $\theta\bigtriangleup_1\omega_{\top}=\theta\bigtriangleup_1\omega_\perp=0$ (see Subsection \ref{sec:KAparallelism}).
Choosing $\theta=\bar{K}$, one has
 $ Y=  Y_\perp +Y_\parallel$ 
 and $\bar{Z}=\bar{Z}_\perp +\bar{Z}_\parallel$. 
 Note that $\nu=\nu_\parallel$ and $\bar{K}=\bar{K}_\parallel$.

 \
 
 Given the decomposition of $ Y$, its Hodge dual must take the form:
 $\ast Y=\ast (Y_\perp)+\ast (Y_\parallel)=\alpha_1 Y_\parallel+\alpha_2 Y_\perp$ for some
 $\alpha_1,\alpha_2\in C^\infty(M,\R)$, since 
the Hodge dual of any component parallel to $\theta$ is orthogonal to $\theta$ while 
the  Hodge dual of any component orthogonal to $\theta$ is parallel to $\theta$ and since 
there are no other four-form valued pinor bilinears that could appear in the right hand side.  
In the current example (see Table \ref{table:AlgClassif}), 
the volume form $\nu$ is twisted central (i.e., we have 
 $\nu\diamond\omega=\pi(\omega)\diamond\nu$ 
for any inhomogeneous differential form $\omega$) and satisfies $\nu\diamond\nu=+1$. The latter
 property amounts to ${\ast}{\ast} Y= Y$, which 
 leads to $\alpha_2=1/\alpha_1$~and implies:
\ben
\label{Yexp}
\ast Y=\alpha_1 Y_\parallel+\frac{1}{\alpha_1} Y_\perp ~~,~~ \mathrm{with}~~\alpha_1\in C^\infty(M,\R)~.
\een

Expanding \eqref{QConstr} into rank components gives the following conditions, 
which are equivalent with the `useful relations' discussed in 
Appendix C of \cite{MartelliSparks}:
\beqan
\label{QConstrExp}
-\frac{1}{6}\iota_F Y +\iota_{\dd\Delta} \bar{K} -2\kappa~b &=& 0~~,\nn\\
-\frac{1}{6}\iota_F \bar{Z} -\frac{b}{3} f +a~ \dd\Delta &=& 0~~,\nn\\
\dd\Delta\wedge\bar{K}-\frac{1}{6}F\bigtriangleup_3 Y +\frac{1}{3}\iota_f \ast \bar{Z} &=& 0~~,\nn\\
-\frac{1}{3}\iota_f \ast Y +2\kappa~\ast \bar{Z}
   +\frac{1}{6}\iota_{\bar{K}} F+\iota_{\dd\Delta} Y-\frac{1}{6}F\bigtriangleup_3\bar{Z}&=& 0~~,\nn\\
\frac{1}{6}F\bigtriangleup_2 Y-2\kappa~\ast Y +\frac{1}{3}f\wedge\ast \bar Z
   -\frac{b}{6}\ast F+\iota_{\dd\Delta} \bar{Z}-\frac{a}{6}F&=& 0~~,\\
-\frac{1}{6}F\wedge \bar{K}+\dd\Delta\wedge Y+\frac{1}{6}F\bigtriangleup_2 \bar{Z} 
   -\frac{1}{3}f\wedge\ast Y &=& 0~~,\nn\\
\dd\Delta\wedge \bar{Z}+\frac{1}{3} \iota_f\ast\bar{K}+\frac{1}{6}F \bigtriangleup_1 Y &=& 0~~,
\nn\\
-\frac{a}{3}~\ast f+2\kappa~\ast \bar{K}
   +\frac{1}{6}F\bigtriangleup_1 \bar Z+~b~\ast (\dd\Delta) &=& 0~~,\nn\\
2a~\kappa~\nu-\frac{1}{3}f \wedge \ast\bar{K}+\frac{1}{6}Y\wedge F  &=& 0~~.\nn
\eeqan
\paragraph{Remark.} 
It will be useful to Hodge dualize relations \eqref{QConstrExp}, 
which gives:
\beqan
\label{QHodgeDuals}
-\frac{1}{6}F\wedge\ast Y+\dd\Delta\wedge\ast \bar K-2\kappa ~b~\nu &=& 0~~,\nn\\
-\frac{1}{6} F\wedge\ast \bar Z-\frac{b}{3}\ast f+a\ast\dd\Delta &=& 0~~,\nn\\
-\iota_{\dd\Delta}\ast \bar K-\frac{1}{6}\ast(F\btu_3 Y)-\frac{1}{3}f\wedge \bar Z &=& 0~~,\nn\\
~~\frac{1}{3}f\wedge Y-2\kappa~\bar Z-\frac{1}{6}\bar K\wedge\ast F-\dd\Delta\wedge\ast Y
-\frac{1}{6}\ast(F\btu_3 \bar Z) &=& 0~~,\\
~~ \frac{1}{6}\ast(F\btu_2 Y) -2 \kappa Y+\frac{1}{3}\iota_f \bar Z
-\frac{b}{6}F+\dd \Delta\wedge\ast\bar Z-\frac{a}{6}\ast F &=& 0 \nn\\
 -\frac{1}{6}\iota_{\bar K}\ast F+\iota_{\dd \Delta}\ast Y-\frac{1}{6}\ast (\bar F\btu_2 Z) 
- \frac{1}{3}\iota_f Y &=& 0 \nn\\
 -\frac{1}{6}\ast(F\btu_1 Y) -\iota_{\dd \Delta}\ast \bar Z-\frac{1}{3}f\wedge \bar K &=& 0 \nn\\
-\frac{a}{3} f+2\kappa \bar K+\frac{1}{6} \iota_{\ast\bar Z} F +b~\dd\Delta &=& 0 \nn\\
 -\frac{1}{3} \iota_f \bar K +\frac{1}{6}\iota_Y \ast F +2\kappa ~a &=& 0~~, \nn
\eeqan
where we used \eqref{Contractions} and \eqref{IdentGeomProd}.

\

\noindent Similarly, the rank expansion of \eqref{NablaConstr} gives:
\beqan
\label{NablaConstrExp}
&&\partial_m b=2\kappa~ \iota_{(e_m)_\sharp} \bar{K}-\frac{1}{2}\ast[(\iota_{(e_m)_\sharp} F)\wedge\bar{Z}]~~,\\
&&\nabla_m \bar{K}=-2\kappa~b~ {(e_m)_\sharp}-\frac{1}{2}\ast((e_m)_\sharp\wedge f\wedge\bar{Z}) 
+\frac{1}{2}{(\iota_{{(e_m)_\sharp}} F)}\bigtriangleup_3Y~~,\nn\\
&&\nabla_m Y=2\kappa~ {(e_m)_\sharp}\wedge\ast\bar{Z}+\frac{1}{2}({(e_m)_\sharp}\wedge f)\bigtriangleup_1\ast{Y}
    +\frac{1}{2}\bar K\wedge(\iota_{(e_m)_\sharp}F)+\frac{1}{2}(\iota_{(e_m)_\sharp} F)\bigtriangleup_2 \bar{Z}~,\nn\\
&&\nabla_m Z=-2\kappa~ {(e_m)_\sharp}\wedge\ast{Y}+\frac{1}{2}\ast ((e_m)_\sharp\wedge \bar K\wedge f)
   +\frac{1}{2}{(e_m)_\sharp}\wedge f\wedge\ast \bar{Z}-\frac{1}{2}(\iota_{(e_m)_\sharp} F)\bigtriangleup_1 Y \nn\\
  &&~~~~~~~~~ -\frac{~b}{2}~{(e_m)_\sharp}\wedge\ast F~,\nn
\eeqan
which in turn implies the following constraints representing the rank components of \eqref{DiffConstr}:
\beqan
\label{DiffConstrExp}
\dd b  &=&  2\kappa~\bar{K}+\frac{1}{2}\iota_{\ast\bar Z }F,\nn\\
\dd \bar{K} &=& -\frac{1}{2}F\bigtriangleup_3 Y+\iota_f\ast \bar{Z}~~,\nn\\
\dd Y &=& F\bigtriangleup_2\bar{Z}-2 f\wedge\ast Y-2F\wedge \bar{K}~~,\\
\dd \bar{Z} &=& \frac{3}{2}F\bigtriangleup_1Y +3\iota_f\ast \bar{K}~~.\nn
\eeqan
The Hodge duals of \eqref{NablaConstrExp}: 
\beqan
\label{NablaAst}
&&\partial_m b ~\nu=2\kappa ~(e_m)_\sharp\wedge \ast \bar K-\frac{1}{2}(\iota_{(e_m)_\sharp}F)\wedge \bar Z \nn\\
&&\nabla_m \ast \bar K=-2\kappa ~ b \ast (e_m)_\sharp+\frac{1}{2}(e_m)_\sharp\wedge f\wedge \bar Z
-\frac{1}{2}(\iota_{(e_m)_\sharp} F)\wedge\ast Y\\
&&\nabla_m\ast Y =2 \kappa ~\iota_{(e_m)_\sharp}\bar Z+\frac{1}{2}((e_m)_\sharp\wedge f)\btu_1 Y
+\frac{1}{2}\iota_{\bar K}((e_m)_\sharp\wedge \ast F)
-\frac{1}{2}(\iota_{(e_m)_\sharp } F)\btu_1\ast \bar Z \nn\\
&&\nabla_m\ast \bar Z=-2 \kappa~ \iota_{(e_m)_\sharp}Y-\frac{1}{2}(e_m)_\sharp\wedge \bar K\wedge f
+\frac{1}{2}\iota_{((e_m)_\sharp\wedge f)}\bar Z
+\frac{1}{2}(\iota_{(e_m)_\sharp}F)\btu_2\ast Y -\frac{1}{2}b ~\iota_{(e_m)_\sharp}F \nn
\eeqan
lead to another set of differential constraints:
\beqan
\label{DiffAst}
\dd(\ast{\bar K}) &=&-16 \kappa ~b ~\nu-2F\wedge\ast Y, \nn\\
\dd(\ast Y) &=&10\kappa~\bar Z-2f\wedge Y+\frac{1}{2}\bar K\wedge \ast F+\frac{3}{2}\ast(F\btu_3\bar Z) ~~,\nn\\
\dd(\ast{\bar Z}) &=&- 2 ~b ~F-8\kappa~Y+\ast(F \btu_2 Y) +2\iota_f {\bar Z}~~.
\eeqan
Finally, expanding \eqref{FierzConstr} and using the properties of $\nu$, we find:
\beqan
\label{FierzExp}
&& \bar{K}\diamond\bar{K}+\bar{K}\diamond Y+ Y\diamond\bar{K}+\bar{K}\diamond\bar{Z}
+\bar{Z}\diamond\bar{K}+Y\diamond Y+\bar{Z}\diamond\bar{Z}+Y\diamond\bar{Z}+\bar{Z}\diamond Y+2b\ast Y+b^2  \nn\\
&&~~~~~~~~~~~~=15 a^2+14 a\bar{K}+14 aY+14 a\bar{Z}+14ab\nu~~.
\eeqan
We write below the expansions for all those geometric products (see \eqref{starprod}) that 
appear in \eqref{FierzExp}. Let us first list the geometric products involving $\bar{K}$:
\beqan
\label{syst1}
&&\bar{K}\diamond\bar{K}=||\bar{K}||^2~~,\nn\\
&&\bar{K}\diamond~Y=\bar{K}\wedge~Y_\perp+
\bar{K}\bigtriangleup_1~Y_\parallel~~,\nn\\
&&~Y\diamond\bar{K}=\bar{K}\wedge~Y_\perp-
\bar{K}\bigtriangleup_1~Y_\parallel~~,\\
&&\bar{K}\diamond\bar{Z}=\bar{K}\wedge\bar{Z}_\perp+
\bar{K}\bigtriangleup_1\bar{Z}_\parallel
=\bar{K}\bigtriangleup_1\bar{Z}~~,\nn\\
&&\bar{Z}\diamond\bar{K}=-\bar{K}\wedge\bar{Z}_\perp+
\bar{K}\bigtriangleup_1\bar{Z}_\parallel
=\bar{K}\bigtriangleup_1\bar{Z}~~.\nn
\eeqan
Notice that $\bar{K}\wedge\bar{Z}_\perp=0$, since there is no 
nontrivial six-form pinor bilinear that can be constructed in this case. Thus, 
we must have $\bar{Z}_\perp=0$, which means $\bar{Z}=\bar{Z}_\parallel$, a
relation which we use when performing the following expansions:
\beqan
\label{syst2}
&&~Y\diamond~Y=2~Y_\parallel\wedge~Y_\perp-
~Y_\parallel\bigtriangleup_2~Y_\parallel - 2~Y_\parallel\bigtriangleup_2~Y_\perp - ~Y_\perp\bigtriangleup_2~Y_\perp+||~Y_\parallel||^2 + ||~Y_\perp||^2~~,\nn\\
&&~Y\diamond\bar{Z}=
 - ~Y_\parallel\bigtriangleup_1\bar{Z}_\parallel
 - ~Y_\perp\bigtriangleup_2\bar{Z}_\parallel +~Y_\parallel\bigtriangleup_3\bar{Z}_\parallel
+~Y_\perp\bigtriangleup_4\bar{Z}_\parallel~~,\nn\\
&&\bar{Z}\diamond~Y=
  ~Y_\parallel\bigtriangleup_1\bar{Z}_\parallel
 - ~Y_\perp\bigtriangleup_2\bar{Z}_\parallel -~Y_\parallel\bigtriangleup_3\bar{Z}_\parallel
+~Y_\perp\bigtriangleup_4\bar{Z}_\parallel~~,\\
&&\bar{Z}\diamond\bar{Z}=
  \bar{Z}_\parallel\bigtriangleup_1\bar{Z}_\parallel
 - \bar{Z}_\parallel\bigtriangleup_3\bar{Z}_\parallel +||\bar{Z}||^2~~.\nn
\eeqan
We mention here that expansions \eqref{QConstrExp}-\eqref{syst2} were obtained 
using a package of procedures which we created using 
 {\tt Ricci} \cite{Ricci}. We have also verified these relations (in tensorial form) using {\tt
Cadabra} \cite{Cadabra}.

We deduce from the Fierz identities in normal cases that any generalized product of form-valued pinor bilinears can be expressed as a form-valued pinor bilinear. 
In \eqref{syst2}, we omitted to write down those terms which are tautologically zero due to the 
graded antisymmetry of the wedge product.  We also omitted writing some other vanishing terms such as $~Y_\perp\wedge~Y_\perp$,
which has rank 8 and thus must be proportional to the volume form, but vanishes since the volume form does not have an orthogonal component. 
Similarly, $~Y_\parallel\bigtriangleup_1\bar{Z}_\parallel=0$ since it should be a parallel 7-form and thus proportional to
  $(\ast\bar{K})_\parallel=\ast(\bar{K}_\perp)$, but
  $\bar{K}_\perp=0$. Furthermore, $~Y_\perp\bigtriangleup_1\bar{Z}_\parallel=0$  since 
  it is an orthogonal 5-form, thus proportional to $\bar{Z}_\perp=0$. For similar reasons,  
$~Y_\perp\bigtriangleup_3\bar{Z}_\parallel=0$ and $~Y_\parallel\bigtriangleup_4\bar{Z}_\parallel=0$. 

Given expansions \eqref{syst1} and \eqref{syst2}, identity \eqref{FierzExp} gives the following relations 
when separated into rank components:
\beqan
||\bar{K}||^2+||Y_\perp||^2+||Y_\parallel||^2+||\bar{Z}||^2+~b^2 &=& 15 ~a^2~~,\nn\\
Y_\perp\bigtriangleup_4\bar{Z} &=& 7~a~\bar{K}~~,\nn\\
2\bar{K}\bigtriangleup_1 \bar{Z}-Y_\parallel\bigtriangleup_2Y_\parallel
  -2Y_\parallel\bigtriangleup_2Y_\perp-Y_\perp\bigtriangleup_2Y_\perp
  -\bar{Z}\bigtriangleup_3\bar{Z}+2~b\ast Y &=& 14 ~a~Y~~,\label{FierzConstrExp}\\
\bar{K}\wedge Y_\perp-Y_\perp\bigtriangleup_2\bar{Z} &=& 7 ~a~\bar{Z}~~,\nn\\
2Y_\parallel\wedge Y_\perp+\bar{Z}\bigtriangleup_1\bar{Z} &=& 14 ~a~b~\nu~~. \nn
\eeqan
In order to solve the system \eqref{FierzConstrExp}, 
 we introduce the notations:
\beqan
\label{notations}
&&\bar{K}\wedge Y_\perp=e \bar{Z}~~,
~~\bar{K}\bigtriangleup_1 Y_\parallel=f \ast \bar{Z}~~,
~~\bar{K}\bigtriangleup_1\bar{Z}=g  Y_\perp~~,\nn\\
&& Y_\parallel\wedge Y_\perp=h\nu~~,~~
~Y_\parallel\bigtriangleup_2 Y_\parallel=y_1  Y_\perp
~~, ~~Y_\parallel\bigtriangleup_2 Y_\perp=y_2 ~Y_\parallel~~,~
~Y_\perp\bigtriangleup_2 Y_\perp=y_3  Y_\perp~,\nn\\
 && Y_\parallel\bigtriangleup_1\bar{Z}=n \ast\bar{K}~~,~~ Y_\perp\bigtriangleup_2\bar{Z}= r \bar{Z}~~,~~
 Y_\parallel\bigtriangleup_3\bar{Z}=s \ast\bar{Z}~~,\\
&& Y_\perp\bigtriangleup_4\bar{Z}=t \bar{K}~~,~~
 \bar{Z}\bigtriangleup_1\bar{Z}=u\nu~~,~~
 \bar{Z}\bigtriangleup_3\bar{Z} =x  Y_\perp~~,\nn
\eeqan
where $e,~f,~g,~h,~y_1,~y_2,~y_3,~n,~r,~s,~t,~u,~x$ are smooth functions on $M$ 
which we want to determine in terms of $a$ and $b$.

Using \eqref{notations}, the system \eqref{FierzConstrExp} leads to four 
independent relations for ranks $0,1,5,8$ and two independent 
identities for the parallel and perpendicular components of rank $4$:
\beqan
\label{Fierz1}
&&b^2+||\bar{K}||^2+||~Y_\parallel||^2+||~Y_\perp||^2+||\bar{Z}||^2=15a^2~~,\nn\\
&& t=7a~~~, ~~~e-r=7a~~~,~~~2h+u=14ab~~,\\
&&b\alpha_1-y_2=7a~~~,~~~2b/\alpha_1+2g-(y_1+y_3)-x=14a~~.\nn
\eeqan
Using associativity of the geometric product: 
\beqa
&&(\bar{K}\diamond\bar{K})\diamond~Y=
\bar{K}\diamond(\bar{K}\diamond~Y)~~~,
~~~(\bar{K}\diamond~Y)\diamond~Y=
\bar{K}\diamond(~Y\diamond~Y)~~,\\
&&(\bar{K}\diamond\bar{Z})\diamond\bar{Z}=
\bar{K}\diamond(\bar{Z}\diamond\bar{Z})~~~,
~~~(\bar{K}\diamond~Y)\diamond\bar{Z}=
\bar{K}\diamond(~Y\diamond\bar{Z})~~,\\
&&(~Y\diamond~Y)\diamond\bar{Z}=
~Y\diamond(~Y\diamond\bar{Z})~~~,
~~~(\bar{K}\diamond\bar{Z})\diamond~Y=
\bar{K}\diamond(\bar{Z}\diamond~Y)~~
\eeqa
as well as \eqref{syst1}, \eqref{syst2} and \eqref{notations}, we find the relations:
\beqan
\label{Fierz2}
&&||\bar{K}||^2=eg~~,~~e=f\alpha_1~~,~~en+ft=2h~~,~~er+fs=e(y_1+y_3)~~,~~es+rf=2y_2f~ ,\nn\\
&&||~Y_\parallel||^2+||~Y_\perp||^2=et+fn~~,~~||\bar{Z}||^2=tg~~,
~~u=0~,~~gr=xe~~,~fx=sg~,\nn\\
&& f||\bar{Z}||^2=n||\bar{K}||^2~~,
~~e||\bar{Z}||^2=t||\bar{K}||^2~~,~~2h+2sy_2=ne+2rs+ft~~,\nn\\
&& ||~Y_\parallel||^2+||~Y_\perp||^2+ r(y_1+y_3)=nf+r^2+s^2+te~~~,
~~~t(y_1+y_3)=rt+ns~~,\\
&& n||\bar{K}||^2=gh~~,
~~g||~Y_\perp||^2=t||\bar{K}||^2~~,~~ 2y_2n=nr+ts~~,~~y_3=r~~,~~y_2=s\alpha_1~~,\nn
\eeqan
which need not be independent. We thus find that the Fierz identities are 
equivalent to the algebraic system of equations \eqref{Fierz1}-\eqref{Fierz2}, which can be solved using {\tt Mathematica}\textsuperscript{\textregistered} and give:
\beqa
&& e=a~~,~~f=b~~,~~g=\frac{a^2-b^2}{a}~~,~~h=7ab~~,
~~y_1=-\frac{6b^2}{a}~~,~~y_2=-6a~~,~~y_3=-6a~~,\\
&& n=7b~~,~~r=-6a~~,~~s=-6b~~,~~t=7a~~,~~u=0~~,
~~x=-\frac{6(a^2-b^2)}{a}~~,~~\alpha_1=\frac{a}{b}~~,\\
&& ||\bar{K}||^2=a^2-b^2~~,~~||~Y_\parallel||^2=7b^2~~,~~
||~Y_\perp||^2=7a^2~~,~~||\bar{Z}||^2=7(a^2-b^2)~~.
\eeqa
Substituting this solution into \eqref{notations} gives:
\beqan
\label{FierzSol}
&&\bar{Z}=\frac{1}{a}\bar{K}\wedge Y_\perp~~~,
~~~\ast\bar{Z}=\frac{1}{b}\bar{K}\bigtriangleup_1 Y_\parallel~~~,
~~~ \ast\bar{K}=\frac{1}{7b} Y_\parallel\bigtriangleup_1\bar{Z}~~,\nn\\
&&  Y_\parallel\wedge Y_\perp=7ab\nu~~,~~
 Y_\parallel\bigtriangleup_2 Y_\parallel=-\frac{6b^2}{a} Y_\perp
~~, ~~ Y_\parallel\bigtriangleup_2 Y_\perp=-6a  Y_\parallel~~,\nn\\
&& Y_\perp\bigtriangleup_2 Y_\perp=-6a ~Y_\perp~~~,
~~~\ast Y=\frac{a}{b} Y_\parallel+\frac{b}{a} Y_\perp~~,\\
&& ||\bar{K}||^2=a^2-b^2~~,~~|| Y_\parallel||^2=7b^2~~,~~
|| Y_\perp||^2=7a^2~~,~~||\bar{Z}||^2=7(a^2-b^2)~~,\nn\\
&& Y= Y_\parallel+ Y_\perp=\frac{b}{||\bar{K}||^2}\bar{K}\wedge\ast\bar{Z}+
  \frac{a}{||\bar{K}||^2}~\iota_{\bar{K}} \bar{Z}~~,~~
  \bar Z\btu_3 \bar Z=-6\iota_{\bar K} \bar Z~.\nn
\eeqan
Taking (as in \cite{MartelliSparks}) $a=1$ and $b=\sin\zeta$ 
(since $1-b^2>0$ for the norms to be positive) we find agreement 
with the results in loc. cit. (up to a sign issue which is discussed in \cite{g2}), as for example:
\beqan
&&||K||^2=1~~,~~\mathrm{where}~~K\eqdef (\cos\zeta)^{-1}\bar{K}~~,\label{Results1}\\
&&|| \ast Z||^2=7 ~~,~~\mathrm{where}~~Z\eqdef (\cos\zeta)^{-1}\bar{Z},\label{Results2}\\
&&\iota_K(\ast Z)=0~~,\label{Results3}\\
&&Y=\iota_K Z-(\ast Z)\wedge K\sin\zeta~~,\label{Results4}\\
&&\dd(e^{3\Delta}\bar K)=0~~,\label{Results5}\\
&&K\wedge \dd(e^{6\Delta}\iota_K Z)=0~~,\label{Results6}\\
&&e^{-6\Delta}\dd(e^{6\Delta}\ast\bar Z)=\ast F-F\sin\zeta+4\kappa Y~~,\label{Results7}\\
&&e^{-3\Delta}\dd(e^{3\Delta}\sin\zeta) = f-4\kappa~\bar{K}~~,\label{Results8}\\
&& e^{-12\Delta}\dd (e^{12\Delta}\rm{vol}_7\cos\zeta)=
-8\kappa~\rm{vol}_7\wedge K\sin\zeta~~,\label{Results9}
\eeqan
with $\rm{vol}_7$ defined as in \cite{MartelliSparks}:
\ben
\label{vol7}
\rm{vol}_7
=\frac{1}{7}\phi\wedge\iota_K\ast\phi=-\frac{1}{7}(\ast Z)\wedge\iota_K Z=-\iota_K\nu=-\ast K~.
\een
It is easy to see that the first four of these conditions 
follow directly from the Fierz identities
\eqref{FierzSol}, while the last four can be obtained using 
the algebraic constraints \eqref{QConstrExp} 
and the differential 
constraints \eqref{DiffConstrExp}. 

\paragraph{Proof of  relations \eqref{Results1} and \eqref{Results2}.} From \eqref{FierzSol},
 we have:
\be
||\bar{K}||^2=a^2-b^2~~~\rm{and}~~~||\bar{Z}||^2=7(a^2-b^2)~~.
\ee
For $a=1$, ~$b=\sin\zeta$, ~$K= (\cos\zeta)^{-1}\bar{K}$ and ~
$Z=(\cos\zeta)^{-1}\bar{Z}$, this result becomes:
\be
||K||^2=1~~~\rm{and}~~~||\ast Z||^2=||Z||^2=7~~.
\ee
\paragraph{Proof of  relation \eqref{Results3}.} The second relation in
 \eqref{Contractions} gives $\iota_K(\ast Z)=-\ast(K
\wedge Z)$, where $K \wedge Z=\frac{1}{\cos^2\zeta}\bar{K}\wedge\bar{Z}$ must be a
6-form bilinear in $\xi$ since
$\bar{K}$ and $\bar{Z}$ are bilinears in $\xi$ and since the Fierz identities allow
us to reduce $\bar{K}\wedge\bar{Z}$ to a pinor bilinear. But there is no 
nontrivial 6-form pinor bilinear which
can be constructed in our case, and thus we
must have $\bar{K} \wedge\bar{Z}=0$. This proves relation
\eqref{Results3}.

\paragraph{Proof of relation \eqref{Results4}.} Starting from the expression  
$Y=\frac{b}{||\bar{K}||^2}\bar{K}\wedge\ast\bar{Z}+
  \frac{a}{||\bar{K}||^2}~\iota_{\bar{K}} \bar{Z}$  found in  \eqref{FierzSol},
 taking the normalization used in \cite{MartelliSparks,Tsimpis} 
  and the notations in \eqref{Results1}, \eqref{Results2}, one finds:
  \be
  Y=\frac{1}{\cos^2\zeta}(\sin\zeta~\bar{K}\wedge\ast\bar{Z}+
  \iota_{\bar{K}} \bar{Z})=\iota_K Z-(\ast Z)\wedge K\sin\zeta~~,
  \ee
  thus proving relation \eqref{Results4}.

\paragraph{Proof of relation \eqref{Results5}.}
The third algebraic constraint listed in \eqref{QConstrExp}:
\be
\dd\Delta\wedge\bar{K}-\frac{1}{6}F\bigtriangleup_3 Y +\frac{1}{3}\iota_f \ast \bar{Z} = 0
~\Longleftrightarrow~3~\dd\Delta\wedge\bar{K} = \frac{1}{2}F\bigtriangleup_3 Y
-\iota_f\ast\bar{Z}~~\\
\ee
and the second differential constraint listed in \eqref{DiffConstrExp}:
\be
\dd \bar{K} =-\frac{1}{2}F\bigtriangleup_3 Y+\iota_f\ast \bar{Z}~~
\ee
imply the relation:
\ben
\label{dK}
3~\dd\Delta\wedge\bar{K}+\dd\bar{K}=0\Longleftrightarrow \dd {\bar K}=3{\bar K}\wedge \dd \Delta~~,
\een
which is easily seen to be equivalent with \eqref{Results5}. 

\paragraph{Proof of relation \eqref{Results6}.}
Using \eqref{Results4}, one can write \eqref{Results6} in the equivalent form:
\be
K\wedge[6~\dd\Delta\wedge Y +\dd Y - \sin\zeta (\ast Z)\wedge\dd K]=0~~,
\ee
where we used $K\wedge K=0$. Relations \eqref{dK} and $K\wedge {\bar K}=0$ imply $K\wedge (\sin\zeta (\ast Z)\wedge\dd K)=0$, so 
\eqref{Results6} reduces to:
\ben
\label{Results6reduced}
K\wedge(6~ \dd\Delta\wedge Y+\dd Y)=0~~.
\een
To prove \eqref{Results6reduced}, notice that adding the sixth algebraic constraint listed in \eqref{QConstrExp}: 
\be
6~\dd\Delta\wedge Y = F\wedge \bar{K}-F\bigtriangleup_2 \bar{Z} 
   +2f\wedge\ast Y ~~
\ee
to the third differential constraint listed in \eqref{DiffConstrExp}:
\be
\dd Y =-2F\wedge \bar{K}+F\bigtriangleup_2\bar{Z}-2f\wedge\ast Y 
~~,\nn
\ee
gives:
\be
6~ \dd\Delta\wedge Y+\dd Y=-F\wedge \bar{K} ~~,
\ee
which in turn implies \eqref{Results6reduced} upon using the identity $K\wedge\bar{K}=0$.

\paragraph{Proof of relation \eqref{Results7}.}
Relation \eqref{Results7} is equivalent with:
\ben
\label{Results7red}
6~\dd\Delta\wedge(\ast\bar Z)+\dd(\ast {\bar Z})=\ast F-F\sin\zeta+4\kappa~Y~~
\een
To prove \eqref{Results7red}, notice that the fifth relation in \eqref{QHodgeDuals}, which is the 
Hodge dual of the fifth algebraic relation listed in \eqref{QConstrExp},
when multiplied by $6$ gives:
\ben
\label{Rel6int1}
6~\dd\Delta\wedge\ast \bar{Z}= \ast F+F\sin\zeta+12\kappa~ Y -\ast(F\bigtriangleup_2 Y)-2\iota_f\bar{Z}~~.
\een
On the other hand, the fourth differential relation listed in \eqref{DiffAst} gives:
\ben
\label{Rel6int2}
\dd(\ast\bar{Z}) =- 2 F\sin \zeta -8 \kappa~Y+\ast(F\bigtriangleup_2 Y)+2\iota_f\bar{Z}~~.
\een
Adding \eqref{Rel6int1} and \eqref{Rel6int2} gives \eqref{Results7red}. This finishes the proof of \eqref{Results7}. 

\paragraph{Proof of relation \eqref{Results8}.}
Relation \eqref{Results8} is obviously equivalent with:
\ben
\label{Results8int}
3(\sin \zeta) \dd\Delta+\dd\sin\zeta=f-4\kappa~\bar{K}~~.
\een
To prove \eqref{Results8int}, notice that the Hodge dual of the eighth algebraic 
constraint listed in \eqref{QConstrExp} (i.e. the eight relation in \eqref{QHodgeDuals}),
 after multiplication by $3$, 
gives:
\ben
\label{Results8int1}
3(\sin \zeta) \dd\Delta=f-6\kappa~\bar{K} - \frac{1}{2}\iota_{\ast\bar Z} F~~.
\een
On the other hand, the first differential constraint listed in \eqref{DiffConstrExp} gives:
\ben
\label{Results8int2}
\dd\sin\zeta  =  2\kappa~\bar{K}+\frac{1}{2}\iota_{\ast\bar Z} F~~,
\een
Adding \eqref{Results8int1} and \eqref{Results8int2} gives \eqref{Results8int}, 
which finishes the proof of relation \eqref{Results8}.

\paragraph{Proof of relation \eqref{Results9}.}
Using the first relation in \eqref{DiffAst} and first relation in \eqref{QHodgeDuals} 
one finds 
\be
12 \dd\Delta\wedge\ast \bar{K}+\dd\ast \bar{K}=8\kappa \sin\zeta\nu~~,
\ee
which, when expressing the volume form $\nu$ in terms of $\rm{vol}_7$ defined in
 \eqref{vol7}, leads to \eqref{Results9}.

\section{Conclusions and further directions}
\label{sec:conclusions}

We showed that geometric algebra techniques can be used to give a
highly synthetic, conceptually transparent and computationally
efficient reformulation of the constrained Killing pinor equations,
which constitute the condition that a flux background preserves a
given amount of supersymmetry. This formulation clearly displays the
algebraic and differential structure governing the supersymmetry
conditions, leading to a description which opens the way for unified
studies of flux backgrounds aimed at uncovering their deeper
structure. We showed that our general formalism recovers results and methods which were used in
\cite{MartelliSparks} and therefore that it provides a powerful way to
extend them. Our formulation is highly amenable to implementation in
various symbolic computational packages specialized in tensor algebra,
and we touched on two particular implementations which we have carried
out using {\tt Ricci} \cite{Ricci} in {\tt
Mathematica}\textsuperscript{\textregistered}, as well as {\tt
Cadabra} \cite{Cadabra}.

Here we illustrated our approach with the case of the most general
compactifications of M-theory which preserve ${\cal N}=1$ supersymmetry in three
dimensions \cite{MartelliSparks,Tsimpis}, showing how the results derived
through different methods in loc. cit. can be recovered through our
techniques.  We stress that the methods introduced in this paper have much
wider applicability than the example considered in Section \ref{sec:application},
leading to promising new directions in the study of supergravity backgrounds
and supergravity actions. In particular, we believe that many computationally
difficult issues in the subject could be understood much better by using such
techniques. The connection with a certain form of geometric quantization
(which we have only touched upon) also leads to interesting ideas, problems
and directions for further research, which are currently under investigation.
Further applications of our approach can be found in \cite{g2,g2s,ga2,gf}.

\

\appendix

\section{Identities satisfied by the covariant derivative of pinors}
\label{sec:identities}

Let $(M,g)$ be a pseudo-Riemannian manifold endowed with a local
coordinate system $(x^m)$ and a local pseudo-orthonormal frame $(e_a)$
(vielbein) of $(TM,g)$, both defined above an open set $U\subset
M$. In this Appendix, both $m$ and $a$ run from $1$ to $\dim M$.  As usual, pseudo-orthonormality of $e_a$ means
$g(e_a,e_b)=\eta_{ab}$, where $\eta_{ab}$ is a diagonal matrix all of
whose diagonal entries equal $+1$ or $-1$.  We let $e^a$ denote the
dual coframe of $M$, defined through $e^a(e_b)=\delta^a_b$; it is a
local frame of $T^\ast M$ which is pseudo-orthonormal with respect to the
metric ${\hat g}$ induced on $T^{\ast} M$, i.e. ${\hat
g}(e^a,e^b)=\eta^{ab}$ where $\eta^{ab}\eta_{bc}=\delta^a_c$. We have
$\partial_m\eqdef \frac{\partial}{\partial x^m}=e^a_m(x) e_a$ and
$e_a=e^m_a(x)\partial_m$ for some locally defined functions $e^a_m$,
$e_a^m$ which satisfy $e^a_m e^m_b=\delta^a_b$ and
$e_m^ae_a^n=\delta^n_m$. This implies $g(e_a,e_b)=e_a^m e_b^n
g_{mn}=\eta_{ab}$ and $g(\partial_m,\partial_n)=e_m^a
e_n^b\eta_{ab}=g_{m n}$. Any tensor field $t\in \Gamma(TM^{\otimes
p}\otimes (T^{\ast}M)^{\otimes q})$ of type $(p,q)$ expands as:
\ben
\label{tensor}
t=_Ut^{m_1\ldots m_p}_{n_1\ldots n_q} \partial_{m_1}\otimes \ldots \otimes \partial_{m_p}\otimes
\dd x^{n_1}\otimes \ldots \otimes \dd x^{n_q}=
t^{a_1\ldots a_p}_{b_1\ldots b_q} e_{a_1}\otimes \ldots \otimes e_{a_p}\otimes
e^{b_1}\otimes \ldots \otimes e^{b_q}~~,
\een
where the locally-defined coefficient functions $t^{m_1\ldots m_p}_{n_1\ldots n_q}$ and $t^{a_1\ldots a_p}_{b_1\ldots b_q}$
are related through:
\be
t^{a_1\ldots a_p}_{b_1\ldots b_q}=e^{a_1}_{m_1}\ldots e^{a_p}_{m_p}e^{n_1}_{b_1}\ldots
e^{n_q}_{b_q} t^{m_1\ldots m_p}_{n_1\ldots n_q}\Longleftrightarrow
t^{m_1\ldots m_p}_{n_1\ldots n_q}=e_{a_1}^{m_1}\ldots e_{a_p}^{m_p}e_{n_1}^{b_1}\ldots
e_{n_q}^{b_q} t^{a_1\ldots a_p}_{b_1\ldots b_q}~~.
\ee
Here and below, indices denoted by letters chosen from the middle of the Latin alphabet
refer to the coordinate frame defined by
$(\partial_m)$ while indices denoted by letters chosen from the
beginning of the Latin alphabet refer to the local pseudo-orthonormal frame (vielbein) defined by $(e_a)$.

A differential $k$-form $\omega\in \Omega^k(M)$ expands locally as in
\eqref{FormExpansion}. Similarly, a polyvector field $\alpha\in
\Gamma(M, \wedge^k T M)$ expands locally as:
\ben
\label{beta}
\alpha=_U\frac{1}{k!}\alpha^{a_1\ldots a_k}(x)e_{a_1}\wedge \ldots \wedge e_{a_k}~~,
\een
(with coefficients functions which are totally antisymmetric in the indices).

Let $\nabla$ be the Levi-Civita connection of $(M,g)$. Its Christoffel
symbols $\Gamma^\rho_{mn}$ in the given local coordinates are
defined through
$\nabla_m(\partial_n)=\Gamma_{mn}^\rho\partial_\rho$, while
its coefficients $\w_{m a}^b$ with respect to the given
coordinate system and vielbein are determined by the expansion
$\nabla_m(e_a)=\w_{m a}^be_b$. Here and below, we set $\nabla_m\eqdef \nabla_{\partial_m}$. The two sets of
connection coefficients are related through:
\be
\w_{m b}^a=e^a_n e^\lambda_b\Gamma^n_{m\lambda}-e^\lambda_b \partial_m e^a_\lambda \Longleftrightarrow
\Gamma^n_{m\lambda}=e^n_a e^b_\lambda \w_{m b}^a + e^n_a\partial_m e^a_\lambda ~~.
\ee
The fact that $\nabla$ is torsion-free amounts to the conditions:
\ben
\label{omega_asym}
\Gamma^\rho_{mn}=\Gamma_{n m}^\rho\Longleftrightarrow \w_{m ab}=-\w_{m ba}~~,
\een
where $\w_{m a b}$ is defined through:
\be
\w_{m a b}\eqdef \eta_{ac} \w_{m b}^c=g(e_a,\nabla_m e_b)=-g(\nabla_m e_a, e_b)~~.
\ee
With respect to the vielbein, the covariant derivative of a $(p,q)$-tensor \eqref{tensor} takes the form:
\be
\nabla_m t=(\nabla_m t)^{a_1\ldots a_p}_{b_1\ldots b_q}  e_{a_1}\otimes \ldots \otimes e_{a_p}\otimes
e^{b_1}\otimes \ldots \otimes e^{b_q}~~,
\ee
where:
\be
(\nabla_m t)^{a_1\ldots a_p}_{b_1\ldots b_q}\eqdef t^{a_1\ldots a_p}_{b_1\ldots b_q;m}=\partial_m
t^{a_1\ldots a_p}_{b_1\ldots b_q}
+\sum_{s=1}^p{\w_{m a}^{a_s} t^{a_1\ldots a_{s-1}, a, a_{s+1}\ldots  a_p}_{b_1\ldots b_q} }
-\sum_{t=1}^q{\w_{m b_t}^a t^{a_1\ldots a_p}_{b_1\ldots b_{t-1}, a, b_{t+1}\ldots b_q}}~~.
\ee
For a differential form \eqref{FormExpansion}, this gives
$\nabla_m \omega=(\nabla_m \omega)_{b_1\ldots b_k}e^{b_1\ldots b_k}$, with:
\ben
\label{alpha_cov}
(\nabla_m \omega)_{b_1\ldots b_k}= \omega_{b_1\ldots b_k; m}=\partial_m
\omega_{b_1\ldots b_k}-\sum_{s=1}^k{\w_{m b_s}^a \omega_{b_1\ldots b_{t-1}, a, b_{t+1}\ldots b_k}}~~,
\een
while for a polyvector field \eqref{beta}, we find
$\nabla_m \alpha=(\nabla_m \alpha)^{a_1\ldots a_k}e_{a_1}\wedge \ldots \wedge e_{a_k}$, with:
\ben
\label{beta_cov}
(\nabla_m \alpha)^{a_1\ldots a_k}\eqdef \alpha^{a_1\ldots a_p}_{;m}=\partial_m
\alpha^{a_1\ldots a_k}+\sum_{s=1}^k{\w_{m a}^{a_s} \alpha^{a_1\ldots a_{s-1}, a, a_{s+1}\ldots  a_k}}~~.
\een
Let $S$ be a pin bundle over $M$. Recall that the connection $\nabla^S$ induced by the Levi-Civita connection $\nabla$ takes the form:
\be
\nabla^S_m=\partial_m +\wf_m~~~{\rm where}~~~\wf_m=
\frac{1}{4}\w_{m a b}\gamma^{a b}=\frac{1}{4}g(e_a,\nabla_m e_b)\gamma^{a b}\in \Gamma(M,\End(S))~~;
\ee
this acts on sections of $S$ in the obvious manner. Here, $\gamma^a\in \Gamma(M,\End(S))$ are the gamma operators associated
with the coframe $(e^a)$, which satisfy:
\be
[\gamma^a,\gamma^b]_{+,\circ}=2\eta^{ a b}~~.
\ee
We will also use the operators $\gamma_a\eqdef \eta_{ a b}\gamma^b$, which satisfy 
$[\gamma_a,\gamma_b]_{+,\circ}=2\eta_{a b }$. In what follows,
we recall some basic properties of $\nabla^S$.

\paragraph{Algebraic identities.} Let $V$ be a finite-dimensional $\K$-vector space. For $X,Y\in \End(V)$, we set
$[X,Y]_\epsilon\eqdef XY+\epsilon YX$, where $\epsilon\in \{-1,+1\}$, so that $[X,Y]_{+1}\eqdef [X,Y]_{+,\circ}$ is the
anticommutator of $X$ with $Y$, while $[X,Y]_{-1}\eqdef [X,Y]_{-,\circ}$ is the
commutator. We start with the following trivial observation:

\paragraph{Lemma 1.} For any $A,B,C\in \Mat(n,\K)$ and any $\epsilon\in \{-1,+1\}$, we have:
\be
[AB,C]_\epsilon=A[B,C]_\epsilon-\epsilon [A,C]_\epsilon B~~.
\ee

\paragraph{Proposition 1.} The following identities hold for all $p\neq q$:
\beqan
\left[ \gamma^p\circ \gamma^q, \gamma^{a_1\ldots a_k} \right]_{-,\circ}&=&2\sum_{s=1}^k{\eta^{q a_s}\gamma^{a_1\ldots a_{s-1}~ p~ a_{s+1} \ldots a_k}}
-(p\leftrightarrow q) ~~, \label{prop1rel1}\\
\left[ \gamma^p\circ \gamma^q, \gamma_{a_1\ldots a_k}\right]_{-,\circ}&=&2\sum_{s=1}^k{\delta^q_{a_s}\eta^{pp'}
\gamma_{a_1\ldots a_{s-1}~ p'~ a_{s+1} \ldots a_k}}-
(p\leftrightarrow q)~~.\label{prop1rel2}
\eeqan
{\bf Proof.} We prove only the first identity, since it immediately implies the second upon lowering indices with $\eta$.
Applying the lemma with $\epsilon=(-1)^{k-1}$, we find:
\ben
\label{c1}
[\gamma^p\circ \gamma^q,\gamma^{a_1\ldots a_k}]_{-,\circ}=
\gamma^p\circ [\gamma^q,\gamma^{a_1\ldots a_k}]_{(-1)^{k-1},\circ }+
(-1)^k[\gamma^p, \gamma^{a_1\ldots a_k}]_{(-1)^{k-1},\circ }\circ \gamma^q~~.
\een
A simple computation (or a mathematical induction argument) gives:
\be
[\gamma^p,\gamma^{a_1\ldots a_k}]_{(-1)^{k-1},\circ }=2\sum_{s=1}^k (-1)^{s-1} \eta^{p a_s}\gamma^{a_1\ldots \hat{a_s}\ldots a_k}~~,
\ee
where the hat indicates that the corresponding index is missing. Using this equation and its counterpart with $p$ replaced by $q$
in \eqref{c1} gives:
\ben
\label{c2}
[\gamma^p\circ \gamma^q,\gamma^{a_1\ldots a_k}]_{-,\circ}=2\sum_{s=1}^k (-1)^{s-1} \eta^{q a_s}\gamma^p\circ \gamma^{a_1\ldots \hat{a}_s\ldots a_k}
+2\sum_{s=1}^k (-1)^{s-1} \eta^{p a_s}\gamma^{a_1\ldots \hat{a}_s\ldots a_k}\circ \gamma^q~~.
\een
We next use the following identities, which can be checked by mathematical induction \footnote{ Notice that there
is no Einstein summation over $p$ or $q$.}:
\beqa
\gamma^p\circ \gamma^{a_1\ldots \hat{a}_s\ldots a_k} &=& (-1)^{s-1}\gamma^{a_1\ldots a_{s-1}~p~a_{s+1}\ldots a_k}+
2\sum_{t=1}^{s-1}(-1)^{t-1} \eta^{pa_t} \gamma^p\circ \gamma^{a_1\ldots \hat{a}_t\ldots \hat{a}_s\ldots a_k} ~~,\\
\gamma^{a_1\ldots \hat{a}_s\ldots a_k}\circ \gamma^q &=& (-1)^{k-s}\gamma^{a_1\ldots a_{s-1}~q~a_{s+1}\ldots a_k}+
2\sum_{t=s+1}^k(-1)^{k-t} \eta^{qa_t} \gamma^q\circ \gamma^{a_1\ldots \hat{a}_s\ldots \hat{a}_t\ldots a_k}~~.
\eeqa
Inserting these in \eqref{c2} gives:
\be
[\gamma^p\circ \gamma^q,\gamma^{a_1\ldots a_k}]_{-,\circ}=2\sum_{s=1}^k{\eta^{q a_s}\gamma^{a_1\ldots a_{s-1}~ p~ a_{s+1} \ldots a_k}}
-(p\leftrightarrow q) +T~~,
\ee
where the term $T$ is given by:
\be
T=4(T_1-T_2) ~~,
\ee
with $T_1=\sum_{1\leq t < s\leq k} (-1)^{s+t}\eta^{pa_t}\eta^{qa_s}\gamma^{a_1\ldots \hat{a}_t\ldots \hat{a}_s\ldots a_k}$ and
$T_2=\sum_{1\leq s < t\leq k} (-1)^{s+t}\eta^{qa_t}\eta^{pa_s}\gamma^{a_1\ldots \hat{a}_s\ldots \hat{a}_t\ldots a_k}$. Since
$T_2=T_1|_{p\leftrightarrow q}$, we have $T=0$ and the first relation \eqref{prop1rel1} is proved.

Proposition 1 has the following immediate consequence, which follows by using the antisymmetry property \eqref{omega_asym}
of $\w_{m pq}$:

\paragraph{Proposition 2.} The following identities hold:
\beqan
\left[\nabla^S_m, \gamma^{a_1\ldots a_k}\right]_{-,\circ}= \frac{1}{4}\w_{m b c }[\gamma^{b c}, \gamma^{a_1\ldots a_k}]_{-,\circ} &=&
-\sum_{s=1}^k\w_{m}{}^{a_s}{}_{p}~\gamma^{a_1\ldots a_{s-1} ~p~ a_{s+1}\ldots
a_k}\label{omega_comm1} ~~,\\
\left[\nabla^S_m, \gamma_{a_1\ldots a_k}\right]_{-,\circ}=\frac{1}{4}\w_{m b c }[\gamma^{b c}, \gamma_{a_1\ldots a_k}]_{-,\circ} &=&
+\sum_{s=1}^k\w_{m}{}^{p}{}_{a_s}~\gamma_{a_1\ldots a_{s-1} ~p~ a_{s+1} \ldots a_k}\label{omega_comm2} ~~.
\eeqan
Consider an arbitrary $k$-form $\omega$ as in \eqref{FormExpansion} and an arbitrary
polyvector field $\alpha$ as in \eqref{beta}. We define endomorphisms of the pin bundle $S$ via:
\beqa
\gamma(\omega)=\frac{1}{k!}\omega_{a_1\ldots a_k}\gamma^{a_1\ldots
a_k}~~&,&~~\gamma(\nabla_m
\omega)=\frac{1}{k!}(\nabla_m\omega)_{a_1\ldots a_k}\gamma^{a_1\ldots
a_k} ~~,\\ \gamma(\alpha)=\frac{1}{k!}\alpha^{a_1\ldots
a_k}\gamma_{a_1\ldots a_k}~~&,&~~\gamma(\nabla_m
\alpha)=\frac{1}{k!}(\nabla_m\alpha)^{a_1\ldots a_k}\gamma_{a_1\ldots
a_k}~~.
\eeqa
Proposition 2 implies the following:

\paragraph{Proposition 3.} $\nabla^S$ satisfies the following identities for any differential form $\omega$ and
any polyvector field $\alpha$:
\beqa
\left[\nabla^S_m, \gamma(\omega) \right]_{-,\circ}&=& \gamma(\nabla_m \omega)~~,\nn\\
\left[\nabla^S_m, \gamma(\alpha) \right]_{-,\circ}&=& \gamma(\nabla_m \alpha)\nn~~.
\eeqa
\noindent {\bf Proof.} These identities follow by using Proposition 2, equations \eqref{alpha_cov} and \eqref{beta_cov}
as well as an obvious relabeling of dummy indices.

\paragraph{Observation.} The first identity in Proposition 3 is the well-known statement that $\nabla^S$ is a Clifford
connection on $S$ in the sense typically used in spin geometry (see, for example, \cite{BGV}).

\section{Component approach to pinor bilinears}
\label{sec:direct}

In this Appendix, we show that the abstract equations \eqref{AlgConstraints} and \eqref{FierzFlatness} are equivalent
with a set of equations which were found in \cite{MartelliSparks} via component calculations
pertaining to the particular case considered in loc. cit.

\subsection{Alternate form of the algebraic constraints}
\label{sec:Qdirect}

Taking linear combinations of \eqref{AlgebraicConstraints}
(equivalently, using the fact that $ \cK(L_{\check{Q}})\cap \cK(R_{\tau_\cB(\check{Q})})=
\cK(L_{\check{Q}}+R_{\tau_\cB(\check{Q})})\cap \cK(L_{\check{Q}}-R_{\tau_\cB(\check{Q})})$) shows that the algebraic
constraints derived in Section \ref{sec:fierz} can also be written in
the following form:
\be
\cB(\xi,\left(Q^t\circ \gamma_{a_1\ldots a_k}\pm \gamma_{a_1\ldots a_k}\circ Q\right) \xi') =
\xi^t\left(Q^t\circ \gamma_{a_1\ldots a_k}\pm \gamma_{a_1\ldots a_k}\circ Q\right) \xi'=0~~.
\ee
These equations are equivalent --- for the particular case considered
there --- with a set of conditions used in
\cite{MartelliSparks}. Starting from equation (2.26) of that
reference, let us show that the `useful relations' of \cite[Appendix~C]{MartelliSparks}
are equivalent with our algebraic constraints. Recall that
\cite{MartelliSparks} deals with the case of Majorana spinors $\xi$ in
eight Euclidean dimensions, a case which was also the subject of our
application in Section \ref{sec:application}. As explained in that
Section, we prefer to work directly with $\xi$ rather than with the
quantities $\varepsilon_+=\frac{1}{\sqrt{2}}\xi$ and
$\varepsilon_-=\frac{1}{\sqrt{2}}\gamma^{(9)}\xi$ used in loc. cit.
(which provide a redundant parameterization of $\xi$).

When expressed in terms of $\xi$, equation (2.26) of \cite{MartelliSparks} is equivalent with
$Q\xi=0$, where $Q$ is given in \eqref{Qappl}. Choosing a local frame of $S$, we can think of $Q$ as a locally-defined matrix-valued function and 
of $\xi$ as a column matrix with entries given by locally-defined smooth functions. When $Q\xi=0$, we also
have ${\xi}^t Q^t=0$, where:
\be
Q^t=\frac{1}{2}(\partial_n \Delta)\gamma^n +
\frac{1}{6}f_n \gamma^n\circ  \gamma^{(9)} -\frac{1}{288}F_{npqr}\gamma^{npqr} -\kappa \gamma^{(9)}~.\nn
\ee
The two equations $Q\xi=0$ and ${\xi}^t Q^t=0$ imply the relations:
\beqan
{\xi}^t T \circ Q \xi~ &=& 0~, \label{eq1} \\
{\xi}^t Q^t \circ T \xi &=& 0 ~,\label{eq2}
\eeqan
where $T \in \Mat(16,\R)\approx \Cl(8,0)$ is a general Clifford
matrix. Using relations \eqref{epsxi} and the fact that $\gamma^{(9)}$
anticommutes with $\gamma^1,\ldots,\gamma^8$, it is easy to check that
the `useful relations' (C.1)-(C.3) given in Appendix C of
\cite{MartelliSparks} take the following form when expressed in terms
of $\xi$:
{\footnotesize \beqan
\!\!\!\! \!\!\! \!\!\!\! \frac{1}{288} F_{pqrs}\xi^t[\gamma^{pqrs},T]_{\mp}\xi - \frac{1}{2} (\partial_m\Delta) \xi^t[\gamma^m,T]_{\mp}\xi
    + \kappa \xi^t[\gamma^{(9)},T]_{\mp}\xi -\frac{1}{6}f_m \xi^t[\gamma_m\gamma^{(9)},T]_{\pm}\xi = 0~~,~~~~~~~~~~\label{useful2}~ \\
\!\!\!\! \frac{1}{288} F_{pqrs}\xi^t[\gamma^{pqrs},T\gamma^{(9)}]_{\mp}\xi -\frac{1}{2} (\partial_m\Delta) \xi^t[\gamma^m,T\gamma^{(9)}]_{\mp}\xi +\kappa \xi^t[\gamma^{(9)},T\gamma^{(9)}]_{\mp}\xi -
\frac{1}{6}f_m \xi^t[\gamma_m\gamma^{(9)},T \gamma^{(9)}]_{\pm}\xi = 0~.~ \label{useful3}~~~
\eeqan}
It is now clear that the first identity in \eqref{useful2} is
equivalent with the difference \eqref{eq1}-\eqref{eq2}, the second is
equivalent with the sum \eqref{eq1}+\eqref{eq2}, while the two
identities in \eqref{useful3} are equivalent with $\eqref{eq1} \mp
\eqref{eq2}$ where $T$ is replaced by $T\gamma^{(9)}$.  One can easily
check that the third and fourth identities are not independent, being
equivalent with the first two. We conclude that the useful relations
of \cite{MartelliSparks} are equivalent with the particular
incarnation of our algebraic constraints for the case considered in
loc. cit.

\subsection{Alternate derivation of the differential constraints}
\label{sec:DiffDirect}

For simplicity, let us consider only the case $\K=\R$. Recall that the pin bundle $S$ of a pseudo-Riemannian manifold $(M,g)$ is endowed with
admissible bilinear pairings $\cB$, which, in particular, satisfy \eqref{Bproperties}.

Since $\wf_m^t=- \wf_m$, we find that $ \wf_m$ is
anti-self-adjoint with respect to the pairing $\cB$, which means that
the pin covariant derivative $\nabla^S$ induced by the Levi-Civita connection of $(M,g)$ is
compatible with $\cB$ in the sense that this pairing is $\nabla^S$-flat:
\be
\partial_m \cB(\xi, \xi')=
\cB(\nabla^S_m \xi, \xi') + \cB( \xi, \nabla^S_m \xi')~~,~~\forall \xi, \xi'\in \Gamma(M,S)~~.
\ee
The deformed pin connection takes the form $D_m=\nabla^S_m+A_m=\partial_m+ \wf_m+A_{m}$. Since $A_m^t=-A_m$, it follows that $A_m$ (and thus also $ \wf_m+A_{m}$) is
again anti-self-adjoint with respect to the scalar product on $S$; as a consequence, the deformed pin connection is also
compatible with this pairing:
\be
\partial_m \cB( \xi, \xi')=
\cB( D_m \xi, \xi') + \cB( \xi, D_m \xi')~~,~~\forall
\xi, \xi'\in \Gamma(M,S)~~.
\ee
Replacing $\xi'$ with $T\xi'$ in the last equation (where $T\in \Gamma(M, \End_\R(S))$ is arbitrary) gives:
\be
\partial_m \cB( \xi, T\xi')=
\cB( D_m \xi, T\xi') + \cB( \xi, [D_m, T]_{-,\circ} \xi')+  \cB( \xi, T D_m \xi')~~,~~\forall
\xi, \xi'\in \Gamma(M,S)~~,
\ee
which immediately implies:

\paragraph{Lemma 2.} When $D_m\xi=D_m \xi'=0$, we have:
\be
\partial_m \cB( \xi, T \xi')=  \cB( \xi, [D_m, T]_{-,\circ} \xi')~~,~~\forall T\in \Gamma(M,\End_\R(S))~~.
\ee

\noindent Let now $\bcE^{(k)}\in \Omega^k(M)$ be a $k$-form defined through:
\be
\bcE^{(k)}_{a_1\ldots a_k}\eqdef\cB( \xi, \gamma_{a_1\ldots a_k}\xi') \Longrightarrow \bcE^{(k)}=
\cB( \xi, \gamma_{a_1\ldots a_k}\xi' ) e^{a_1\ldots a_k}~~,
\ee
where $(e_a)$ is a local pseudo-orthonormal frame of $(M,g)$ and $\gamma_a=\gamma((e_a)_\sharp)=\eta_{ab}\gamma^b$.

\paragraph{Proposition 4.} When $D_m \xi=D_m \xi'=0$, we have:
\ben
\label{diffcps}
(\nabla_m \bcE^{(k)})_{a_1\ldots a_k}=\bcE^{(k)}_{a_1\ldots a_k;m}=\cB( \xi, [A_m, \gamma_{a_1\ldots a_k}]_{-,\circ}\xi')~~.
\een

\noindent {\bf Proof.} Applying Lemma2 to $T=\gamma_{a_1\ldots a_k}$ gives:
\beqa
\partial_m \bcE^{(k)}_{a_1\ldots a_k} &=& \cB( \xi, [D _m,\gamma_{a_1\ldots a_k}]_{-,\circ}\xi')
= \cB( \xi, [\nabla^S_m,\gamma_{a_1\ldots a_k}]_{-,\circ}\xi')
+\cB( \xi, [A_m,\gamma_{a_1\ldots a_k}]_{-,\circ}\xi') \nn\\
&=&\cB( \xi,\sum_{s=1}^k\w_{m}{}^{p}{}_{a_s}\gamma_{a_1\ldots a_{s-1}~p~a_{s+1}\ldots a_k}\xi)
+\cB( \xi,[A_m, \gamma_{a_1\ldots a_k}]_{-,\circ}\xi')~~,\nn
\eeqa

\noindent where we used the second identity stated in Proposition 2 of Appendix
\ref{sec:identities}. The conclusion follows upon moving the
first term of the last expression to the left hand side and applying
 \eqref{alpha_cov}.

\subsection*{Acknowledgements}

This work was supported by the CNCS projects
PN-II-RU-TE (contract number 77/2010), PN-II-ID-PCE (contract
numbers 50/2011 and 121/2011) and PN 09 37 01 02 / 2009. 
The work of C.I.L. was also supported by the 
Research Center Program of IBS (Institute for Basic Science) in Korea
(grant CA1205-01).
C.I.L and E.M.B. thank the Center for Geometry and
Physics, Institute for Basic Science and Pohang University of Science
and Technology (POSTECH), Korea and especially Jae-Suk Park for
providing excellent conditions at various stages during the
preparation of this work, through the research visitor program
affiliated with Grant No. CA1205-1. The Center for Geometry and
Physics is supported by the Government of Korea through the Research
Center Program of IBS (Institute for Basic Science).  C.I.L. also thanks
Perimeter Institute for hospitality and for providing an excellent and
stimulating research environment during the last stages of the
preparation of this paper. Research at Perimeter Institute is
supported by the Government of Canada through Industry Canada and by
the Province of Ontario through the Ministry of Economic Development
and Innovation. C.I.L. thanks Lilia Anguelova for interest and
for stimulating discussions as well as for critical input during the
final stages of this project. I.A.C. acknowledges the student scholarship from the Dinu
Patriciu Foundation ``Open Horizons'', which supported part of her
studies.


\end{document}